\documentclass[10pt,prb,twocolumn,showpacs,preprintnumbers,amsmath,aps,longbibliography]{revtex4-1}
\usepackage{graphicx,hyperref}
\usepackage{bm}
\usepackage[caption=false]{subfig}
\usepackage{floatrow}
\usepackage{subfig}

\def\nbC{{\mathchoice {\setbox0=\hbox{$\displaystyle\rm C$}%
\hbox{\hbox to0pt{\kern0.4\wd0\vrule height0.9\ht0\hss}\box0}}
{\setbox0=\hbox{$\textstyle\rm C$}\hbox{\hbox
to0pt{\kern0.4\wd0\vrule height0.9\ht0\hss}\box0}}
{\setbox0=\hbox{$\scriptstyle\rm C$}\hbox{\hbox
to0pt{\kern0.4\wd0\vrule height0.9\ht0\hss}\box0}}
{\setbox0=\hbox{$\scriptscriptstyle\rm C$}\hbox{\hbox
to0pt{\kern0.4\wd0\vrule height0.9\ht0\hss}\box0}}}}
\def\nbQ{{\mathchoice {\setbox0=\hbox{$\displaystyle\rm
Q$}\hbox{\raise
0.15\ht0\hbox to0pt{\kern0.4\wd0\vrule height0.8\ht0\hss}\box0}}
{\setbox0=\hbox{$\textstyle\rm Q$}\hbox{\raise
0.15\ht0\hbox to0pt{\kern0.4\wd0\vrule height0.8\ht0\hss}\box0}}
{\setbox0=\hbox{$\scriptstyle\rm Q$}\hbox{\raise
0.15\ht0\hbox to0pt{\kern0.4\wd0\vrule height0.7\ht0\hss}\box0}}
{\setbox0=\hbox{$\scriptscriptstyle\rm Q$}\hbox{\raise
0.15\ht0\hbox to0pt{\kern0.4\wd0\vrule height0.7\ht0\hss}\box0}}}}
\def\nbT{{\mathchoice {\setbox0=\hbox{$\displaystyle\rm
T$}\hbox{\hbox to0pt{\kern0.3\wd0\vrule height0.9\ht0\hss}\box0}}
{\setbox0=\hbox{$\textstyle\rm T$}\hbox{\hbox
to0pt{\kern0.3\wd0\vrule height0.9\ht0\hss}\box0}}
{\setbox0=\hbox{$\scriptstyle\rm T$}\hbox{\hbox
to0pt{\kern0.3\wd0\vrule height0.9\ht0\hss}\box0}}
{\setbox0=\hbox{$\scriptscriptstyle\rm T$}\hbox{\hbox
to0pt{\kern0.3\wd0\vrule height0.9\ht0\hss}\box0}}}}
\def\nbS{{\mathchoice
{\setbox0=\hbox{$\displaystyle     \rm S$}\hbox{\raise0.5\ht0%
\hbox to0pt{\kern0.35\wd0\vrule height0.45\ht0\hss}\hbox
to0pt{\kern0.55\wd0\vrule height0.5\ht0\hss}\box0}}
{\setbox0=\hbox{$\textstyle        \rm S$}\hbox{\raise0.5\ht0%
\hbox to0pt{\kern0.35\wd0\vrule height0.45\ht0\hss}\hbox
to0pt{\kern0.55\wd0\vrule height0.5\ht0\hss}\box0}}
{\setbox0=\hbox{$\scriptstyle      \rm S$}\hbox{\raise0.5\ht0%
\hboxto0pt{\kern0.35\wd0\vrule height0.45\ht0\hss}\raise0.05\ht0%
\hbox to0pt{\kern0.5\wd0\vrule height0.45\ht0\hss}\box0}}
{\setbox0=\hbox{$\scriptscriptstyle\rm S$}\hbox{\raise0.5\ht0%
\hboxto0pt{\kern0.4\wd0\vrule height0.45\ht0\hss}\raise0.05\ht0%
\hbox to0pt{\kern0.55\wd0\vrule height0.45\ht0\hss}\box0}}}}
\def\nbZ{{\mathchoice {\hbox{$\sf\textstyle Z\kern-0.4em Z$}}
{\hbox{$\sf\textstyle Z\kern-0.4em Z$}}
{\hbox{$\sf\scriptstyle Z\kern-0.3em Z$}}
{\hbox{$\sf\scriptscriptstyle Z\kern-0.2em Z$}}}}

\usepackage{color}
\usepackage[normalem]{ulem}
\definecolor{green}{rgb}{0.0, 0.44, 0.0}
\definecolor{red}{rgb}{1.0, 0.13, 0.32}
\definecolor{blue}{rgb}{0.06, 0.2, 0.65}
\definecolor{magenta}{rgb}{1.0, 0.0, 1.00}

\newcommand{\PF}{\phi}
\newcommand{\prefac}{{\mathcal{N}_{\ell}}}

\newcommand{\bea}{\begin{eqnarray}}
\newcommand{\eea}{\end{eqnarray}}
\def\le{\left}
\def\ri{\right}

\begin{document}

\title{Point-to-set lengths, local structure, and glassiness}

\author{Sho Yaida} \email{sho.yaida@duke.edu}
\affiliation{Department of Chemistry, Duke University, Durham,
North Carolina 27708, USA}
\author{Ludovic Berthier}  \email{ludovic.berthier@univ-montp2.fr}
\affiliation{Laboratoire Charles Coulomb, CNRS-UMR 5221, Universit\'e de Montpellier, Montpellier, France}
\author{Patrick Charbonneau} \email{patrick.charbonneau@duke.edu}
\affiliation{Department of Chemistry, Duke University, Durham,
North Carolina 27708, USA}
\affiliation{Department of Physics, Duke University, Durham,
North Carolina 27708, USA}
\author{Gilles Tarjus} \email{tarjus@lptl.jussieu.fr}
\affiliation{LPTMC, CNRS-UMR 7600, Universit\'e Pierre et Marie Curie,
bo\^ite 121, 4 Pl. Jussieu, 75252 Paris c\'edex 05, France}

\date{\today}

\begin{abstract}
The growing sluggishness of glass-forming liquids is thought to be accompanied by growing structural order. The nature of such order, however, remains hotly debated. A decade ago, point-to-set (PTS) correlation lengths were proposed as measures of amorphous order in glass formers, but recent results raise doubts as to their generality. Here, we extend the definition of PTS correlations to agnostically capture any type of growing order in liquids, be it local or amorphous. This advance enables the formulation of a clear distinction between slowing down due to conventional critical ordering and that due to glassiness, and provides a unified framework to assess the relative importance of specific local order and generic amorphous order in glass formation.
\end{abstract}


\maketitle
\begin{figure*}
\centerline{
\includegraphics[width=0.22\textwidth]{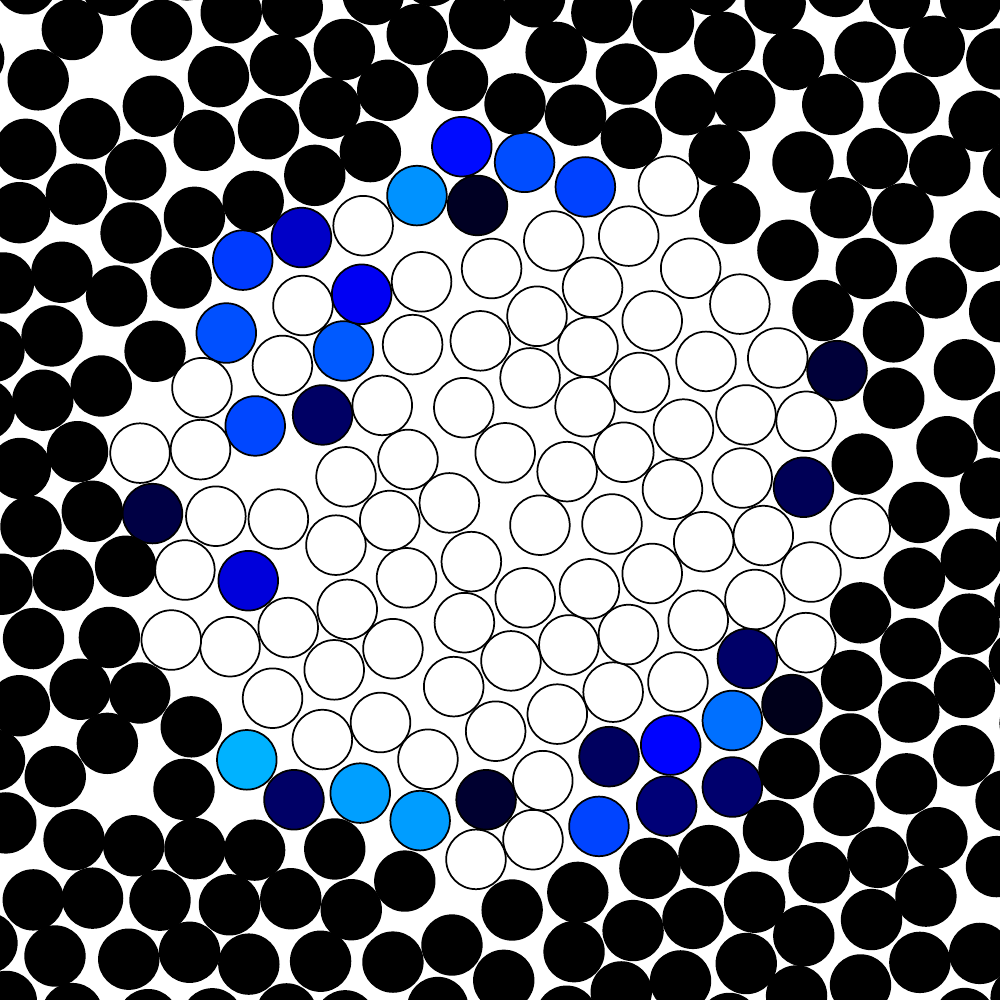}
\includegraphics[width=0.22\textwidth]{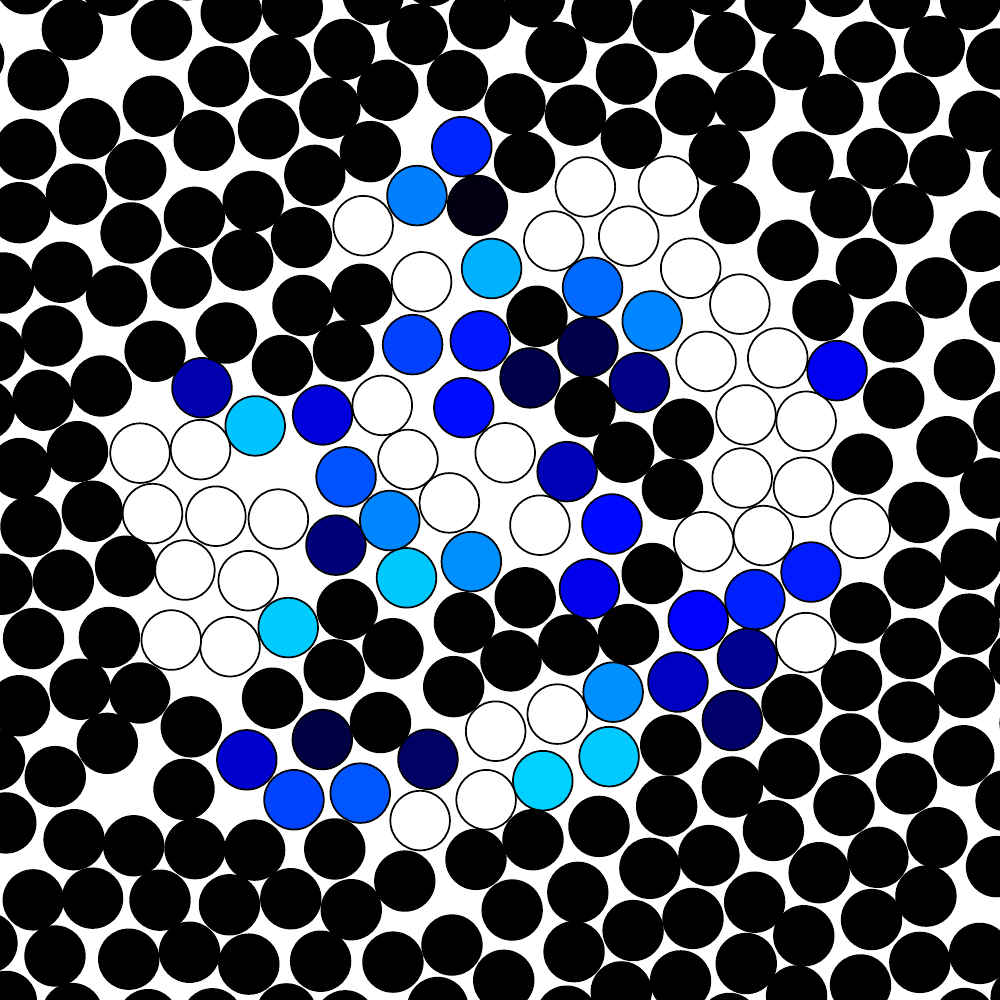}
\includegraphics[width=0.22\textwidth]{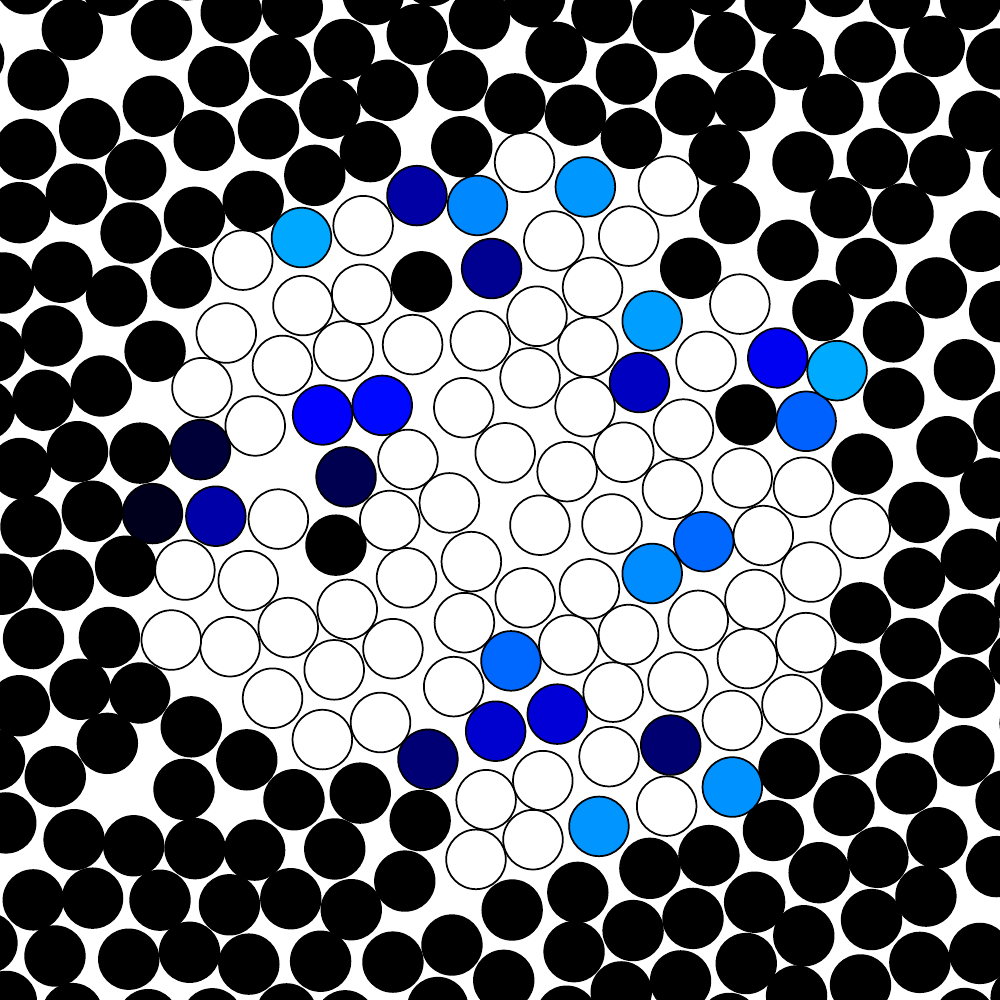}
}
\centerline{\hspace{+0.01in}
\includegraphics[width=0.22\textwidth]{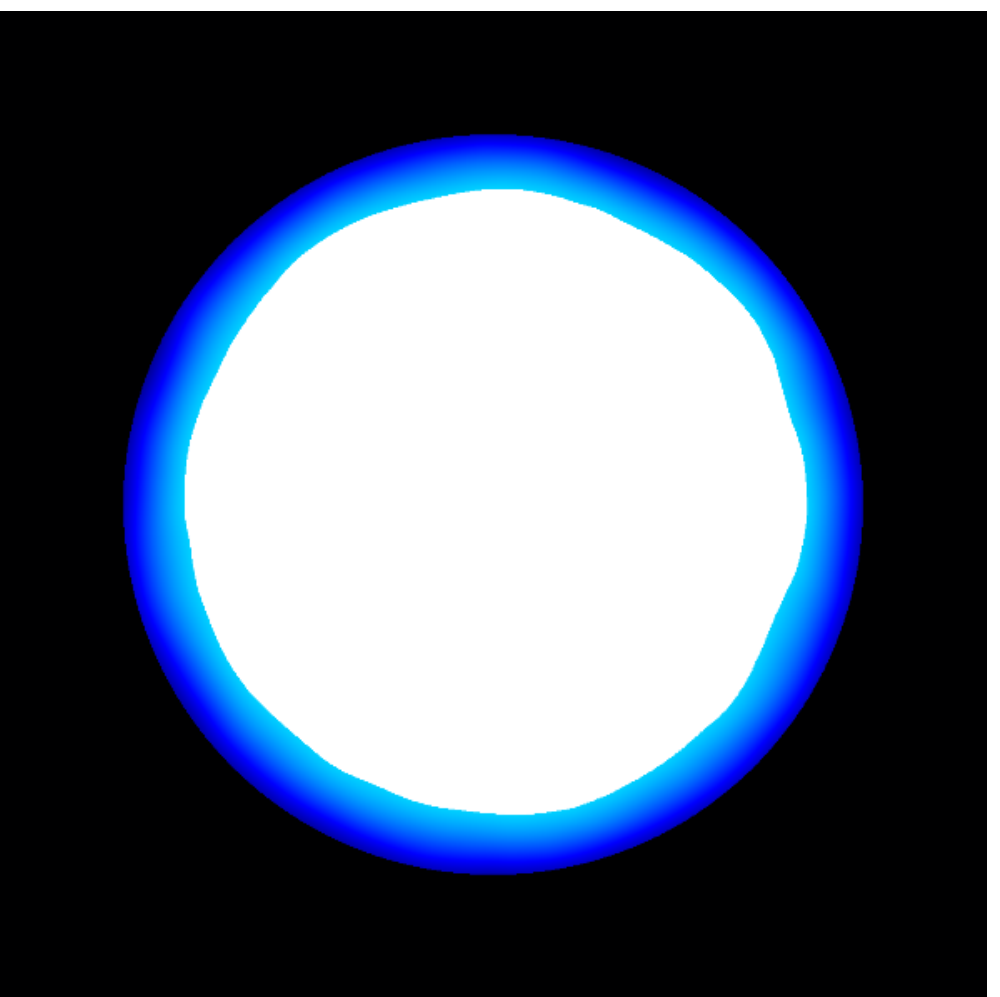}
\includegraphics[width=0.22\textwidth]{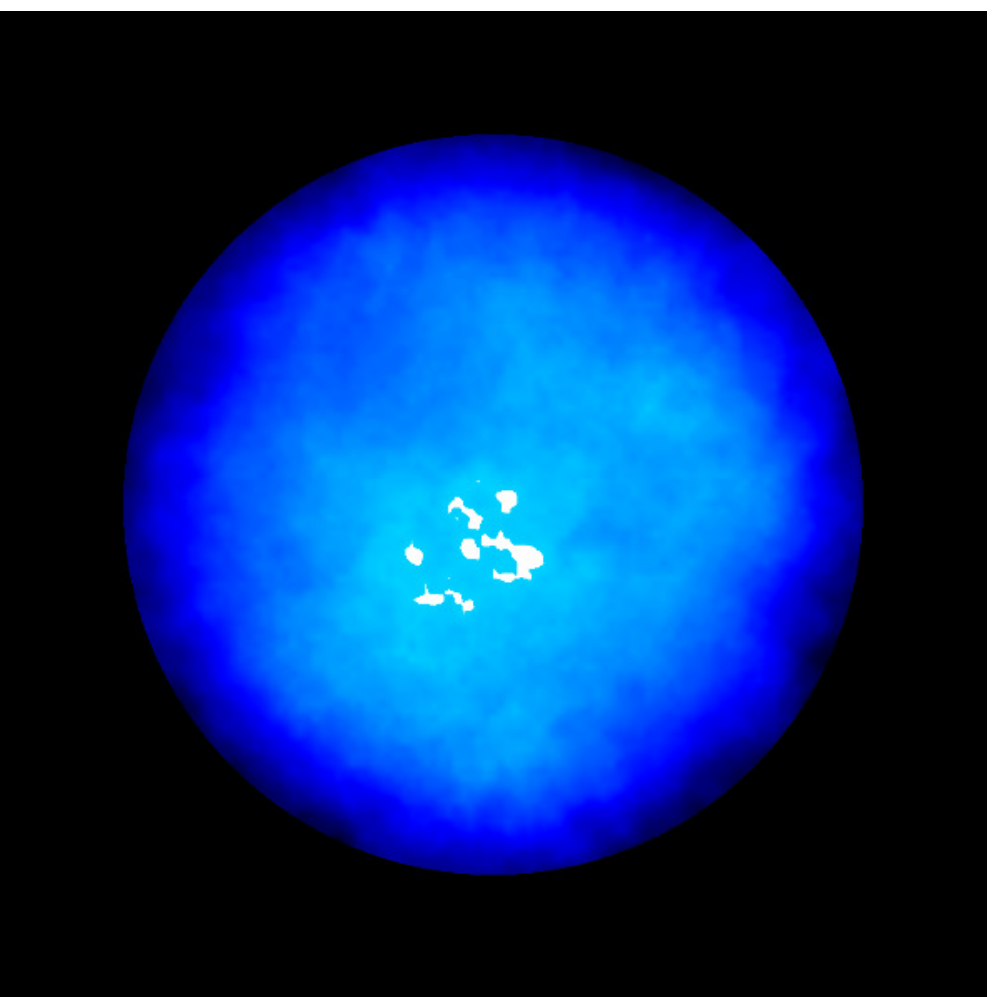}
\includegraphics[width=0.22\textwidth]{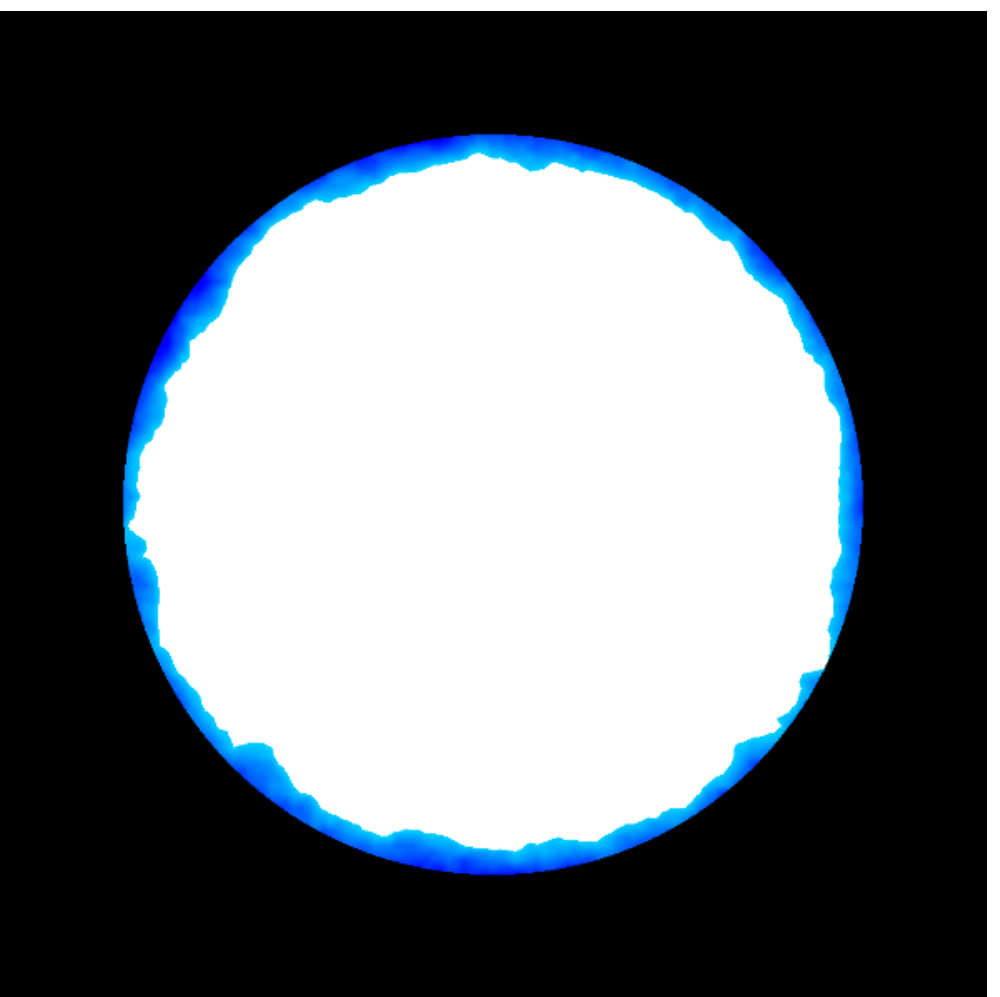}
}
\vspace{-2.7in}
\centerline{\hspace{+6.0in}
\includegraphics[width=0.08\textwidth]{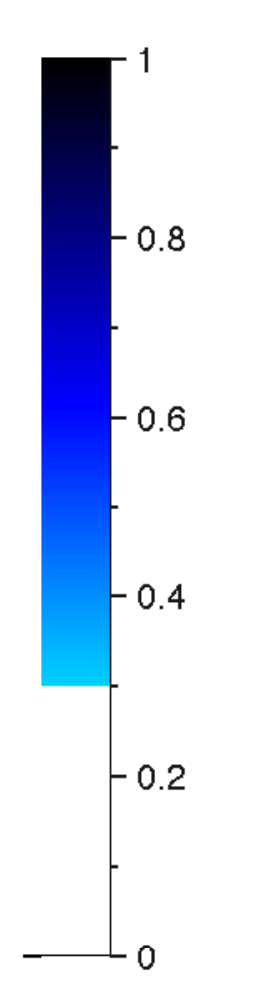}
}
\vspace{+0.5in}
\caption{
Overlaps for monodisperse hard disks at packing fraction $\PF=0.695<\PF_{\rm hexatic}$ and cavity radius $R=6.25\approx\xi_{\ell=6}^{\rm PTS}\approx2\xi_{\rm pos}^{\rm PTS}$.
The positional (left) and bond-orientational overlaps for $\ell=6$ (middle) and $8$ (right) between a specific equilibrated configuration $\textbf{Y}$ with a given reference configuration $\textbf{X}$ are evaluated at the particle centers of the former (top).
Averaging over reference and equilibrated configurations yields smooth overlap profiles (bottom).
To compare different types of overlaps, their average values for two identical configurations are rescaled to unity.
Cavity overlap profiles clearly confirm that the $\ell=6$ overlap detects the incipient hexatic order while positional and other bond-orientational overlaps are blind to it.}
\label{punch_figure}
\end{figure*}

\section{Introduction}
It is tempting to attribute the spectacular dynamical slowdown of glass-forming liquids as one lowers temperature to an increasingly collective behavior characterized by the growth of a static length. The puzzle of glass formation, however, lies in the elusive nature of such a length and of the associated spatial correlations~\cite{BB11,Ta11,WL12}. The structural changes measured by static pair density correlations, as probed by common scattering experiments, stay remarkably weak. Hence, the sought-after static correlations must be quite subtle and, as a result, hard to detect. To make matters worse, whatever the definition of the putative static length, its increase over the dynamical range accessible to computer simulations and experiments is expected to be modest -- by less than a factor of 10 -- due to the activated scaling form between the relaxation time  $\tau$ and the static length $\xi$,
\begin{equation}
\label{eq_activated_dynamic_scaling}
\log\le(\frac{\tau}{\tau_{0}}\ri)\sim B\,  \frac {\xi^\psi}{T}\;,
\end{equation}
with $\tau_0$ and $B$ being liquid-specific constants and $\psi \leq d$ an effective exponent bounded by the spatial dimension~\cite{BB11,Ta11,WL12,footnote_HS}.
It is much harder to detect large growing lengths in glass-forming liquids than in systems approaching standard second-order critical point, whereat the dynamics also slows down, but a power law relates time and length scales,
\begin{equation}
\label{eq_conventional_dynamic_scaling}
\frac{\tau}{\tau_{0}}\sim C\, \xi^z,
\end{equation}
with a dynamical exponent $z={\rm O}(1)$~\cite{HH77}.

Although this difficulty once motivated attempts to avoid making explicit reference to collective changes in static properties of glass-forming liquids~\cite{GC03,footnote_dynamic}, evidence linking static correlations to their sluggishness has since grown prevalent~\cite{KDS14}.
Two main proposals have been formulated to identify static correlations in glass formers.
(i) The first involves the spatial extent of locally preferred structure(s) as obtained, for instance, from multibody correlations associated with bond-orientational order~\cite{SNR83}. In fluids of spherical particles these bond-orientational correlations can detect polytetrahedral or icosahedral local order in $d=3$ and sixfold local order in $d=2$~\cite{Ne02}. An unfortunate drawback of this proposal is that the prevalent local order is a liquid-specific property that may be hard to access in generic molecular glass formers.
(ii)  The second considers the correlations associated with metastability~\cite{BB04,MS06,BBCGV08}, as inspired by the paradigm of a rugged free-energy landscape and by the random first-order transition (RFOT) theory of the glass transition~\cite{KTW89,WL12}. Various ways to access this type of length have been proposed~\cite{KDS09,SL11,GKPP15}, but we focus here on that relying on freezing particles outside a spherical cavity. Simply put, one probes over what distance the amorphous boundary stabilizes a metastable state. Such length scales are expected to capture an incipient \emph{amorphous order} related to the rarefaction of available metastable states--and thus a decrease in the corresponding configurational entropy--as the liquid becomes sluggish. They are associated with point-to-set (PTS) correlations that go beyond standard multibody quantities.

In this work we address two key issues. First, PTS lengths have been claimed to be order agnostic, {\it i.e.}, with no need to specify the type of order potentially growing in the system, be it local or amorphous. If true, this appears as a clear strength of such observables. Yet, recent work by Russo and Tanaka~\cite{RT15} shows that the commonly implemented method of studying PTS correlations is unable to track the growth of sixfold local order in a two-dimensional hard-disk model. They concluded that such correlations are irrelevant to slow dynamics in this system and therefore cannot be order agnostic~\cite{RT15}. Second, behind the two proposals for defining relevant static length scales are often two lines of research that seem largely at odds and often ignore each other. On the one hand, PTS correlations have become an important tool for theorists aiming at assessing the validity of the mean-field description of the glass transition and of the RFOT theory for glass-forming liquids in two and three dimensions~\cite{BB09,BB11}. On the other hand, explanations of glass formation based on the growth of some specific local order are prevalent among soft-matter~\cite{RW15} and metallic-glass scientists~\cite{EPZK07,STNKY16}. Since, for the reasons given above, length scales do not grow large in physical glass formers, it is hard to disentangle the two explanations and ascertain if the collective behavior underlying glass formation is primarily due to specific local order or generic amorphous order.

We resolve the first point by putting forward the following central idea: to agnostically capture incipient ordering {\it PTS correlations must take into account all degrees of freedom that are potentially relevant to describe configurations at a coarse-grained level}, as one would do in crystallography.
For liquids, one should thus extend PTS correlations to (i) positional, (ii) bond-orientational, and (for molecular liquids) (iii) orientational degrees of freedom. (Similarly, a crystalline profile is defined by the equilibrium positions of the atoms on an underlying lattice, modulo small displacements due to vibrations and permutations of identical atoms, and the lattice itself is characterized according to translational and orientational symmetries.) In order to validate this proposal, we perform a computer study of several slowly relaxing liquids, including two-dimensional hard disks heading toward quasi-long-range ordered phases and a canonical three-dimensional glass former (see Appendix~\ref{sec:models}). Our results show that properly defined PTS correlations capture whatever order is growing in a liquid, be it generic amorphous order or a more specific local order (Fig.~\ref{punch_figure}).

Concerning the second issue, we show that, based on the behavior of PTS correlations, one can unambiguously disentangle 
glassiness from critical slowing down due to ordering. This further leads to a natural taxonomy of relaxation slowdown in liquids and provides a unified framework to assess the relative importance of specific local order and generic amorphous orders in glass-forming liquids. 

\begin{figure*}
\vspace{-0.05in}
\centerline{
\vspace{-0.05in}
\subfloat[\ \ \ \ \ \ \ \ \ \ \ \ \ \ \ \ \ \ \ \ \ \ \ \ \ \ \ \ \ ]{\hspace{-0.00in}\includegraphics[width=0.25\textwidth]{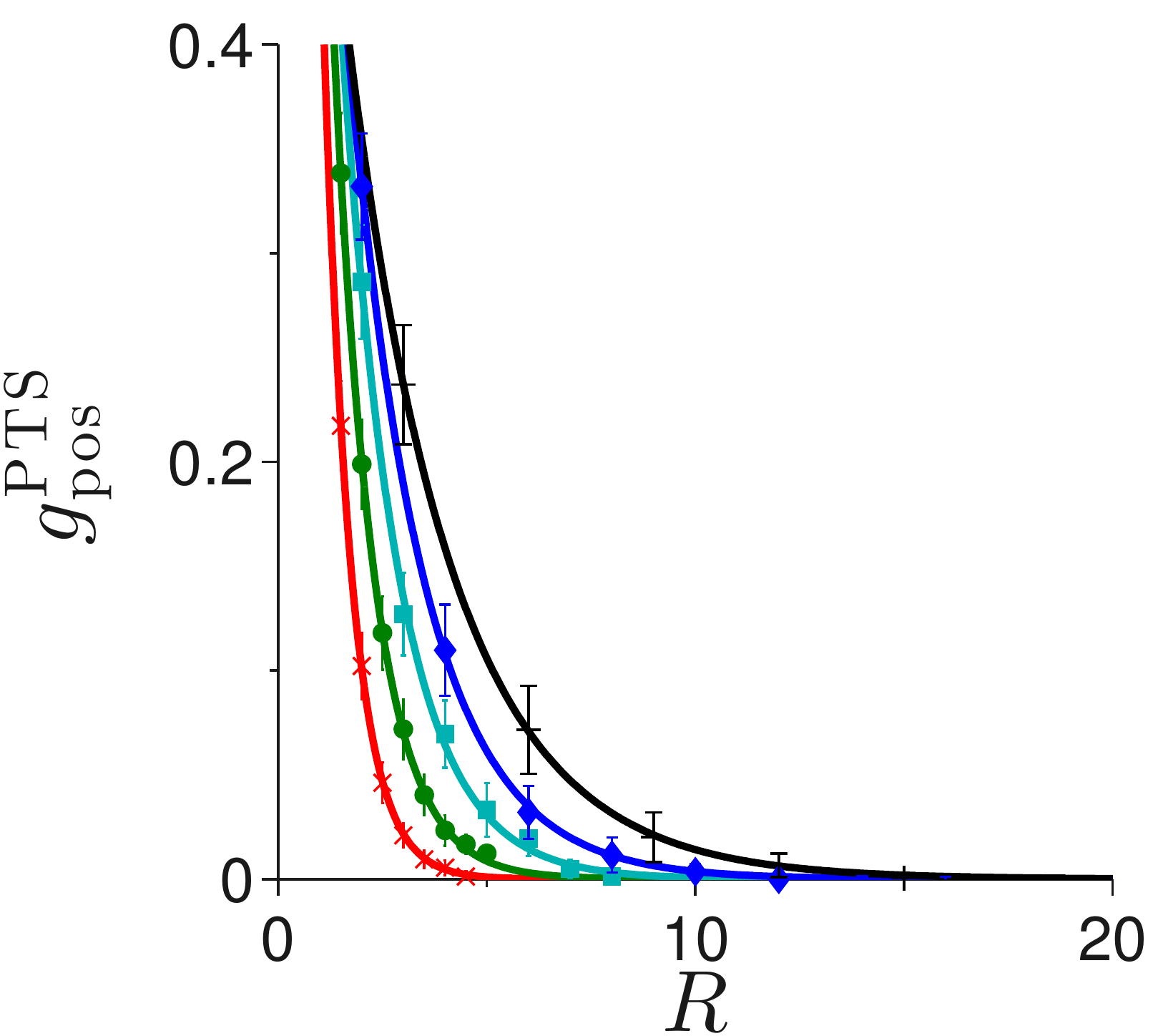}}
\subfloat[\ \ \ \ \ \ \ \ \ \ \ \ \ \ \ \ \ \ \ \ \ \ \ \ \ \ \ \ \ ]{\hspace{-0.00in}\includegraphics[width=0.25\textwidth]{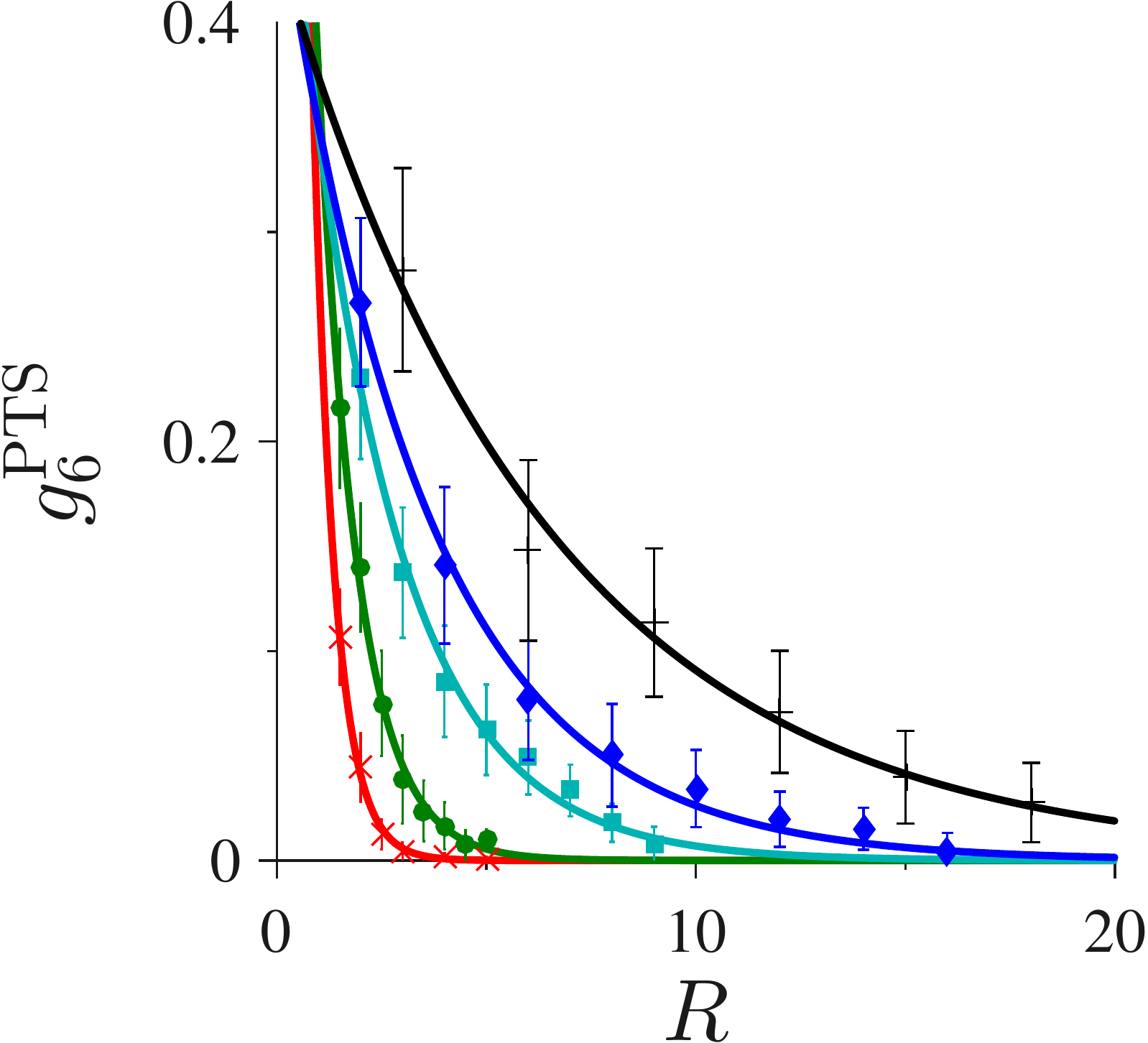}}
\subfloat[\ \ \ \ \ \ \ \ \ \ \ \ \ \ \ \ \ \ \ \ \ \ \ \ \ \ \ \ \ ]{\hspace{-0.00in}\includegraphics[width=0.25\textwidth]{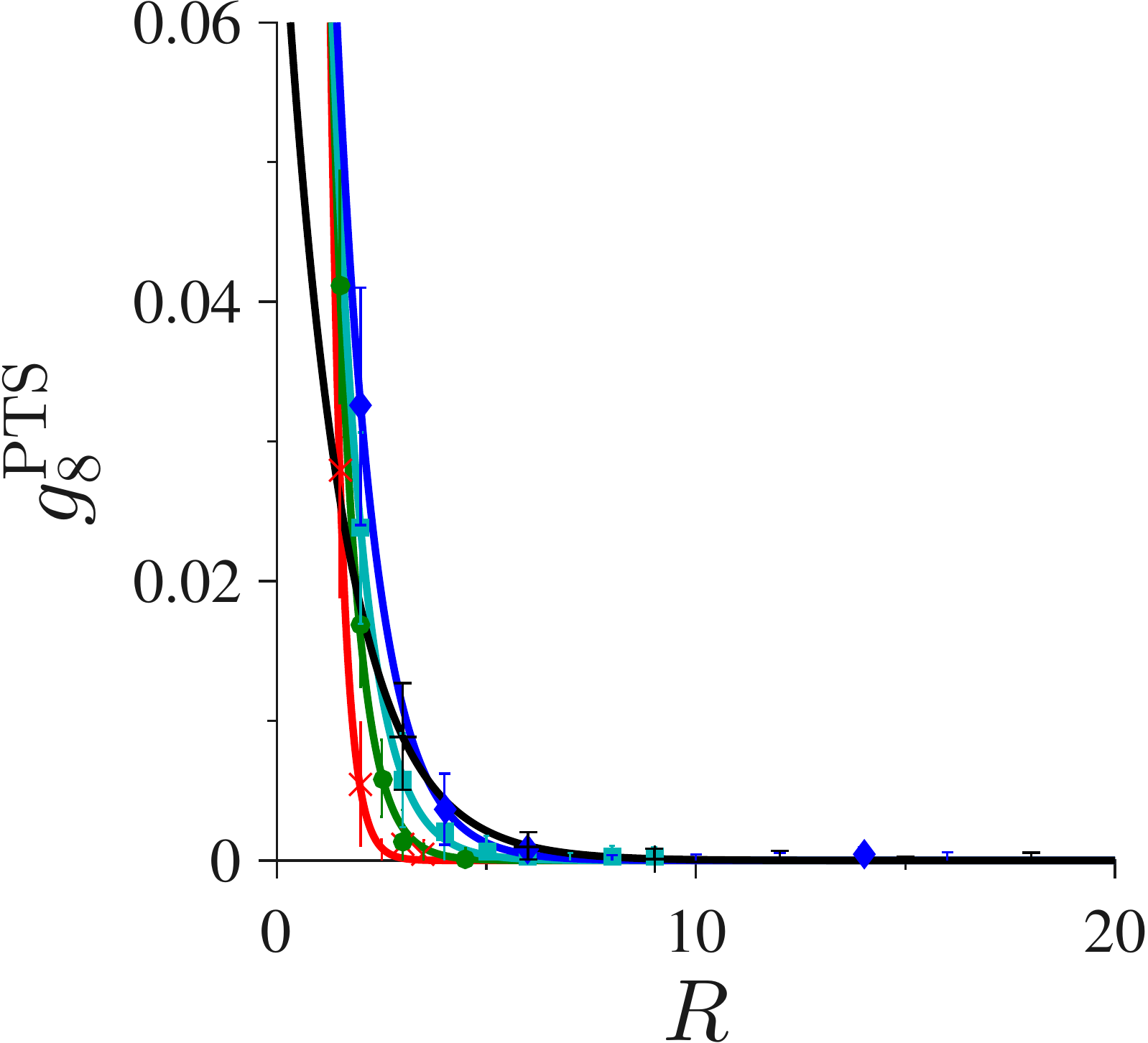}}
\subfloat[\ \ \ \ \ \ \ \ \ \ \ \ \ \ \ \ \ \ \ \ \ \ \ \ \ \ \ \ \ \ \ ]{\hspace{-0.00in}\includegraphics[width=0.25\textwidth]{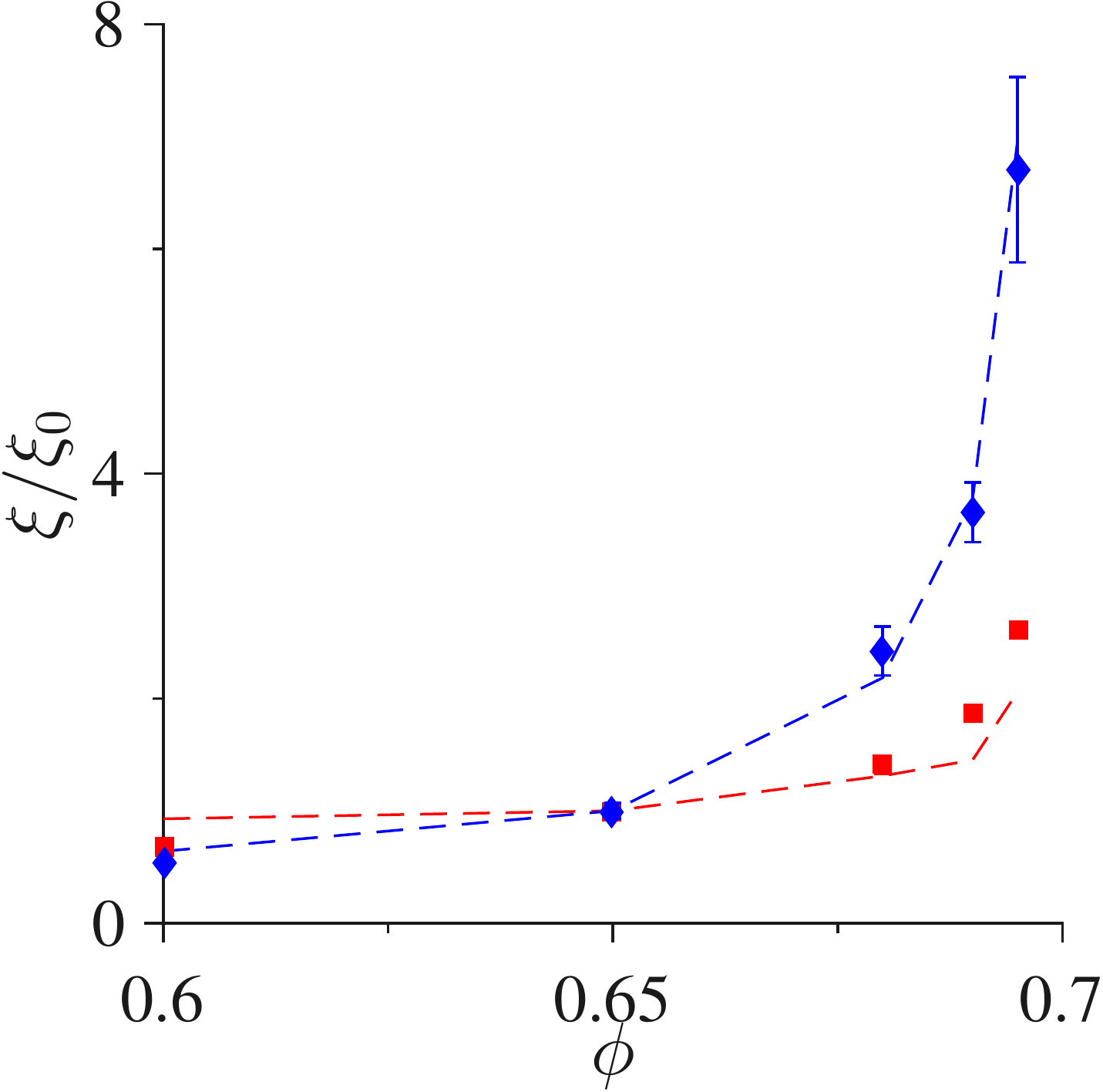}}
\vspace{-0.05in}
}
\caption{Decay of the PTS correlations with cavity radius $R$ for monodisperse hard disks: (a) positional $g_{\mathrm{pos}}^{\mathrm{PTS}}(R)$ and [(b) and (c)] bond-orientational $g_{\ell}^{\mathrm{PTS}}(R)$ for $\ell=6,8$  at packing fractions $\PF = 0.600$ (red-cross), $0.650$ (green-circle), $0.680$ (cyan-square), $0.690$ (blue-diamond), and $0.695$ (black-plus).
Solid lines are exponential fits.
Note that the $y$-axis range is an order of magnitude larger for $\ell = 6$ than for $\ell=8$.
(b) The growth of positional (red-square) and hexatic (blue-diamond) PTS correlation lengths extracted through the exponential fits track the correlation lengths (dashed lines) extracted from the radial and coarse-grained $\ell=6$ correlation functions, respectively. Lengths are relative to the results for $\xi_0\equiv\xi(\PF_0=0.650)$.}
\label{fig_BOoverlap_2D}
\end{figure*}

\section{Extended set of PTS correlations}
PTS correlations are studied by first pinning a fraction of the particles in equilibrium configurations, then letting the rest of the system explore phase space in presence of this constraint, and, finally, measuring the degree of similarity (or overlap) between a new equilibrium configuration and the initial one~\cite{BB04,BBCGV08,BK12,CCT12,HMR12}. What had not been previously appreciated is that PTS correlations are defined not only by a pinning protocol~\cite{BK12} but also by the degrees of freedom considered in assessing the similarity of configurations. For spherical particles studied here, we consider positional as well as bond-orientational overlaps, and freeze particles outside a  cavity of radius $R$ in order to ensure a proper localization of both degrees of freedom within specific states~\cite{footnote_pinning}.

Positional overlap is defined in terms of the particle density by computing the average overlap $\le[\langle Q_{\rm pos}\rangle\ri](R)$. Bond-orientational overlap could be defined using a bond density, but it is more convenient to project that bond density onto circular (in $d=2$) or spherical (in $d=3$) harmonics of rank $\ell$ and study overlaps defined for a range of $\ell$ ($\ell=1$-$16$ is typically sufficient; higher harmonics get increasingly noisy).
A bond-orientational overlap field of rank $\ell$ between the reference configuration $\mathbf X$ and a configuration $\mathbf Y$ equilibrated in the presence of the frozen particles is then 
\begin{equation}
\label{eq_BOO_PTS}
Q_\ell^{\mathbf X \mathbf Y}(\mathbf r)\equiv\prefac\sum_{m} \le\{\psi_{\ell,m}^{\mathbf X}(\mathbf r)\ri\}^*\psi_{\ell,m}^{\mathbf Y}(\mathbf r)\, ,
\end{equation}
where $\psi_{\ell,m}^{\mathbf X}(\mathbf r)$ is the local bond-orientational order parameter, and both the summation over $m$ and the normalization $\prefac$ are $d$-dependent (see Appendix~\ref{sec:PTSdef}).
The mean overlap of rank $\ell$, $\le[\langle Q_\ell\rangle\ri](R)$, is then the average of $Q_\ell^{\mathbf X \mathbf Y}$ over the configuration $\mathbf Y$, and over the reference configuration $\mathbf X$. We also define $g^{\mathrm{PTS}}(R)\equiv \le[\langle Q\rangle\ri]( R)-\le[\langle Q\rangle\ri](\infty)$ for both positional and bond-orientational PTS correlations.

Note that what enters our definition of PTS correlations are not microscopic configurations {\it per se}, but {\it spatially and orientationally coarse-grained} configurations, because approximating the notion of states or density profiles requires averaging over vibrations~\cite{BC14}.
Hence, while bond-orientational and positional degrees of freedom are entirely tied up at the purely microscopic level, upon coarse-graining they become partly independent fields.

\section{Local order through PTS correlations}
In order to show that this extended set of PTS correlations can indeed detect growing local order, we first consider a system of monodisperse hard disks. This system is known to order first into a hexatic phase at a packing fraction $\PF_{\rm hexatic}=0.700(1)$ and then into a hexagonal phase at $\PF_{\rm hexagonal}=0.716(2)$~\cite{HN78,NH79,Yo79,BK11}.
The local order in the liquid is the sixfold bond-orientational order, which can be measured from the decay of the standard correlation function of the local hexatic order parameter, $g_6( r)=\langle\psi_{6}(\mathbf 0)^*\psi_{6}(\mathbf r)\rangle$.  Although the liquid-hexatic transition has been recently shown to be weakly first order for this model~\cite{BK11}, the associated correlation length, $\xi_6$, nonetheless grows rapidly and significantly upon approaching $\PF_{\rm hexatic}$.~\cite{footnote_KTNHY} (By weakly first-order transition, we mean a transition with a small jump of the order parameter and a large yet finite correlation length.) Meanwhile, the positional length $\xi_{\rm pos}$ extracted from the two-point radial correlation function $g(r)$ -- and bond-orientational length $\xi_{\ell}$ extracted from the ``two-point'' bond-orientational correlation function $g_{\ell}(r)$ for $\ell$ incompatible with the hexatic order -- grows much more mildly (see Appendix~\ref{sec:2ptfunct}).

Let us now examine the behavior of positional and bond-orientational PTS correlations in this liquid (Fig.~\ref{fig_BOoverlap_2D} for $\ell=6,8$ and Fig.~\ref{fig_BOoverlap_2D_additional} in Appendix for $\ell=1$-$16$). As also observed by Russo and Tanaka~\cite{RT15}, we find that the positional PTS correlation length grows only slightly, staying of the order of $\xi_{\rm pos}$. However, bond-orientational PTS correlations for $\ell=6$ extend over longer distances as $\PF$ increases, and perfectly track the growth of the hexatic order. Interestingly, one also notes that bond-orientational PTS correlations for $\ell\neq 6,12$ show no distinctive features of the growing hexatic order. When systematically investigated, there is thus no need to \textit{a priori} know or guess the symmetry of the incipient local order. In other words, when properly extended to account for all degrees of freedom, PTS correlations capture the full extent of the local order in a liquid (see Fig.~\ref{punch_figure}).

\section{Amorphous order through PTS correlations}
\label{sec:glassvscritical}

We next investigate a canonical $d=3$ glass-forming liquid, the Kob-Andersen binary Lennard-Jones (KABLJ) model (see Appendix~\ref{sec:models}). Its local order, based on arrangements in the form of  bicapped square antiprisms (also denoted as $(0,2,8)$ polyhedra~\cite{CP07,CTR15}), is known to be strongly frustrated and does not correlate over large distances as temperature decreases~\cite{RMDP15}.
One expects that the symmetry of the preferred local arrangement again leads to nonzero projections on at least some of the bond-orientational local order parameters.
We therefore systematically compute the standard bond-orientational correlation functions $g_\ell(r)$ and contrast them with both positional and bond-orientational PTS correlations.

\begin{figure*}
\vspace{-0.05in}
\centerline{\hspace{-2.2in}
\sidesubfloat[]{\hspace{-0.06in}\includegraphics[width=0.2\textwidth]{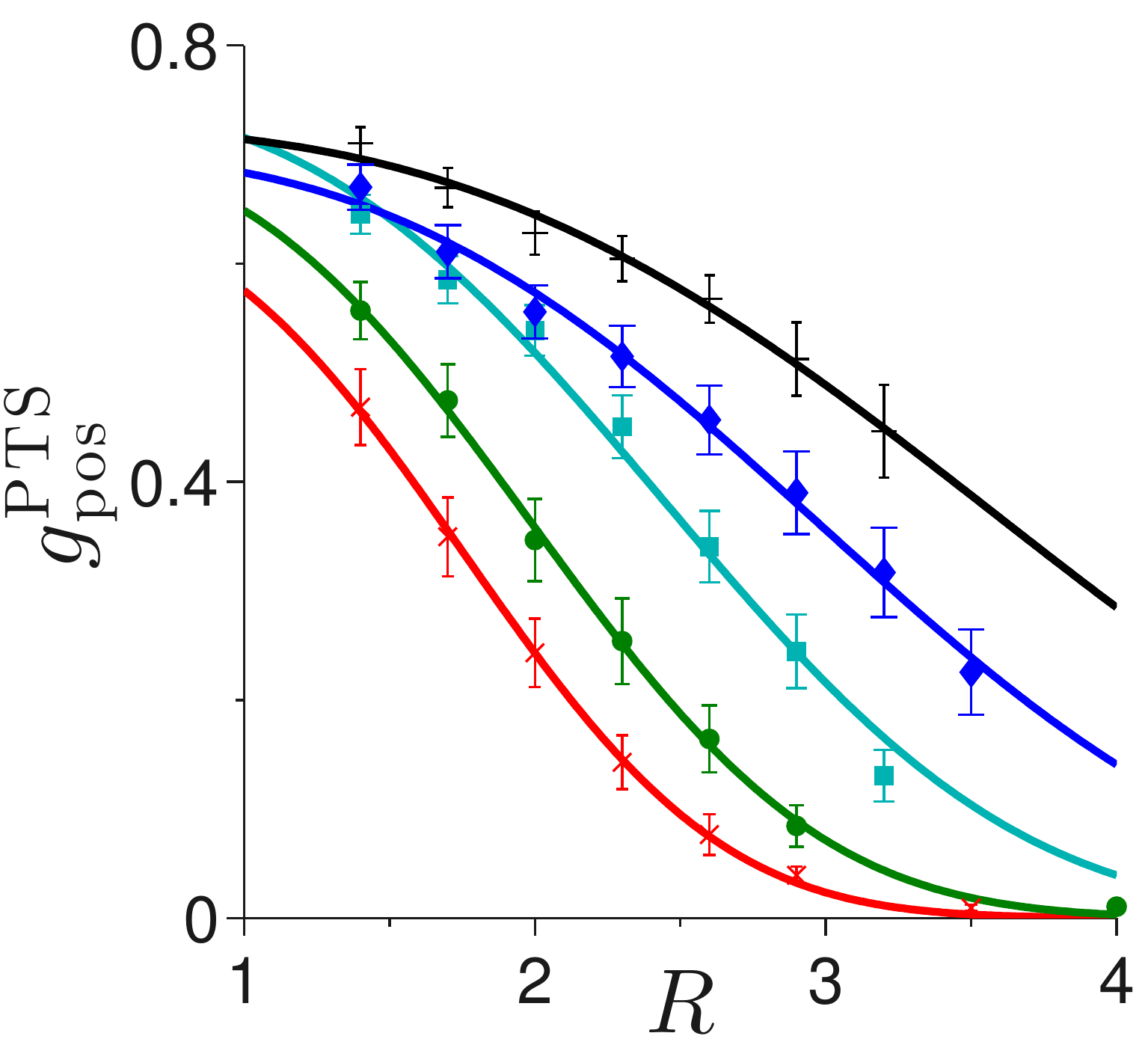}}\hspace{+0.01in}
\sidesubfloat[]{\hspace{-0.06in}\includegraphics[width=0.2\textwidth]{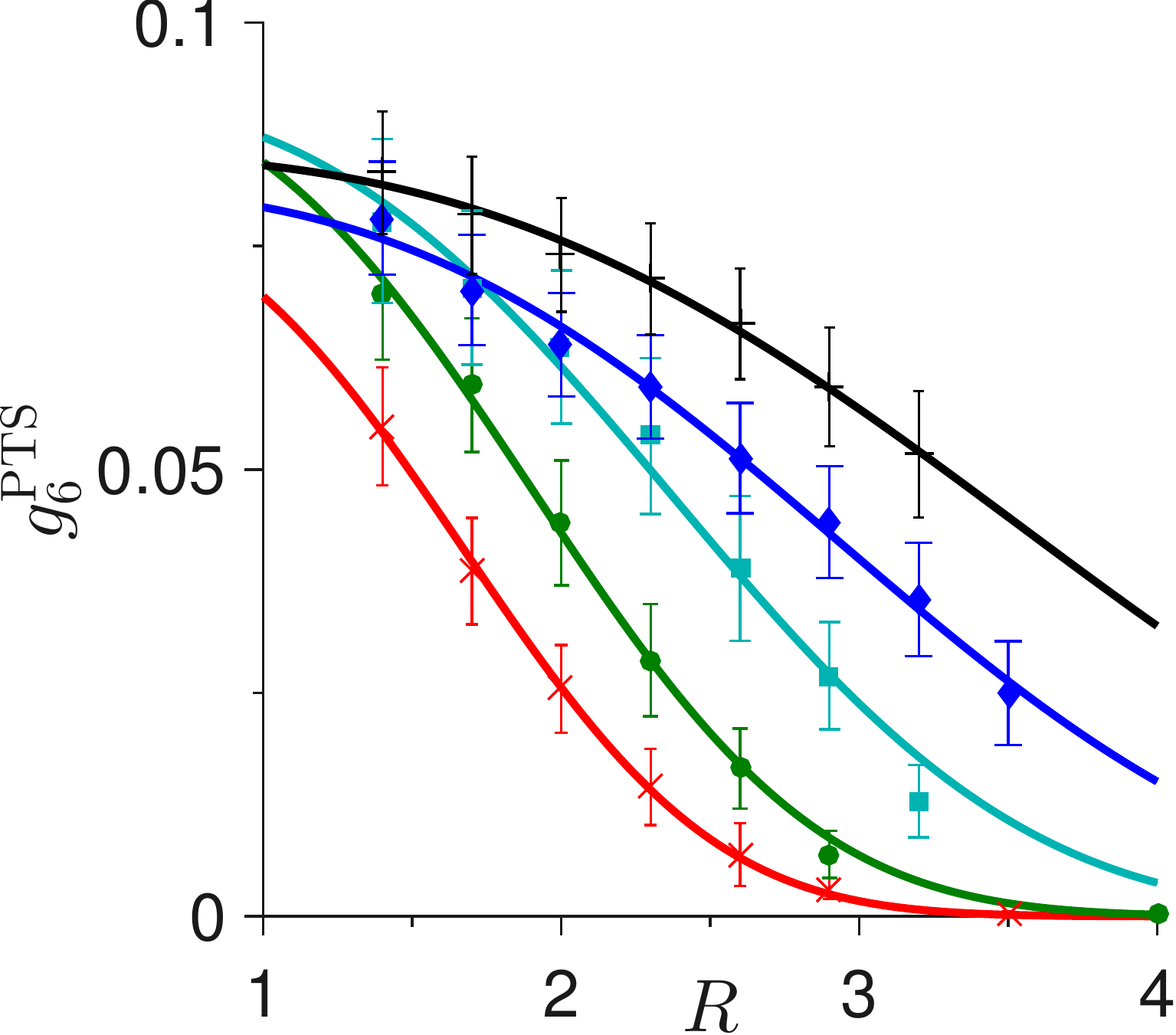}}\hspace{+0.01in}
\sidesubfloat[]{\hspace{-0.06in}\includegraphics[width=0.2\textwidth]{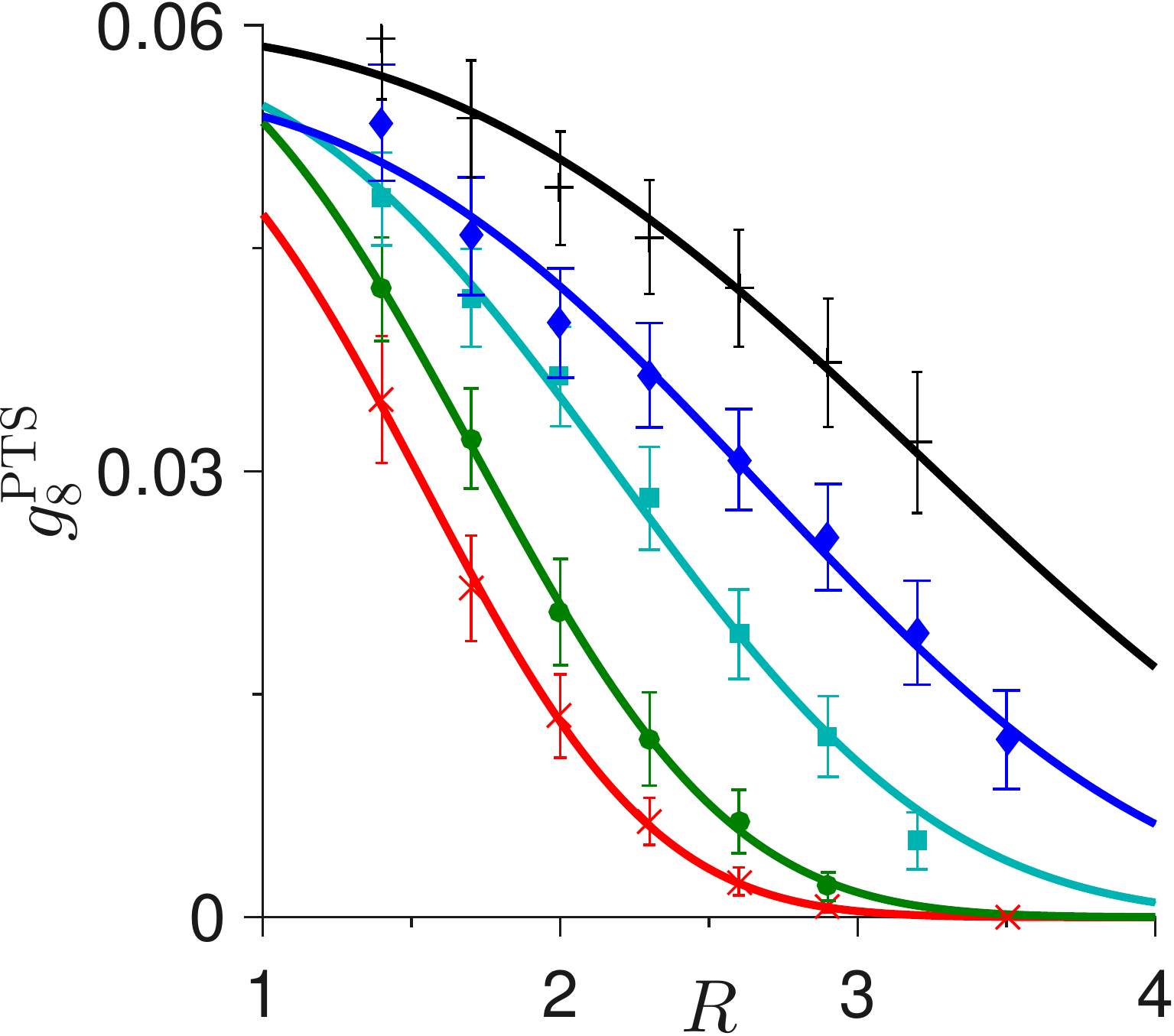}}\hspace{+0.01in}
}
\centerline{\hspace{-2.2in}
\sidesubfloat[]{\hspace{-0.06in}\includegraphics[width=0.2\textwidth]{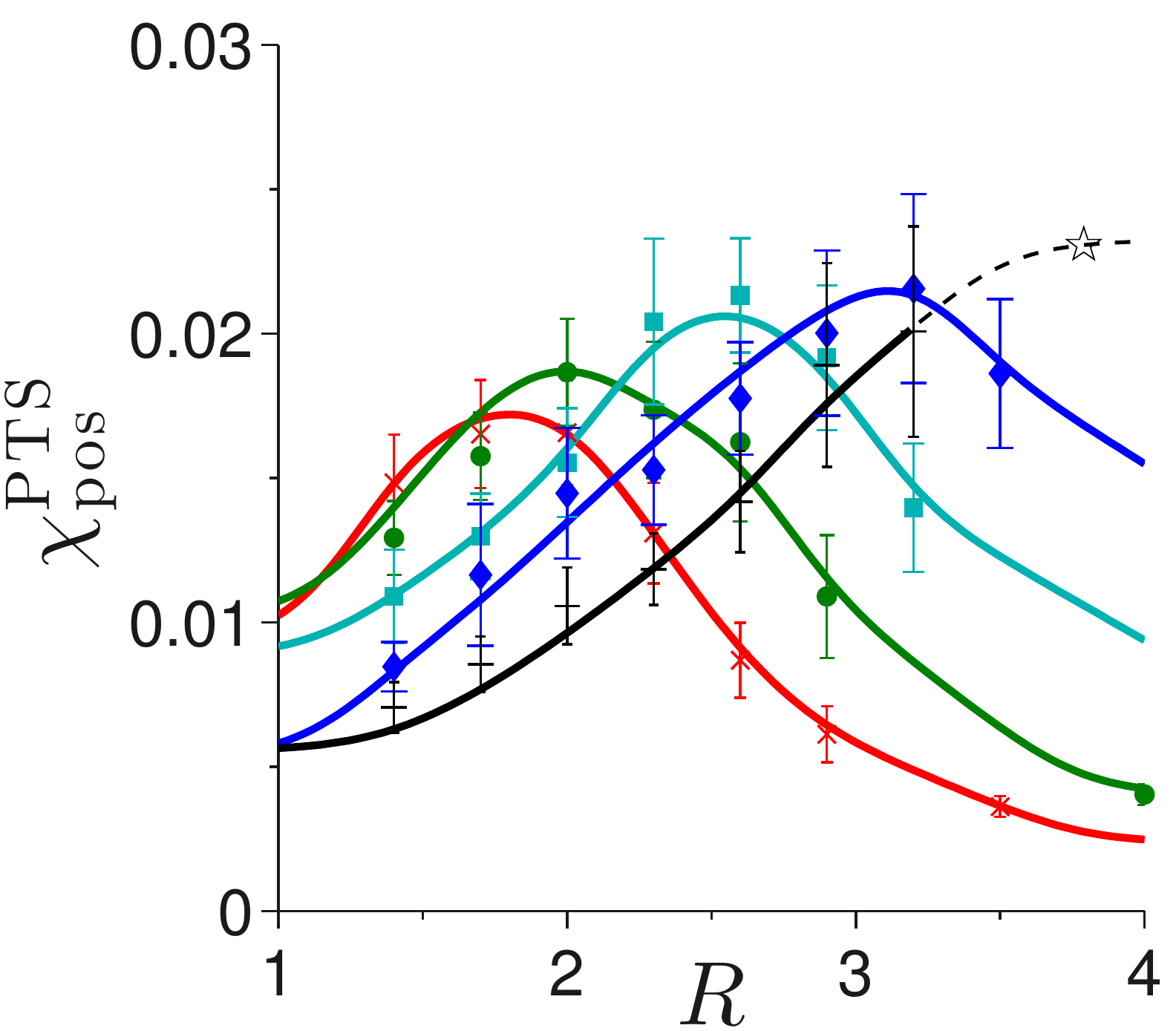}}\hspace{+0.01in}
\sidesubfloat[]{\hspace{-0.06in}\includegraphics[width=0.2\textwidth]{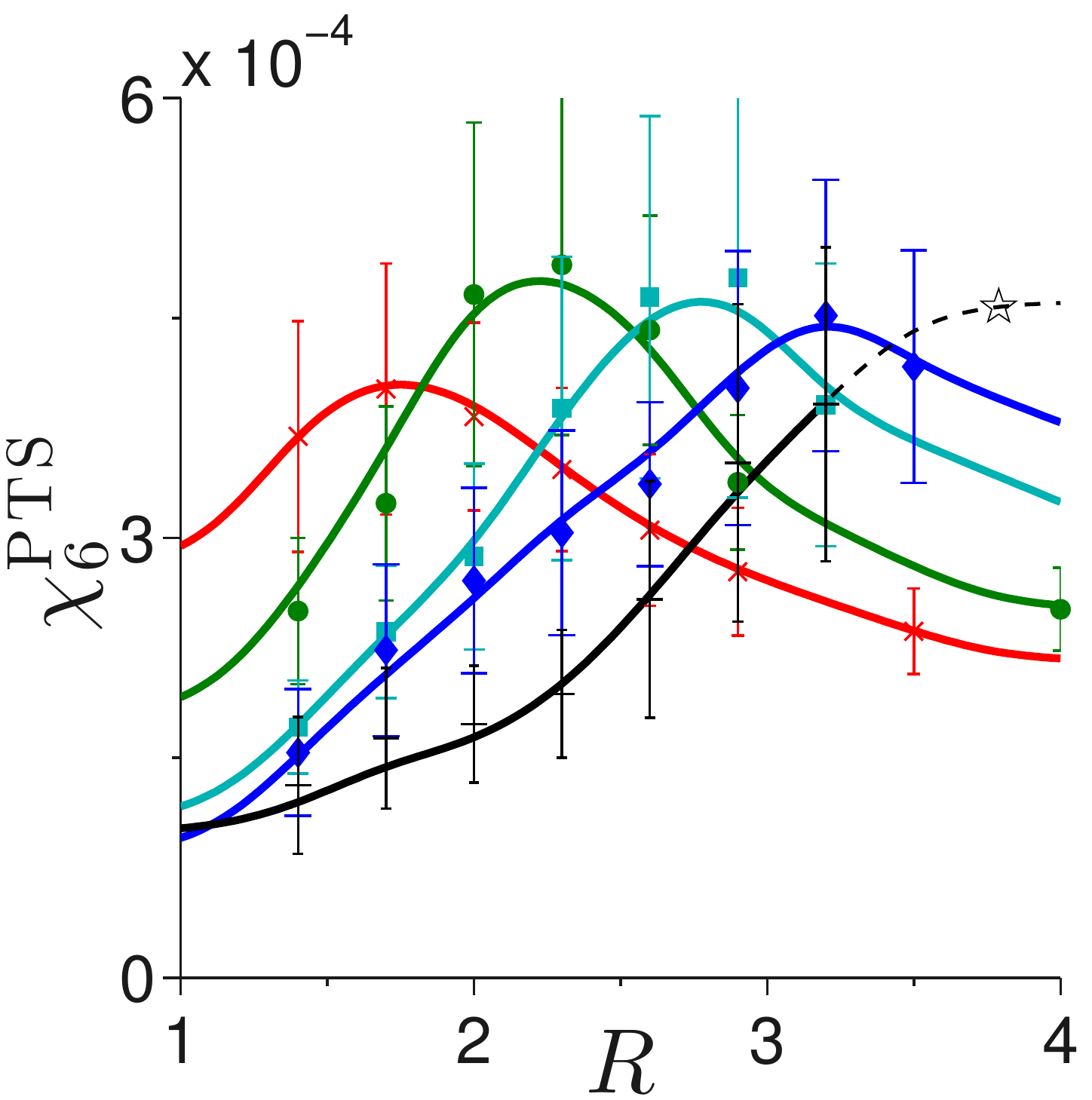}}\hspace{+0.01in}
\sidesubfloat[]{\hspace{-0.06in}\includegraphics[width=0.2\textwidth]{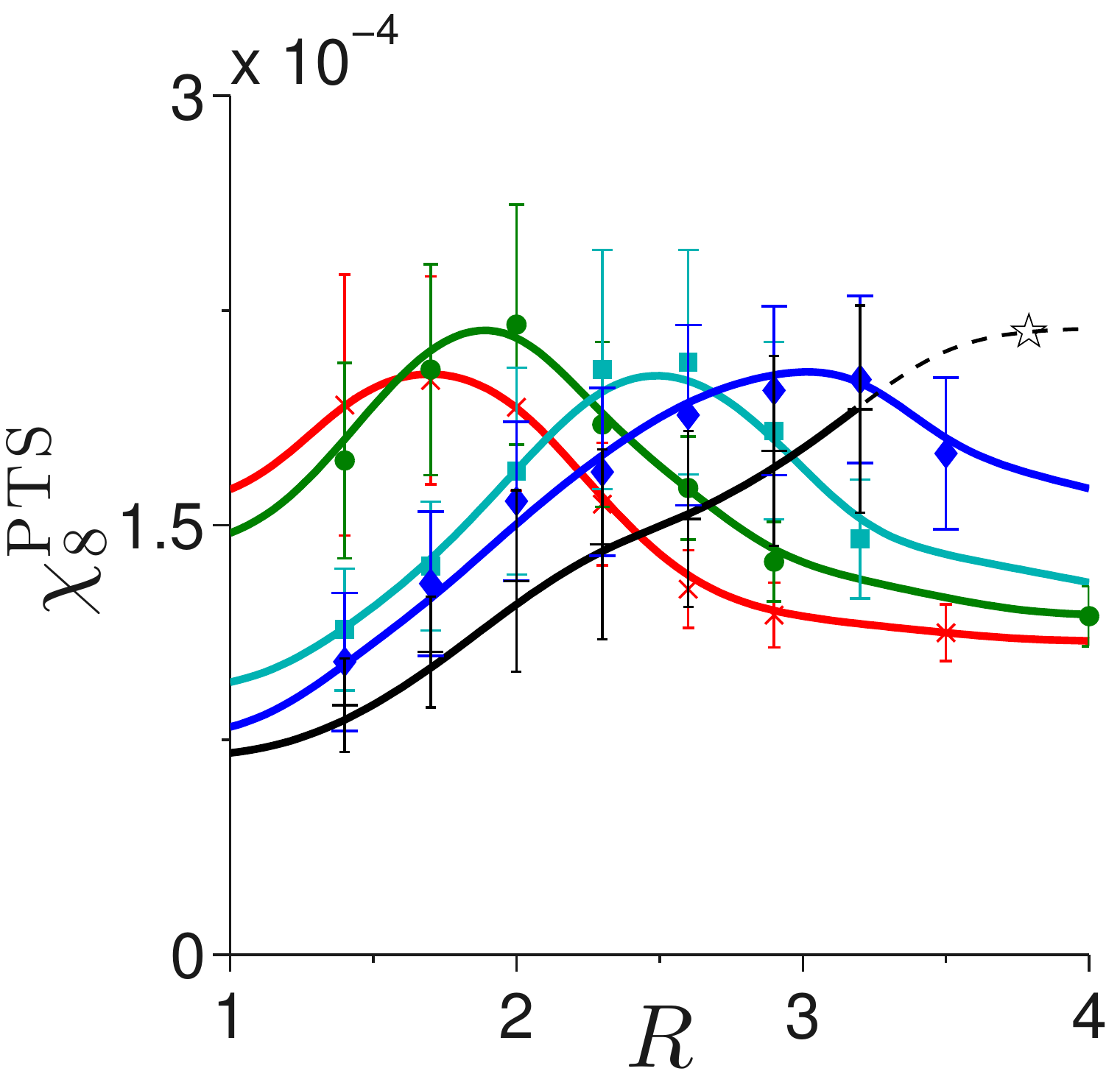}}\hspace{+0.01in}
}
\vspace{-2.2in}
\centerline{\hspace{+5.0in}
\sidesubfloat[]{\hspace{-0.06in}\includegraphics[width=0.27\textwidth]{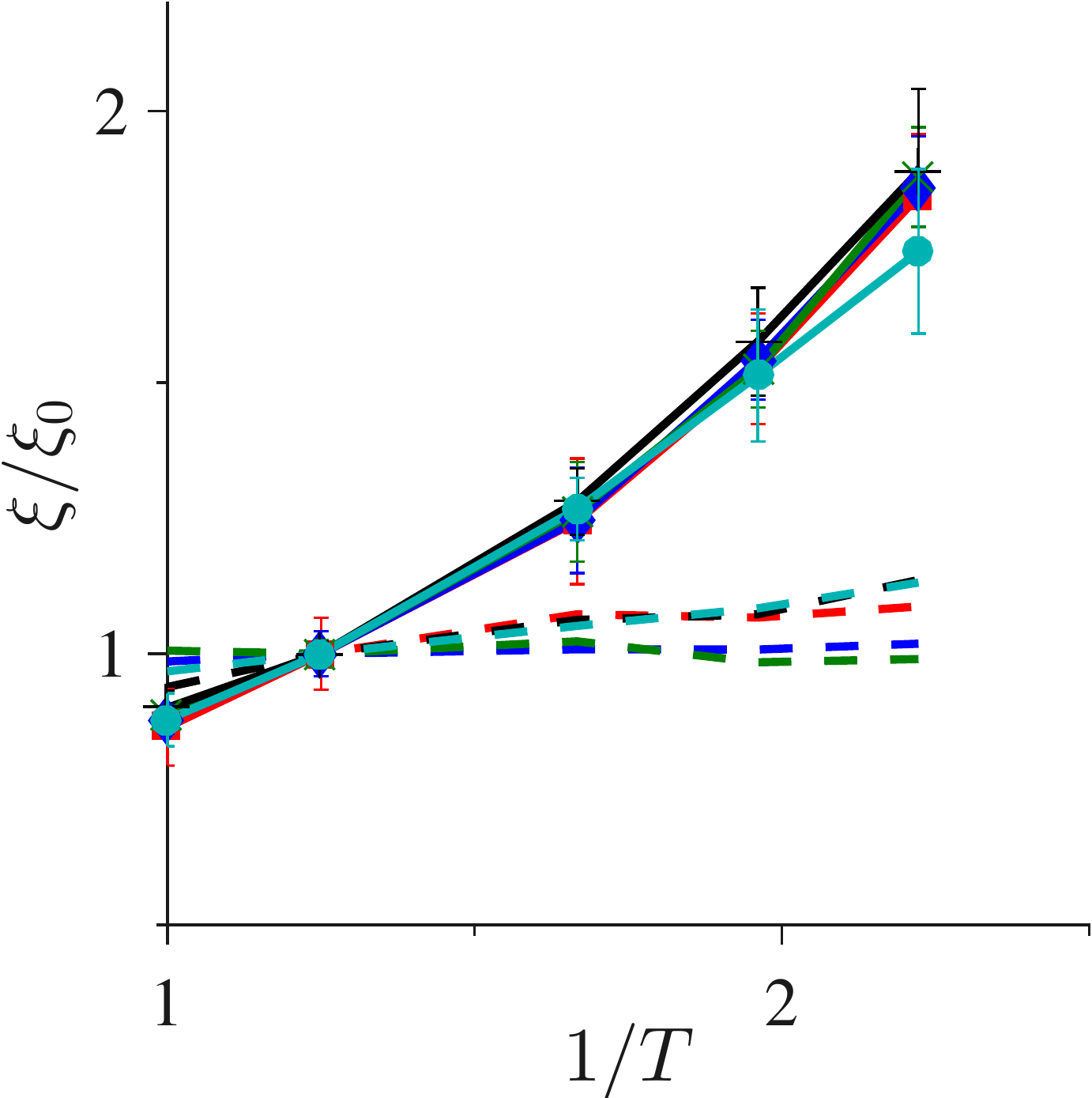}}
}
\vspace{+0.3in}
\caption{PTS correlations [(a)-(c)], susceptibilities [(d)-(f)], and lengths (g) for the $d=3$ KABLJ model as a function of cavity radius $R$: positional [(a) and (d)] and bond-orientational results for $\ell=6$ [(b) and (e)] and $8$ [(c) and (f)] at $T = 1.00$ (red-cross), $0.80$ (green-circle), $0.60$ (cyan-square), $0.51$ (blue-diamond), and $0.45$ (black-plus). For PTS correlations, solid lines are compressed exponential fits $g^{\mathrm{PTS}}=A\exp[-(R/\xi^{\rm PTS}_{\rm fit})^{\eta}]$ with $\eta=3$; for susceptibilities, solid lines are guides for the eye, and dashed lines for $T=0.45$ are extrapolated from $R=3.2$ to the extrapolated peak (star) at $R\approx3.8$~\cite{BCY16}. (g) The bond-orientational PTS lengths $\xi_{\ell}^{\rm PTS}$ for $\ell=6$ (blue-diamond), $7$ (green-cross), $8$ (black-plus), and $12$ (cyan-circle) follow the positional PTS length $\xi_{\rm pos}^{\rm PTS}$ (red-square). Dotted lines depict the standard pair positional length $\xi_{\rm pos}$ and bond-orientational lengths $\xi_\ell$ for $\ell=6,7,8,12$, all of which increase at a slower pace with decreasing temperature, in agreement with Ref.~\onlinecite{RMDP15}.
Each length is relative to $\xi_0\equiv\xi(T_0=0.800)$.}
\label{fig_BOoverlap_3D}
\end{figure*}

Results down to $T=0.45$~\cite{BCY16}, which is near the mode-coupling crossover, show no striking differences between the various overlap functions (Fig.~\ref{fig_BOoverlap_3D} for $\ell=6,8$ and Fig.~\ref{KAPTScatalog} in Appendix B for $\ell=1$-$16$).
In this model, positional PTS correlations and bond-orientational PTS correlations thus go hand in hand.
In addition, although the growth of the associated length scale over the accessible regime is moderate, it is nonetheless larger than either the two-body positional length or the lengths associated with the local structure, as measured by standard bond-orientational correlation functions [Fig.~\ref{fig_BOoverlap_3D}(g)].

The behavior of the KABLJ model appears typical of a glass-forming phenomenology in which the growth of generic amorphous order captured by PTS correlations comes without any significant contribution from the locally preferred order. The latter stays relatively short ranged, which can be taken as a manifestation of its frustration. This frustration has the dual effect of preventing the extension of locally preferred structures -- here partly through compositional constraints~\cite{CTR15} -- and of generating a rugged landscape with a multitude of equivalent low-energy metastable states above the ground state. The only significantly growing length scale is that associated with PTS correlations, which neatly capture the growing amorphous order.

\section{Glassiness versus critical slowing down}

Studying PTS correlations associated with all relevant degrees of freedom has allowed us to contrast two quite different kinds of slowing down in liquids: that of systems heading toward a continuous or weakly first-order transition and that due to a glassy behavior controlled by a multiplicity of metastable states with no significant extension of some locally preferred order.

The monodisperse hard-disk system studied above is an example of slowly relaxing liquids approaching a continuous (or weakly first-order) transition. In these systems, bond-orientational PTS and conventional correlations -- for some spherical harmonics of rank $\ell$ that detect the local order -- track the growth of the incipient order as one approaches the transition. By contrast, positional PTS correlations (as well as the bond-orientational PTS for $\ell$ incompatible with the local order) merely follow standard pair correlations. These correlations do not extend over large distances if, as in the above examples of hexatic phases, the (quasi-)ordered phase does not involve additional breaking of translational symmetry. The relaxation slowdown is then controlled by the incipient local order, which grows because a phase transition is approached. Sluggish dynamics is thus here a form of critical slowing down that should not be confused with the glassy relaxation of liquids approaching a glass transition.   

Quite differently, for some glass-forming systems such as the KABLJ model, essentially all PTS correlation functions, be they bond-orientational or positional, track the same length. This PTS correlation length can thus be associated with growing generic amorphous order and the rarefaction of available disordered metastable states, without any notable competition from the extension of the specific local order. This length increases modestly over the dynamical range accessible to computer simulations but nonetheless does so more than other structural lengths associated with pair positional correlations or local order. The matching of (generic) bond-orientational and positional PTS correlations and the parallel decoupling between the latter and other structural lengths are characteristic of this type of systems~\cite{CGV07,BBCGV08,SL11,HMR12,CCT12,CB12,JB12,KVB12,CGV12,BK12,CCT13,BKP13,CT13,GTCGV13,HBKR14,OKIM15,BCY16}. The relaxation slowdown then appears to be compatible with the RFOT scenario~\cite{WL12}.

Additional information can be obtained from the fluctuations of the cavity overlaps~\cite{BCY16}.
One can define an overlap \emph{susceptibility} $\chi^{\rm PTS}$ from the variance of the overlap fluctuations, averaged over a small region around the cavity center: $\chi^{\rm PTS}(R)\equiv \le[\langle Q^2\rangle-\langle Q\rangle^2\ri](R)$.
For glassy systems characterized by rugged local free-energy landscapes, we expect overlap susceptibilities to develop a peak at $\xi^{\rm PTS}$ (Fig.~\ref{fig_BOoverlap_3D}), as observed numerically also 
in other types of constrained glassy systems~\cite{KB13,B13}.
By contrast, for $d=2$ hard disks, the overlap susceptibility for $\ell=6$ is devoid of such distinctive peak structure and instead monotonically approaches the bulk value, much like simple two-point functions (see Fig.~\ref{trends} in Appendix B).
There also seems to be a difference in the cavity-size dependence of the mean overlap, with the appearance of a nonconvex dependence at the lowest accessible temperatures for the KABLJ model, while no such structure is observed for $d=2$ hard disks (see also Ref.~\onlinecite{BBCGV08}).

Yet another striking difference between critical and glassy slowdowns appears in sampling small cavities.
For the KABLJ model, barriers in the local free-energy landscape at the PTS scale are so high that nonlocal Monte Carlo schemes (such as parallel tempering) are needed to ensure equilibration~\cite{HMR12,KB13,BCY16}. The small cavity sizes make the physical lifetime of the metastable states 
extremely large.
For monodisperse hard disks, by contrast, barriers decrease 
for small cavities as a result of the weakening influence of correlations. A local Monte Carlo scheme suffices. 
The latter dynamical behavior is consistent with the slowing down being due to the proximity of a weakly first-order transition and inconsistent with typical glass phenomenology.

\begin{figure}
\centerline{\includegraphics[width=0.55\textwidth]{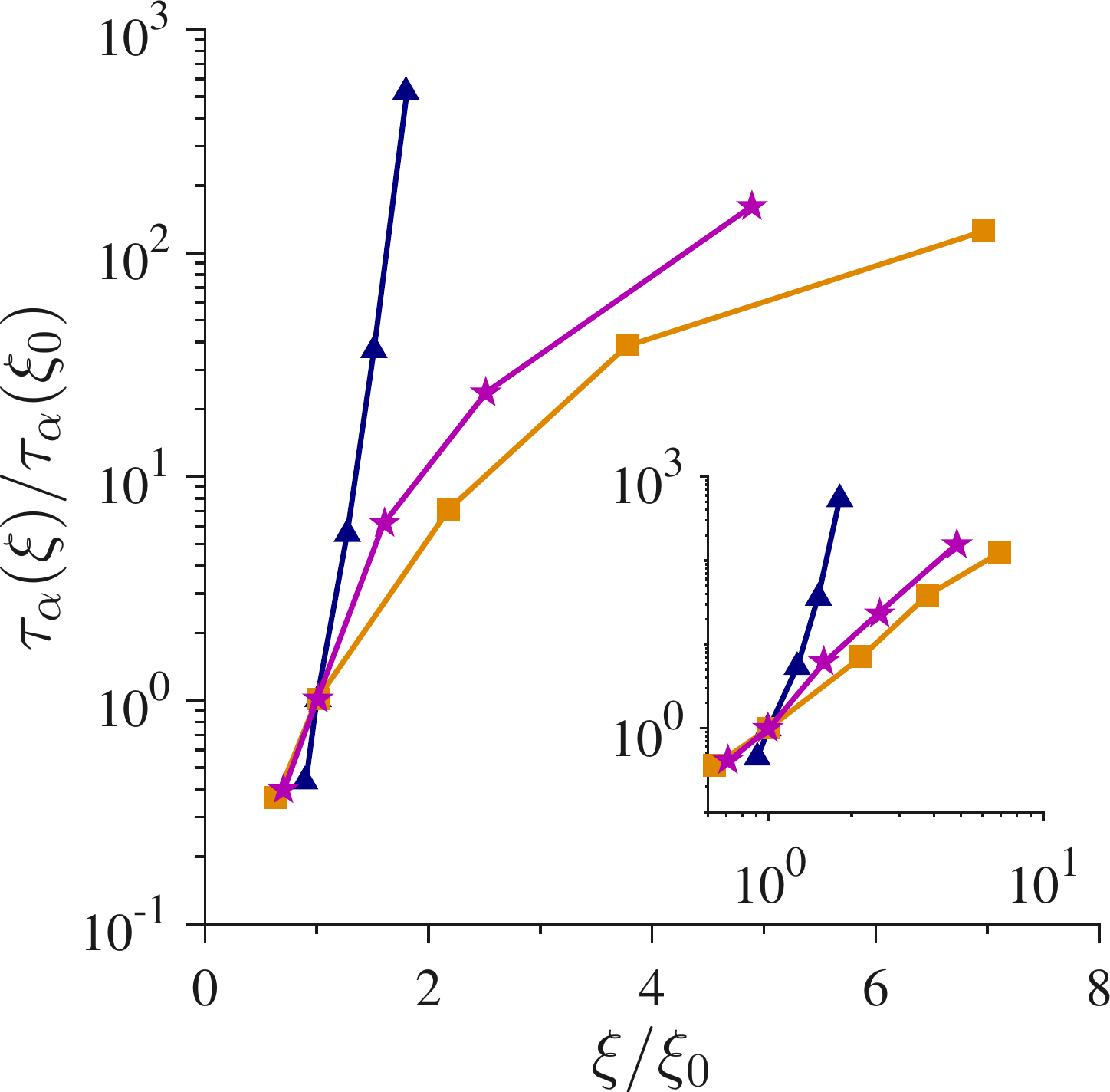}}
\caption{Growth of the structural relaxation time,  $\tau_{\alpha}$, with the dominant length, normalized with $\xi_0\equiv\xi_{\rm pos}^{\rm PTS}(T_0)$ at high temperature $T_0=0.80$ for the KABLJ model (blue triangles) and with $\xi_0\equiv\xi_{\ell=6}(\phi_0)$ at low packing fraction $\phi_0=0.650$ for monodisperse (orange squares) and $\phi_0=0.700$ for 11$\%$ polydisperse hard disks (purple stars). For the KABLJ model, $\tau_\alpha$ increases exponentially while the PTS length grows modestly. By contrast, for hard disks, $\tau_{\alpha;\ell=6}$ as a function of $\xi_{\ell=6}$ (see Appendix C) increases much more mildly, compatible with the power-law scaling of Eq. (\ref{eq_conventional_dynamic_scaling}) (see the inset for the log-log plot).}
\label{versus}
\end{figure}

This difference in dynamical behavior is further corroborated by looking at the parametric evolution of the bulk relaxation time against the dominant growing correlation length in the system (Fig.~\ref{versus}). Whereas the slowing down in the KABLJ model is qualitatively well described by the glassy activated scaling in 
Eq.~(\ref{eq_activated_dynamic_scaling}), that of the hard-disk model are remarkably close to the critical power-law scaling in Eq.~(\ref{eq_conventional_dynamic_scaling}).

\section{Polydisperse hard disks, revisited}

Armed with these observations, we finally take a look at $d=2$ systems with polydispersity  $\Delta=3\%, 6\%, 9\%$, and $11\%$. This last system was first investigated by Russo and Tanaka~\cite{RT15}, and we reproduce here all their results over the range $\PF=0.73$-$0.77$ they studied. In addition, we compute the bond-orientational PTS correlations and find that the $\ell=6$ component perfectly tracks the growth of the hexatic order, while the positional PTS length grows only mildly, just as in the monodisperse ($\Delta=0\%$) case (see Appendix~\ref{sec:2ptfunct}).
For intermediate $\Delta$, we also know $\PF_{\rm hexatic}\le(\Delta\ri)$~\cite{PF04}.
Interestingly, when studied at appropriately rescaled $\PF$, the various static lengths grow similarly at all $\Delta$ (see Fig.~\ref{trends} in Appendix B).
As far as static lengths go, $\Delta=11\%$ is thus smoothly connected and qualitatively similar to the monodisperse system, suggesting that the growth of structural correlations is due to the approach of a weakly first-order transition, and not to glassiness.
In addition, the PTS correlation functions are convex and the overlap susceptibilities display no peak structure, which provide yet more evidence that this system is not glassy.
Finally, we encounter no high barrier to sampling configurations in small cavities; a (semi-)local Monte Carlo scheme suffices to equilibrate them.
Barriers are in fact so low that the hard disks even fail to develop a plateau in the self-intermediate scattering function~\cite{RT15}, which is usually taken as a canonical feature of glassiness. Our results thus indicate that the conclusions drawn by Russo and Tanaka about the role of PTS correlations near the glass transition are obtained from numerical observations performed in a model that does not behave as canonical glass-forming liquids.

\section{Conclusion}

Our study of PTS correlations in slowly relaxing liquids shows that a full characterization of these correlations requires, besides a proper imposition of constraining boundary conditions, an account of all degrees of freedom relevant to describing the configurations at a coarse-grained level. For spherical particles, this should include positional and bond-orientational degrees of freedom. When both are taken into account, PTS correlations prove to be a powerful tool to investigate two- and three-dimensional liquids. They can track both the spatial extent of a specific local order -- possibly associated with an underlying ordering transition -- and the growth of some generic amorphous order associated with a reduction in the number of available disordered metastable states. Furthermore, in the case of molecular liquids, provided orientational degrees of freedom are also included, PTS correlations can provide a useful tool to detect a putative growing local order. Contrary to systems of spherical particles, symmetry classifications of local arrangements are remarkably difficult for these systems, and no systematic investigation of locally preferred structures or motifs has so far been undertaken.

Through the extended PTS correlations we have also been able to disentangle conventional critical slowing down from glass-forming behavior. The liquids studied in this work, namely monodisperse and weakly polydisperse  hard disks on the one hand and the KABLJ model on the other,  appear as two extreme classes in a possible taxonomy of slowly relaxing liquids: the formers display growing local order triggered by an incipient critical ordering but virtually no glassiness, while the latter shows growing amorphous order and glassiness but with a strongly frustrated locally preferred order. One can envisage intermediate classes for which weakly frustrated local order and amorphous order compete to generate a glassy slowdown of relaxation~\cite{SL11,ST10,CP07,HCIR14}. We propose that in these more challenging cases the study of PTS correlations as developed in the present work should allow a useful taxonomic characterization with an assessment of  the relative importance of specific local order versus generic amorphous order  in the dynamical slowdown.
For instance, existing work on a liquid in the hyperbolic (\textit{i.e.}, negatively and uniformly curved) plane provides evidence that for weak frustration, controlled by a small curvature~\cite{STV08,ST10}, the length associated with the growing local sixfold order grows hand in hand with the positional PTS length~\cite{SL11} (and much more than the two-point positional correlation length) in the regime accessible to computer simulations. We can therefore anticipate that in this regime bond-orientational PTS correlations would grow concomitantly. Hexatic and hexagonal ordering being strictly suppressed due to the constant curvature~\cite{footnote_grain}, the system is clearly a glass former, yet it also displays a significant growth of the sixfold local order that saturates at a length scale of the order of the radius of curvature. An open question then concerns the behavior of PTS correlations beyond this saturation. Although this regime may be hard to reach in practice because of the large structural relaxation times, we speculate that there may be a crossover to a regime where PTS correlations decouple from the local order and track a length describing the further reduction of metastable states. It would be particularly interesting to know  the shape of the overlap susceptibilities in this case. Similar studies could further be undertaken on $d=2$ hard-disk models in flat space, if cranking up the polydispersity, $\Delta$, can indeed mimic the effect of negative curvature in suppressing the hexatic order, and in weakly frustrated three-dimensional atomic and molecular glass-forming models.

In any case, the proposed conceptual framework provides a constructive approach to resolving the controversy on the role of static correlations and associated length scales in the slowdown of relaxation leading to glass formation.

\begin{acknowledgments}
We acknowledge fruitful exchanges with H.~Tanaka and  J.~Russo.
We also acknowledge the Duke Compute Cluster for computational times, without which this research would have been impossible to carry out.
The research leading to these results has received funding from the European Research Council under the European Union Seventh Framework Programme
(FP7/2007-2013)/ERC Grant agreement No.~306845 (LB).
PC and SY acknowledge support from the National Science Foundation 
Grant No.~NSF DMR-1055586, the Sloan Foundation Grant No.~BR2013-022, and the Simons Foundation (\#454937, Patrick Charbonneau). Data relevant to this work have been archived and can be accessed at http://doi.org/10.7924/G8BG2KWP.
\end{acknowledgments}

\appendix

\section{Models and simulations}
\label{sec:models}
In this Appendix, we describe the models simulated as well as the simulation procedure employed to obtain the data describe in the main text.
\subsection{2D Monodisperse}
Configurations with $N=10,000$ monodisperse hard disks of equal diameter $\sigma=1.0$ in a periodic box of linear size $L$ are prepared at packing fractions $\PF=0.600,0.650,0.680,0.690, 0.695$, using a Monte Carlo (MC) sampling scheme that includes only local displacements.
Both equilibration- and production-run lengths are $100N_{\tau}$ Monte Carlo sweeps, where $N_{\tau}(\PF=0.600,0.650,0.680,0.690,0.695)=(2,2,5,10,30)\times 10^3$ MC sweeps.
A total of $100$ independent configurations are generated.

In order to measure PTS correlations,  for each of these $100$ configurations, we randomly pick a position within the simulation box as the cavity center.
Phase space within a cavity is again explored through simple MC sampling scheme.
An important change with respect to bulk equilibration is that particles outside the cavity are fixed and we put a hard wall at the edge of the cavity, \textit{i.e.}, all moves that take a mobile particle outside the cavity are rejected.

In order to check for proper equilibration within the cavity, we employ two initialization schemes~\cite{BCY16,CGV12}: (1) from the original configuration and (2) from a randomized configuration, obtained by shrinking the disk diameters from $\sigma$ to $0.5\sigma$, performing $10^4$ MC sweeps to randomize their positions, and re-growing the disks back to $\sigma$ sufficiently slowly to avoid local jamming.
Positional and bond-orientational overlaps between the original and the equilibrated configurations for the two schemes are monitored.
After $N_{\tau}$ MC sweeps, (disorder-averaged) overlap values for the two schemes converge.
For both of these schemes, we run simulations for $100N_{\tau}$ MC sweeps, discarding configurations from the first  $20N_{\tau}$ sweeps and sampling $200$ cavity configurations, equally spaced in time, from the remaining production runs.

In total, $100$ cavities are attained for each $R$ at each packing fraction $\PF$, and two sets of $200$ configurations for each cavity are obtained from the two schemes.

\subsection{Two-dimensional Polydisperse}
\begin{figure}
\centerline{\includegraphics[width=0.6\textwidth]{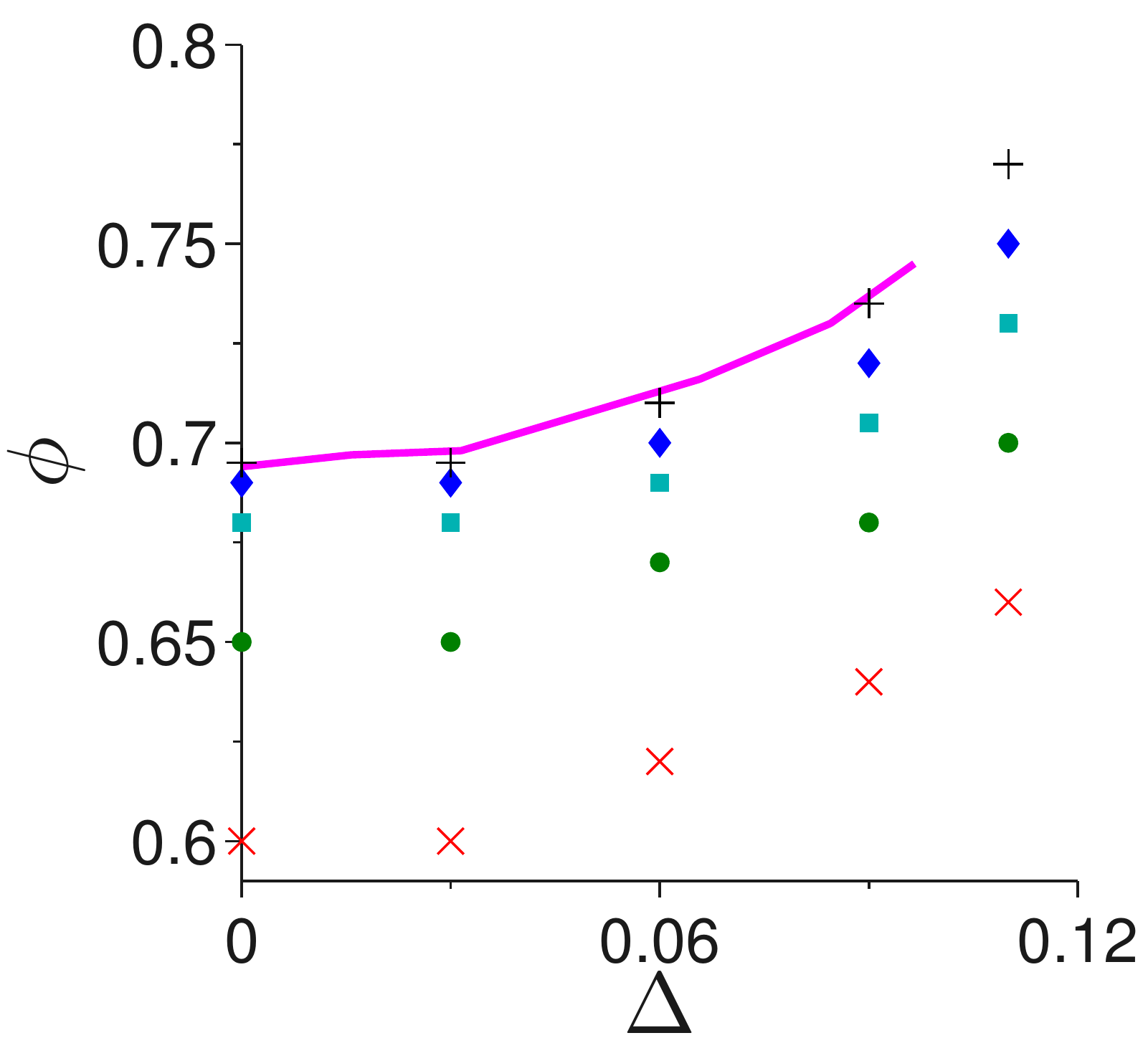}}
\caption{Evolution of weakly first-order transition line (magenta) with polydispersity from a semigrandcanonical ensemble calculation~\cite{PF04}. Dots indicate packing fractions at which PTS measurements are performed for each polydispersity, $\PF = \PF_1$ (red-cross), $\PF_2$ (green-circle), $\PF_3$ (cyan-square), $\PF_4$ (blue-diamond), and $\PF_5$ (black-plus).}
\label{phase_diagram}
\end{figure}

The $d=2$ polydisperse hard-disk system is akin to the monodisperse system but with disk diameters $\{\sigma_i\}_{i=1,...,N}$ that are drawn from a Gaussian distribution.
Through a simple affine transformation on $\{\sigma_i\}_{i=1,...,N}$, we adjust the mean diameter $\bar{\sigma}\equiv\frac{\sum_{i=1}^{N}\sigma_i}{N}=1.0$ and polydispersity $\Delta\equiv\sqrt{\frac{\sum_{i=1}^{N}(\sigma_i-\bar{\sigma})^2}{N\bar{\sigma}^2}}=3\%,6\%,9\%,11\%$.
In addition to local displacements, we employ particles identity swaps ($20\%$ of the MC moves).
We prepare samples at packing fractions $\PF=\PF_1, ..., \PF_5$, where (Fig.~\ref{phase_diagram})
\bea
&&(\PF_1,\PF_2,\PF_3,\PF_4,\PF_5)\nonumber\\
&=&(0.600,0.650,0.680,0.690,0.695)\ {\rm for}\ \Delta=3\%\nonumber\\
&=&(0.620,0.670,0.690,0.700,0.710)\ {\rm for}\ \Delta=6\%\nonumber\\
&=&(0.640,0.680,0.705,0.720,0.735)\ {\rm for}\ \Delta=9\%\nonumber\\
&=&(0.660,0.700,0.730,0.750,0.770)\ {\rm for}\ \Delta=11\%\, ,\nonumber
\eea
and $N_{\tau}(\PF=\PF_1,\PF_2,\PF_3,\PF_4,\PF_5)=(2,2,5,10,30)\times 10^3$,
using the same preparation protocol as for monodisperse disks.

\subsection{Kob-Andersen Binary Lennard-Jones}
The $d=3$ Kob-Andersen binary Lennard-Jones (KABLJ) mixture is a $80:20$ mixtures of $A:B$ atoms with interatomic pair potentials
\begin{equation}
\label{eq_potentials}
\begin{aligned}
v_{\alpha \beta}(r) &= 4\epsilon_{\alpha \beta}\left[\left( \frac{\sigma_{\alpha \beta}}{r}\right)^{12}- \left( \frac{\sigma_{\alpha \beta}}{r}\right)^{6}\right ],
\end{aligned}
\end{equation}
where $\alpha, \beta = A$  or $B$; the potential is truncated at the conventional cutoff of $2.5 \sigma_{\alpha \beta}$ and is shifted so that it vanishes at the cutoff. Molecular dynamics simulations are performed at a density $\rho=1.2$. (Lengths and temperatures are given in Lennard-Jones units of $\sigma_{AA}$ and $\epsilon_{AA}/k_B$, respectively.) More details are given in Ref.~\onlinecite{BCY16}.

We emphasize here that, while for mono/polydisperse hard disks we attained good equilibration in all cavities without resorting to parallel tempering, for the KABLJ model the scheme was essential~\cite{BCY16}. This difference provides additional evidence of glassiness for this model.

\section{PTS Definitions}
\label{sec:PTSdef}
In this section we define both positional and orientational measures of PTS correlations.

\begin{figure*}
\includegraphics[width=0.2\textwidth]{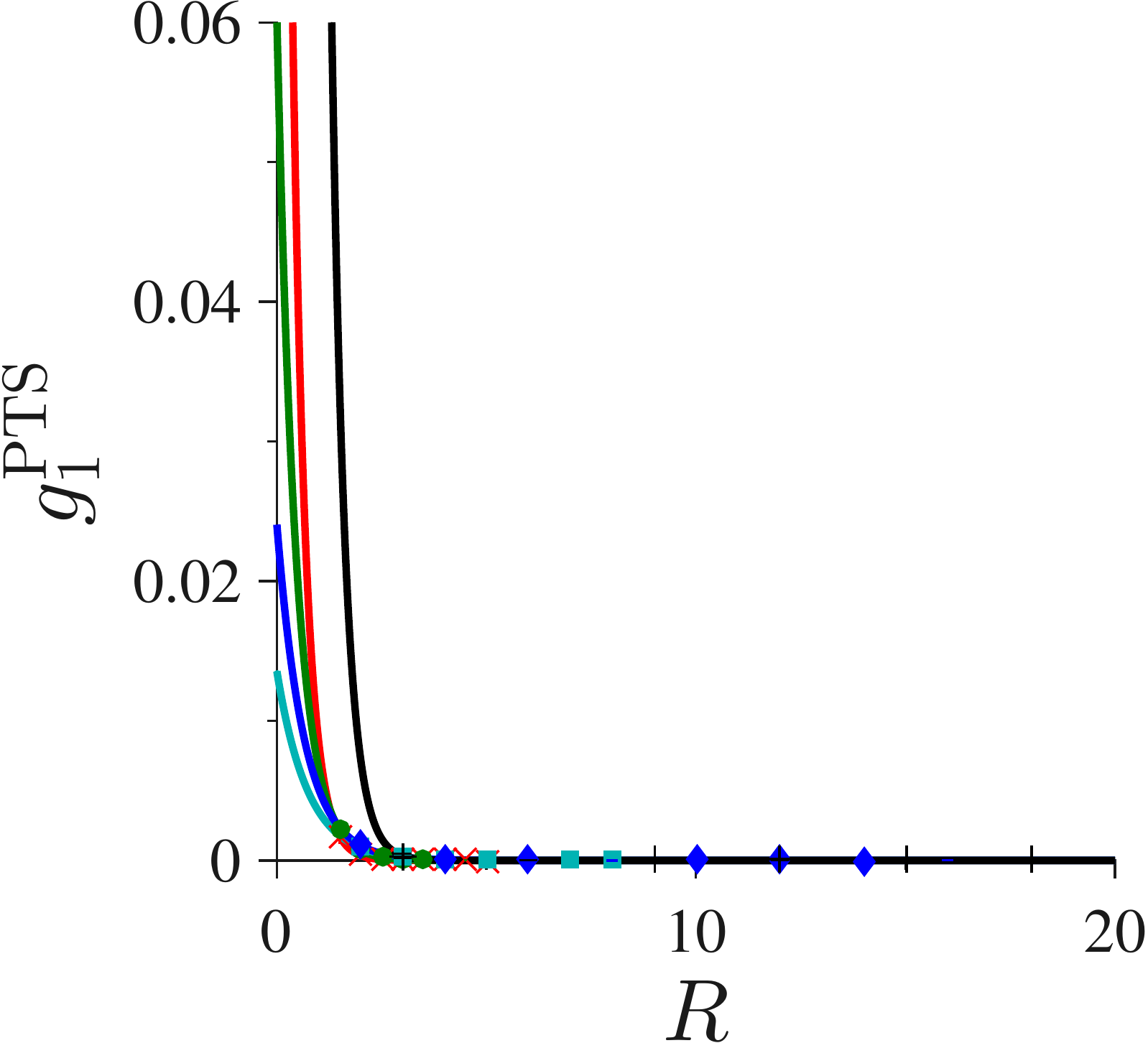}
\includegraphics[width=0.2\textwidth]{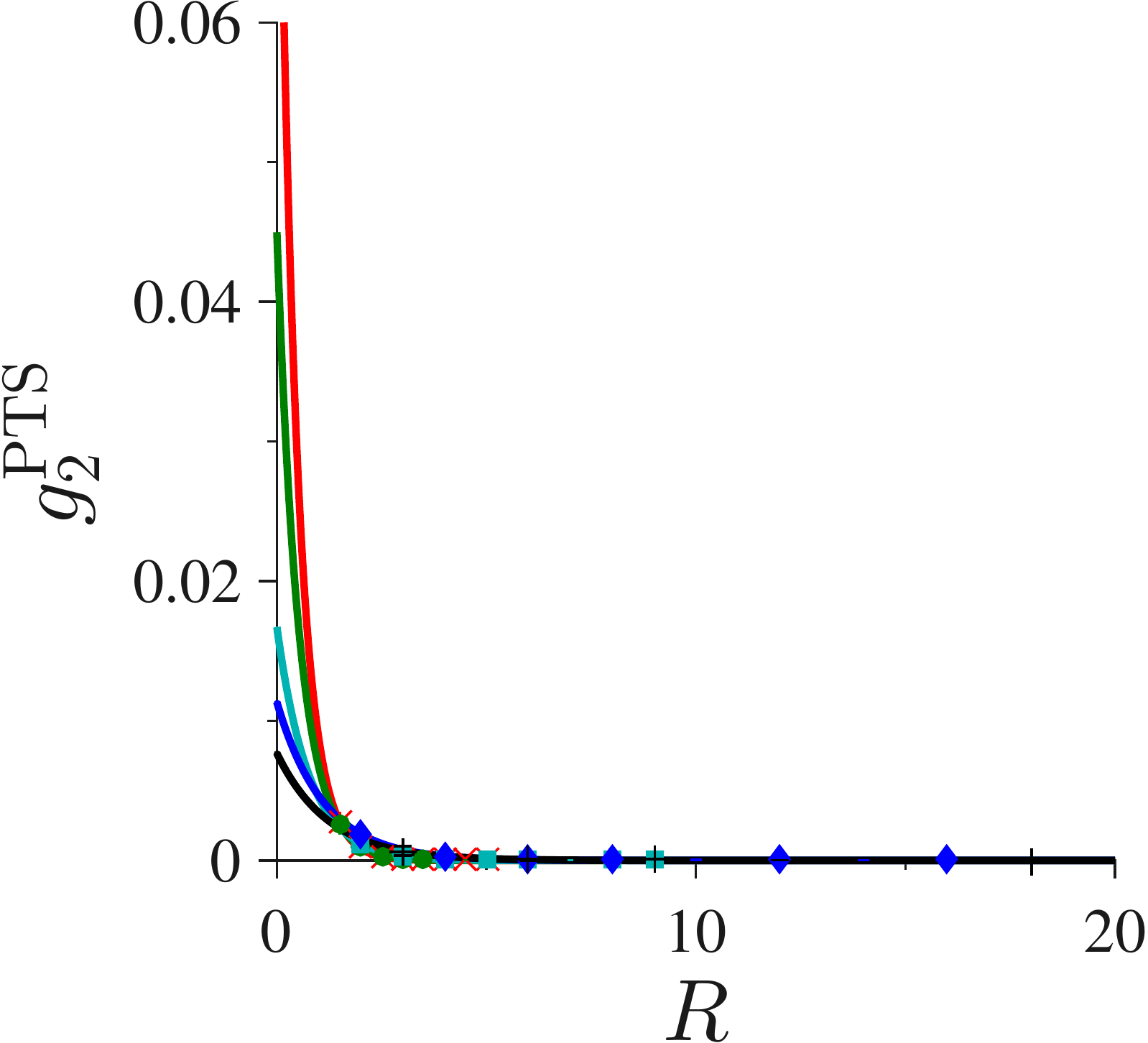}
\includegraphics[width=0.2\textwidth]{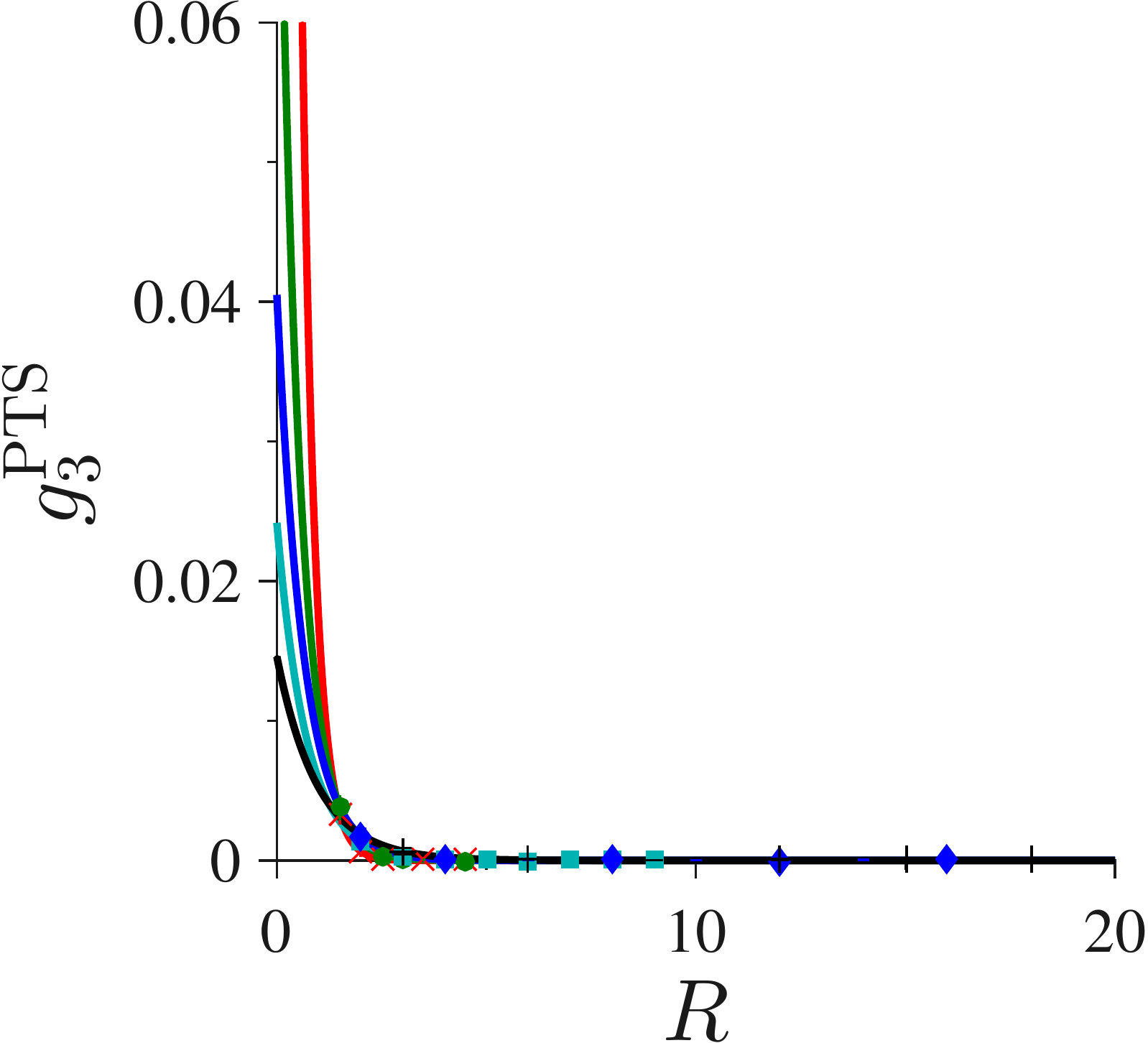}
\includegraphics[width=0.2\textwidth]{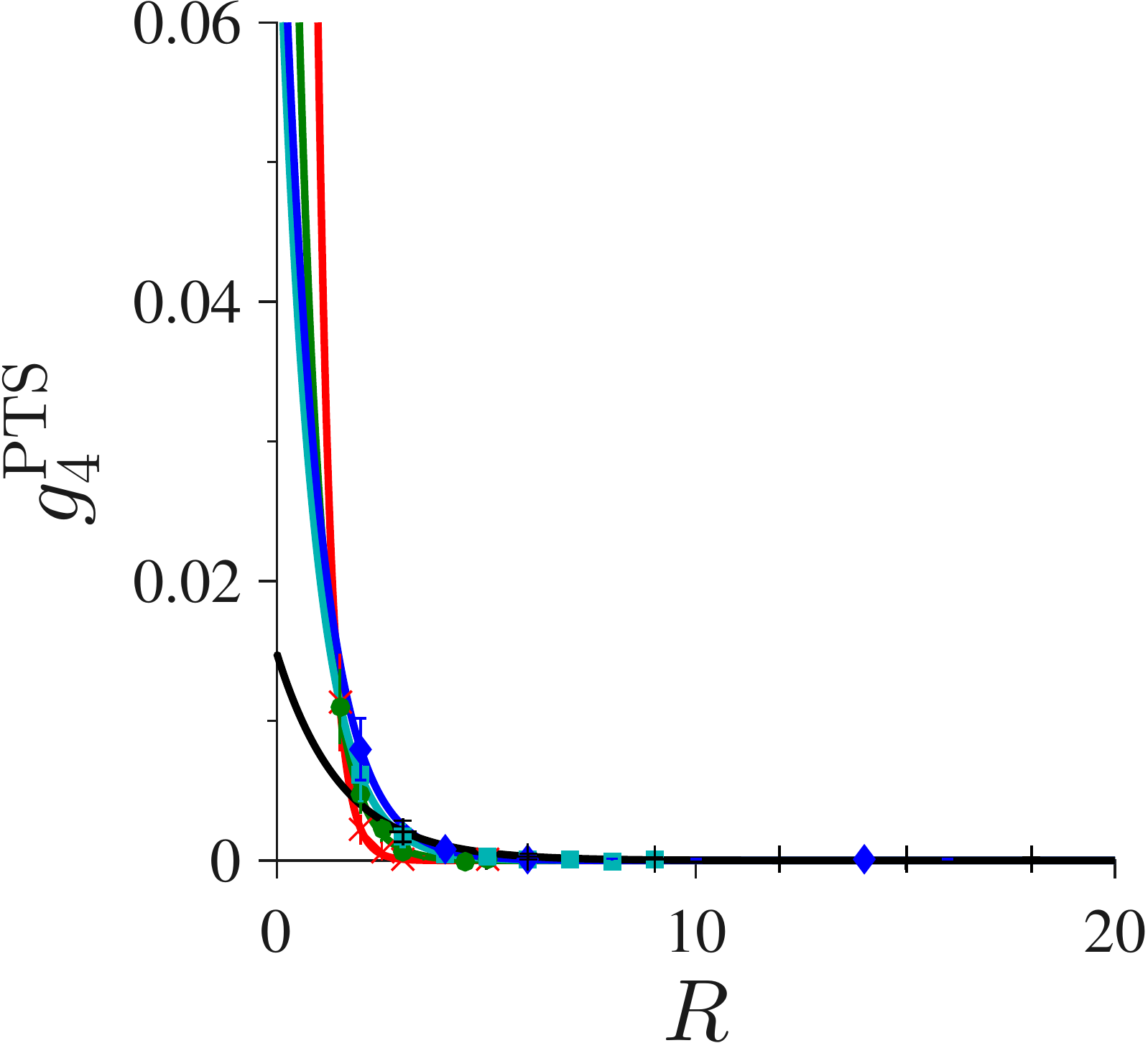}
\includegraphics[width=0.2\textwidth]{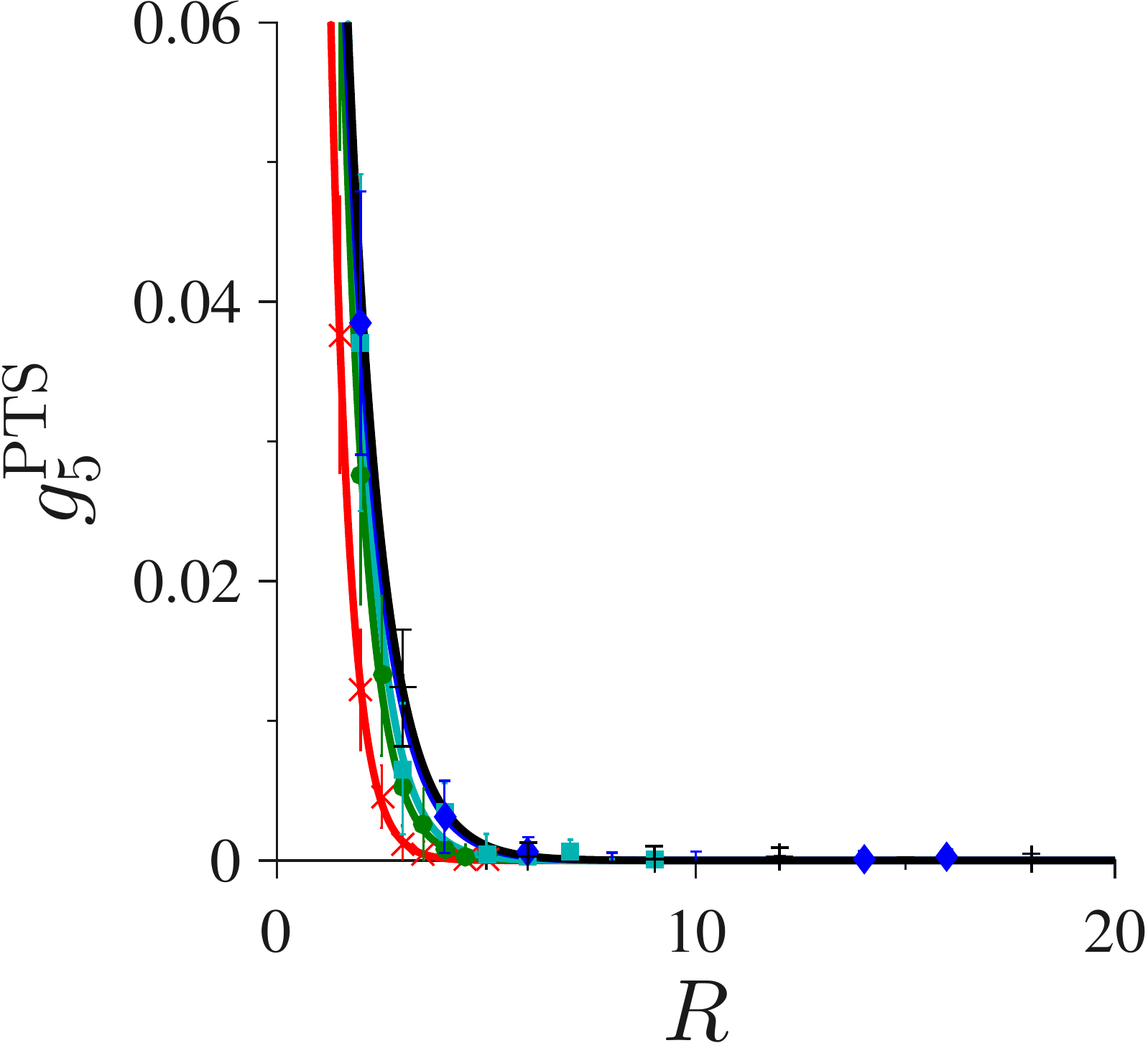}
\includegraphics[width=0.2\textwidth]{PTS_BO0_6.pdf}
\includegraphics[width=0.2\textwidth]{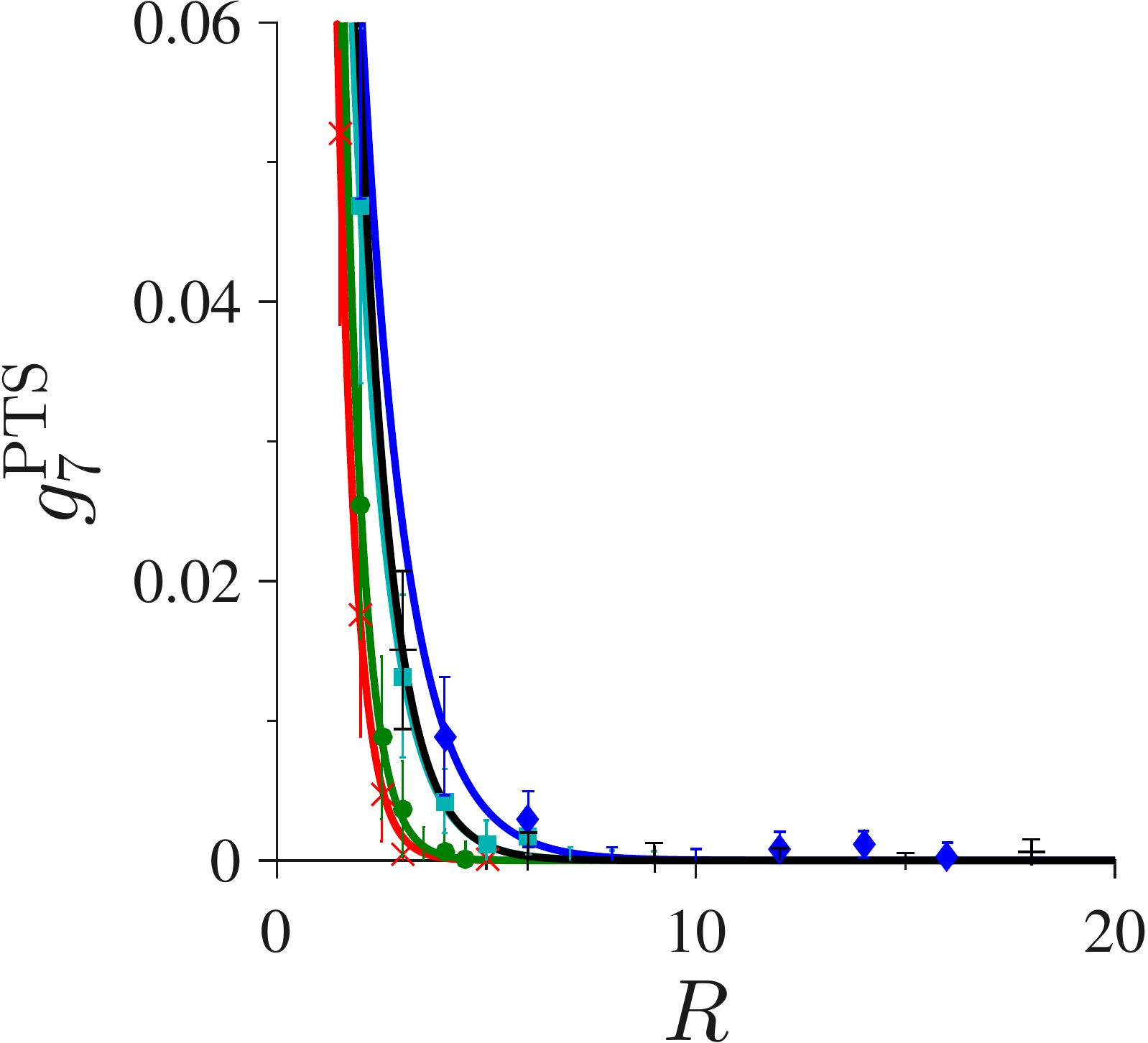}
\includegraphics[width=0.2\textwidth]{PTS_BO0_8.pdf}
\includegraphics[width=0.2\textwidth]{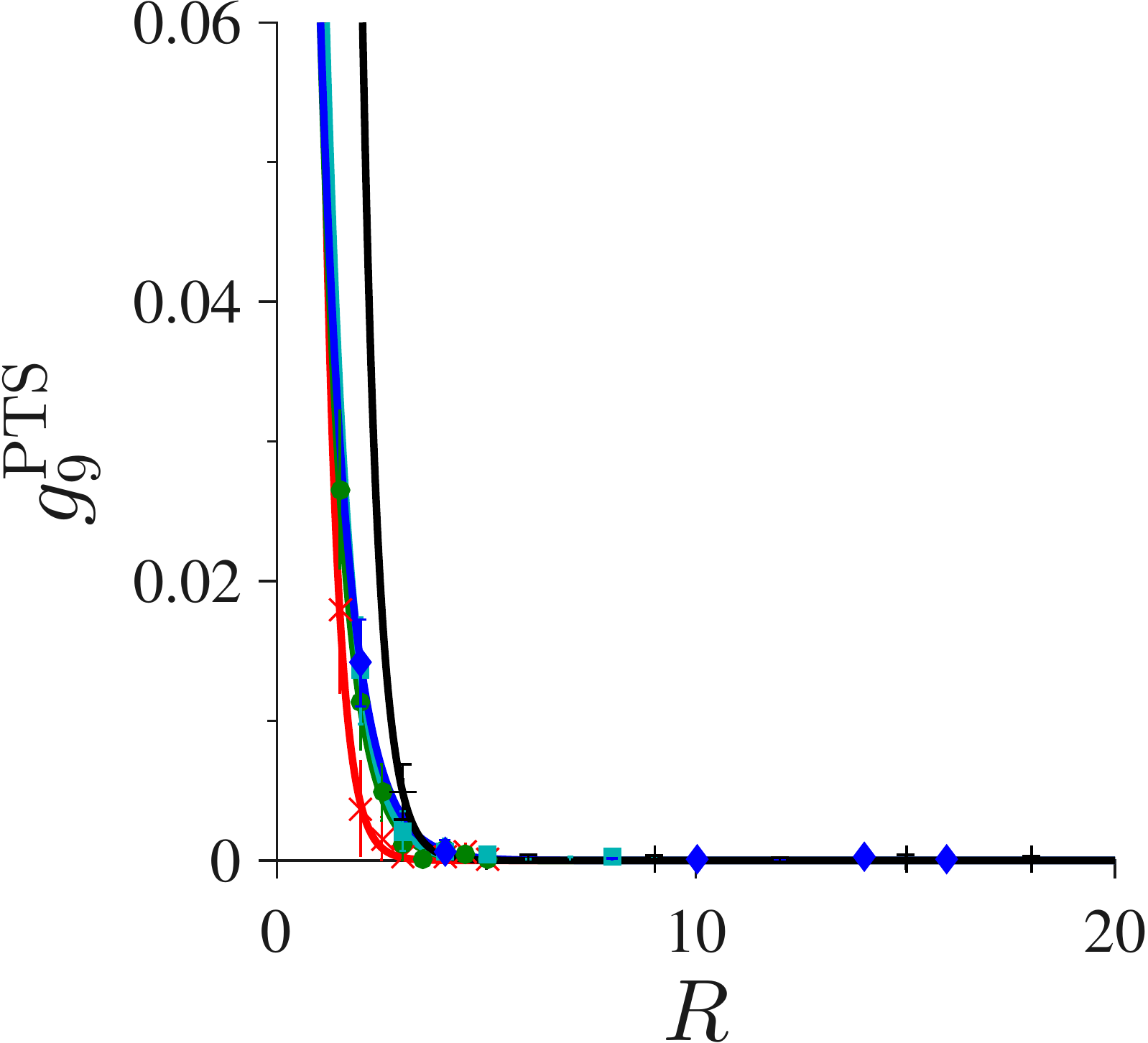}
\includegraphics[width=0.2\textwidth]{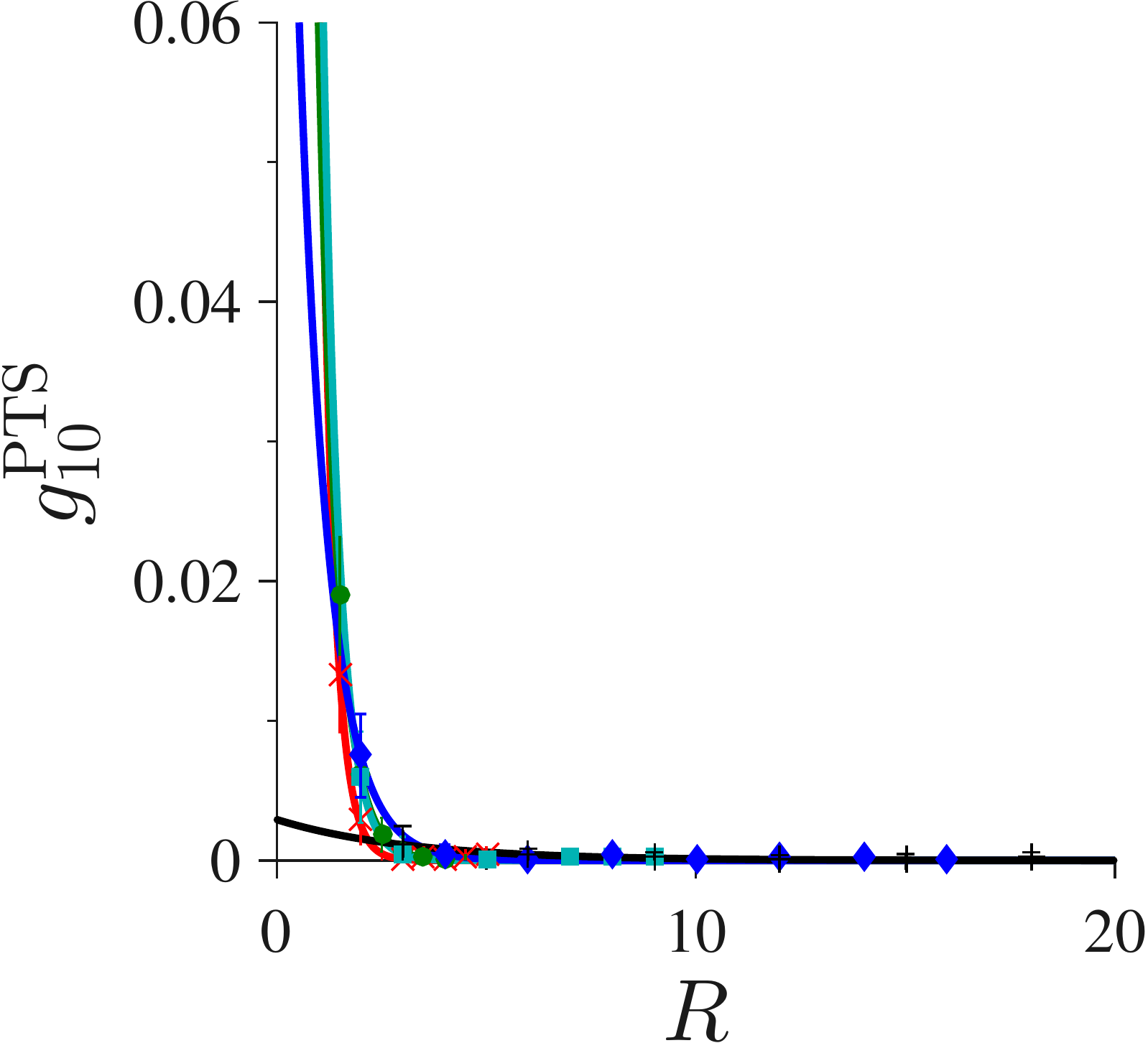}
\includegraphics[width=0.2\textwidth]{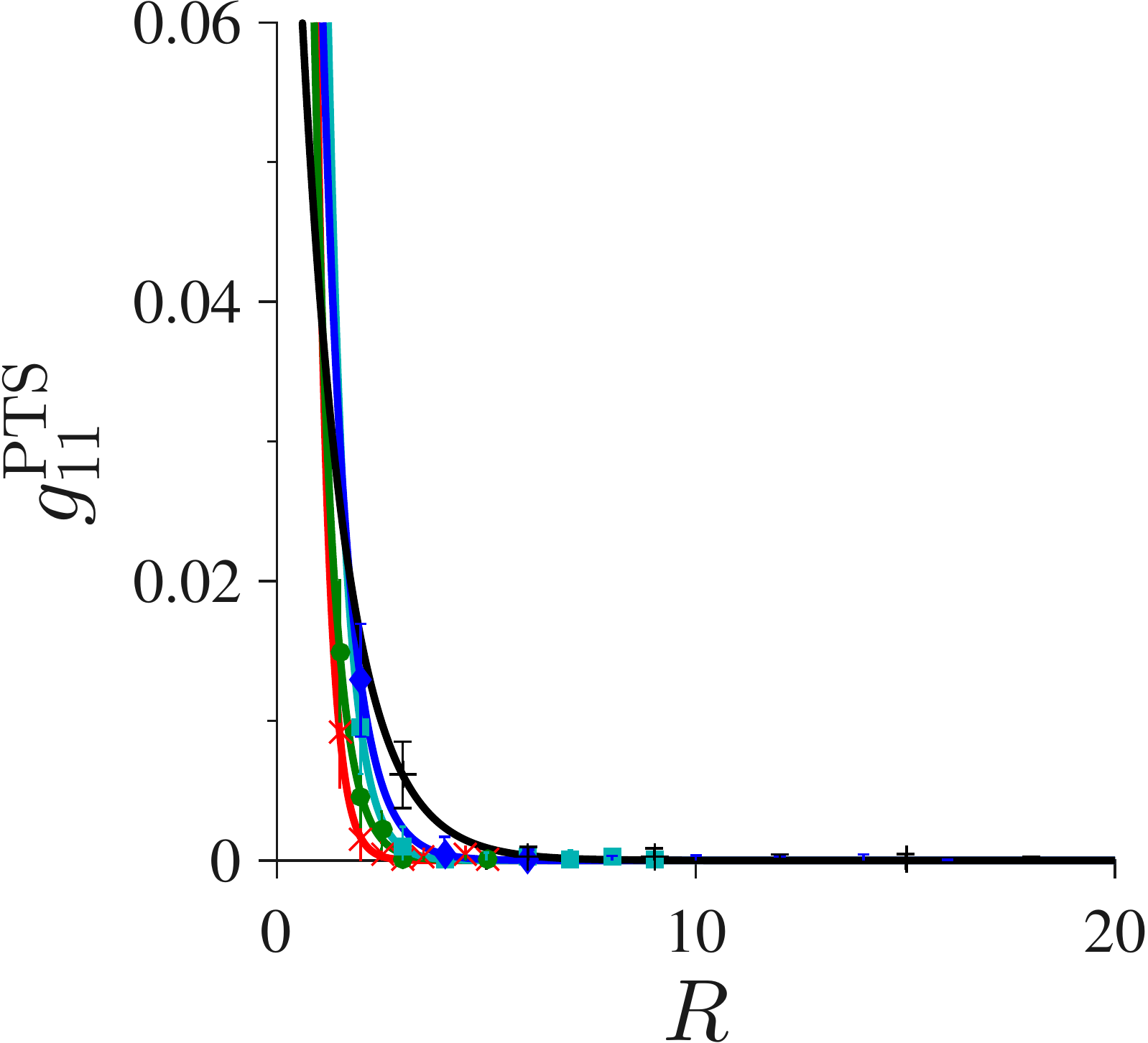}
\includegraphics[width=0.2\textwidth]{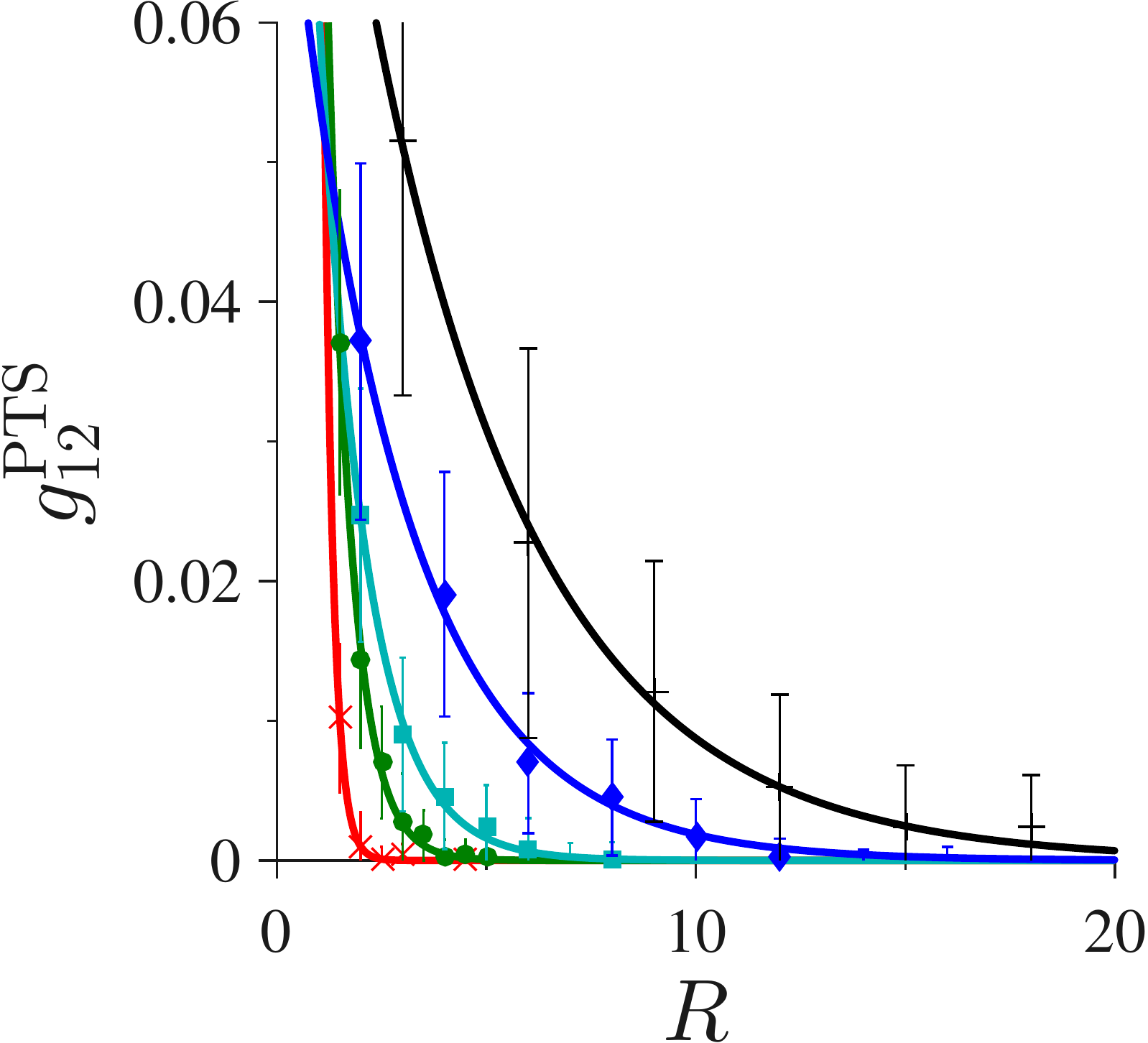}
\includegraphics[width=0.2\textwidth]{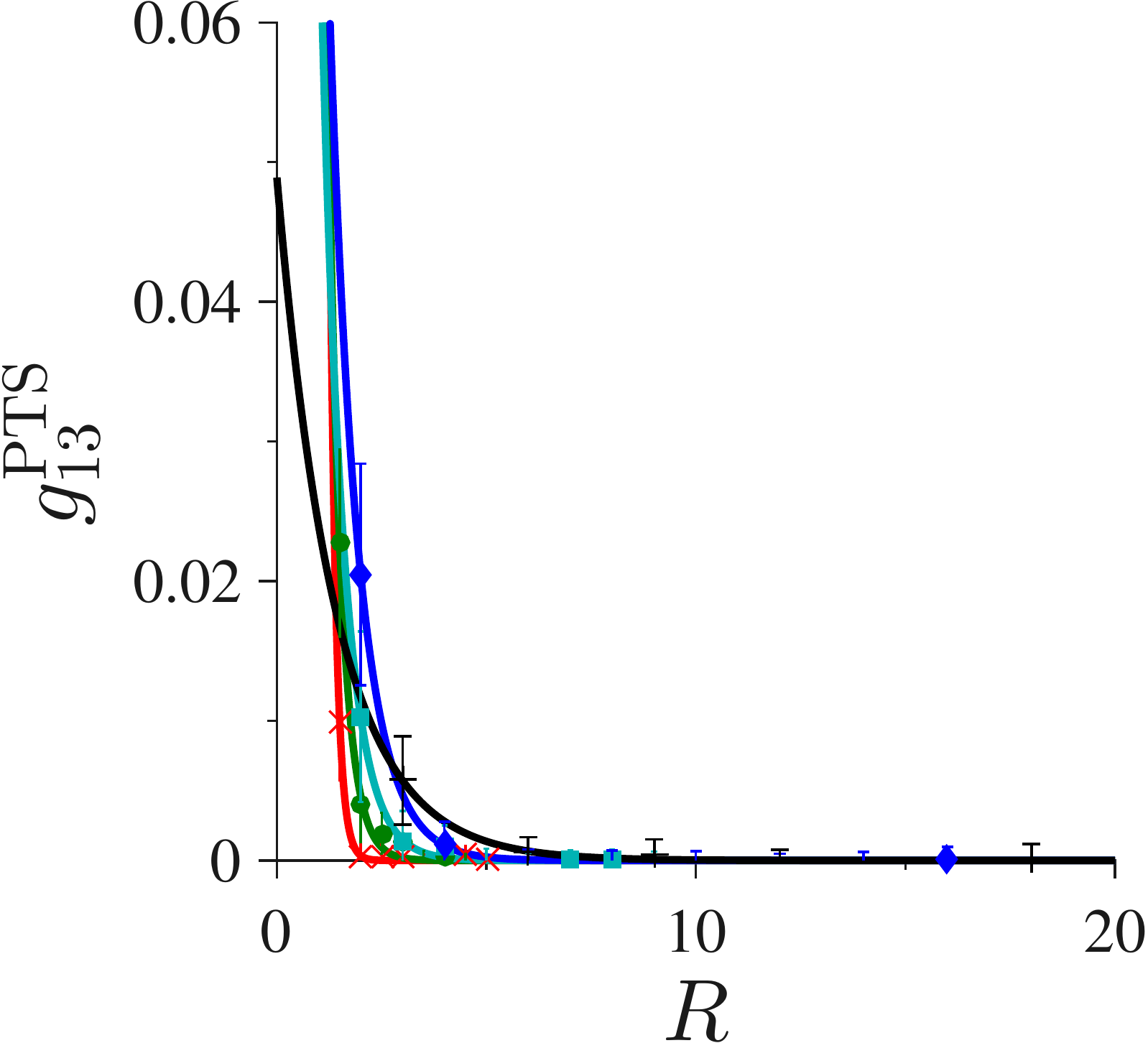}
\includegraphics[width=0.2\textwidth]{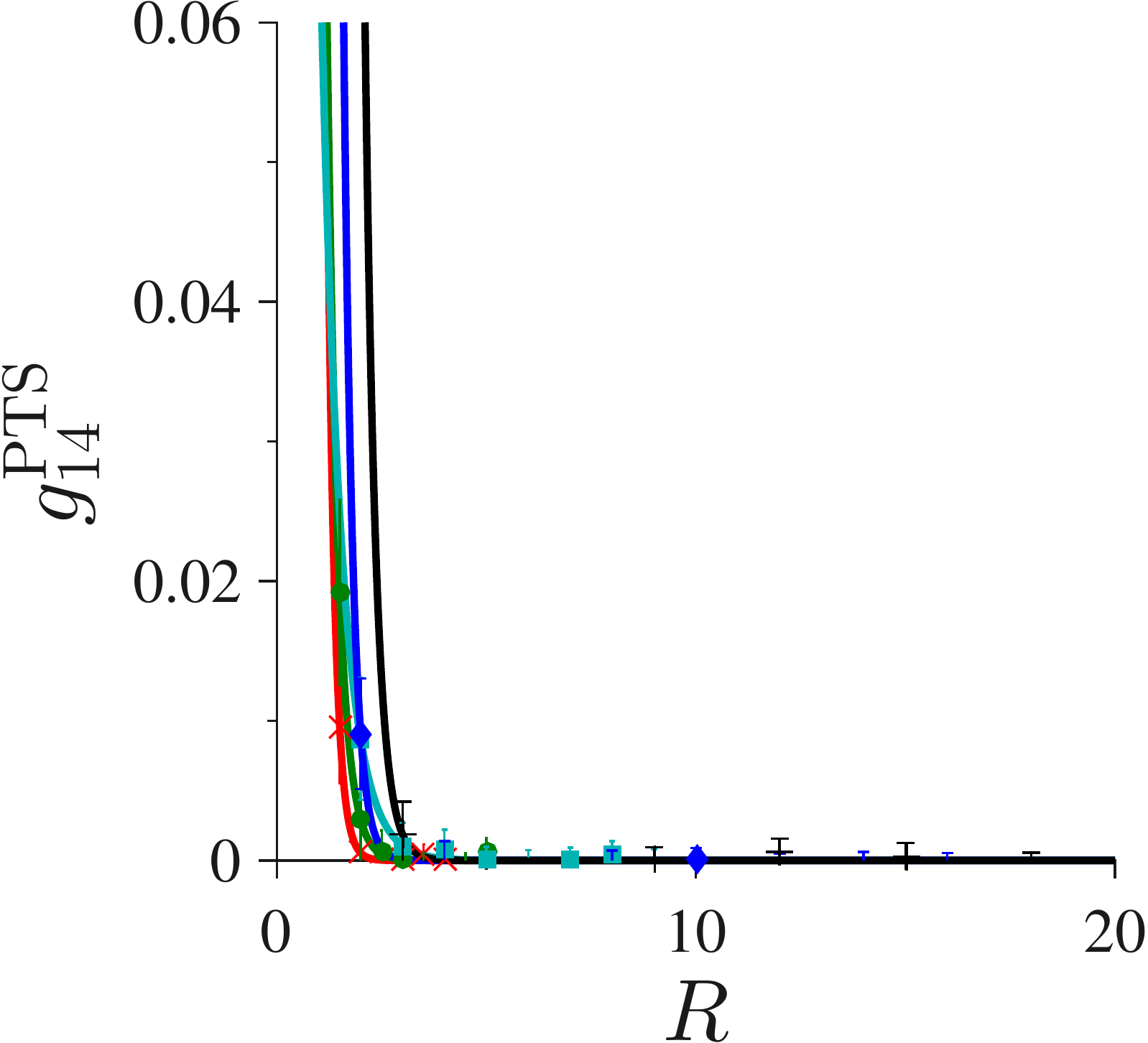}
\includegraphics[width=0.2\textwidth]{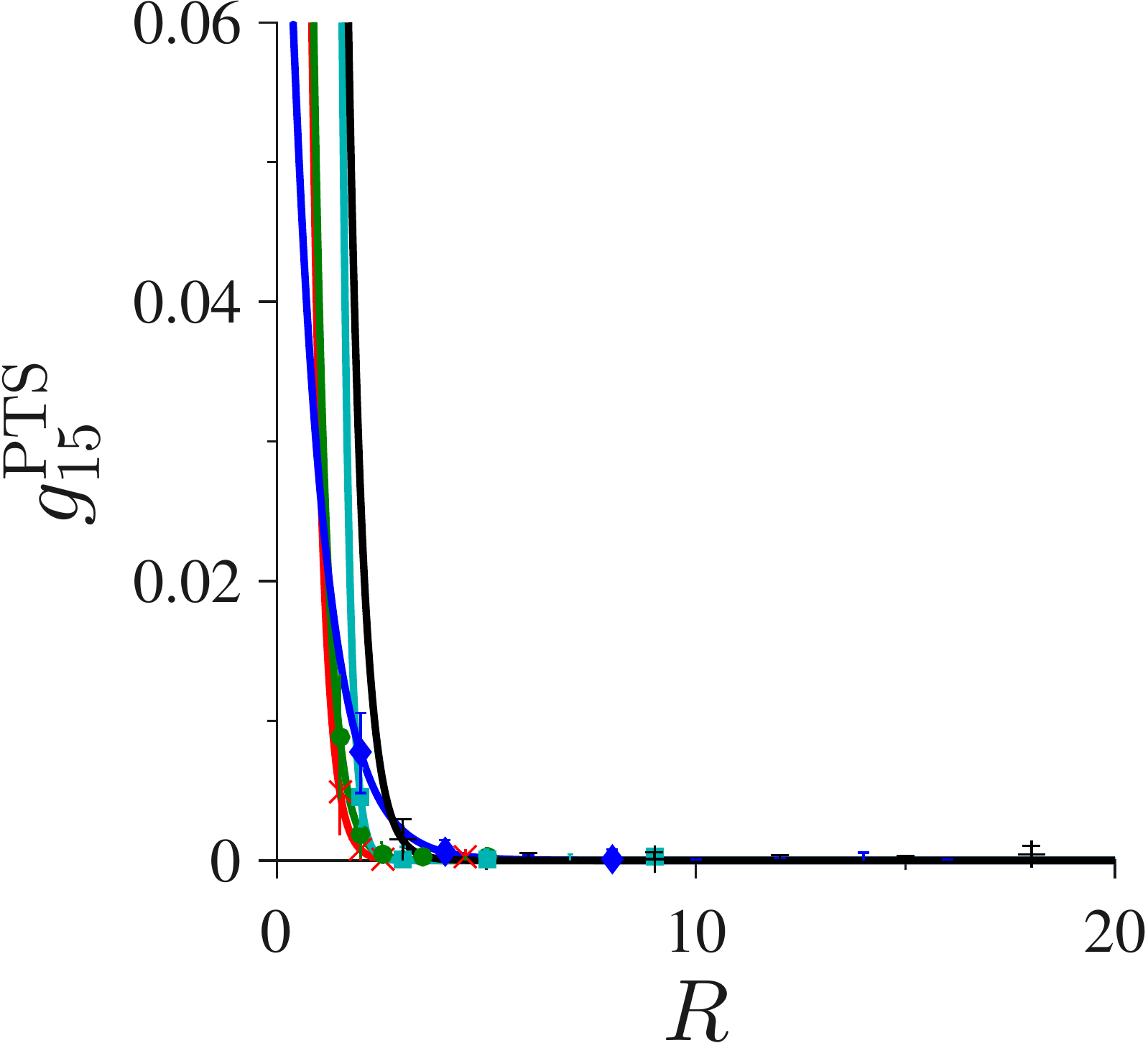}
\includegraphics[width=0.2\textwidth]{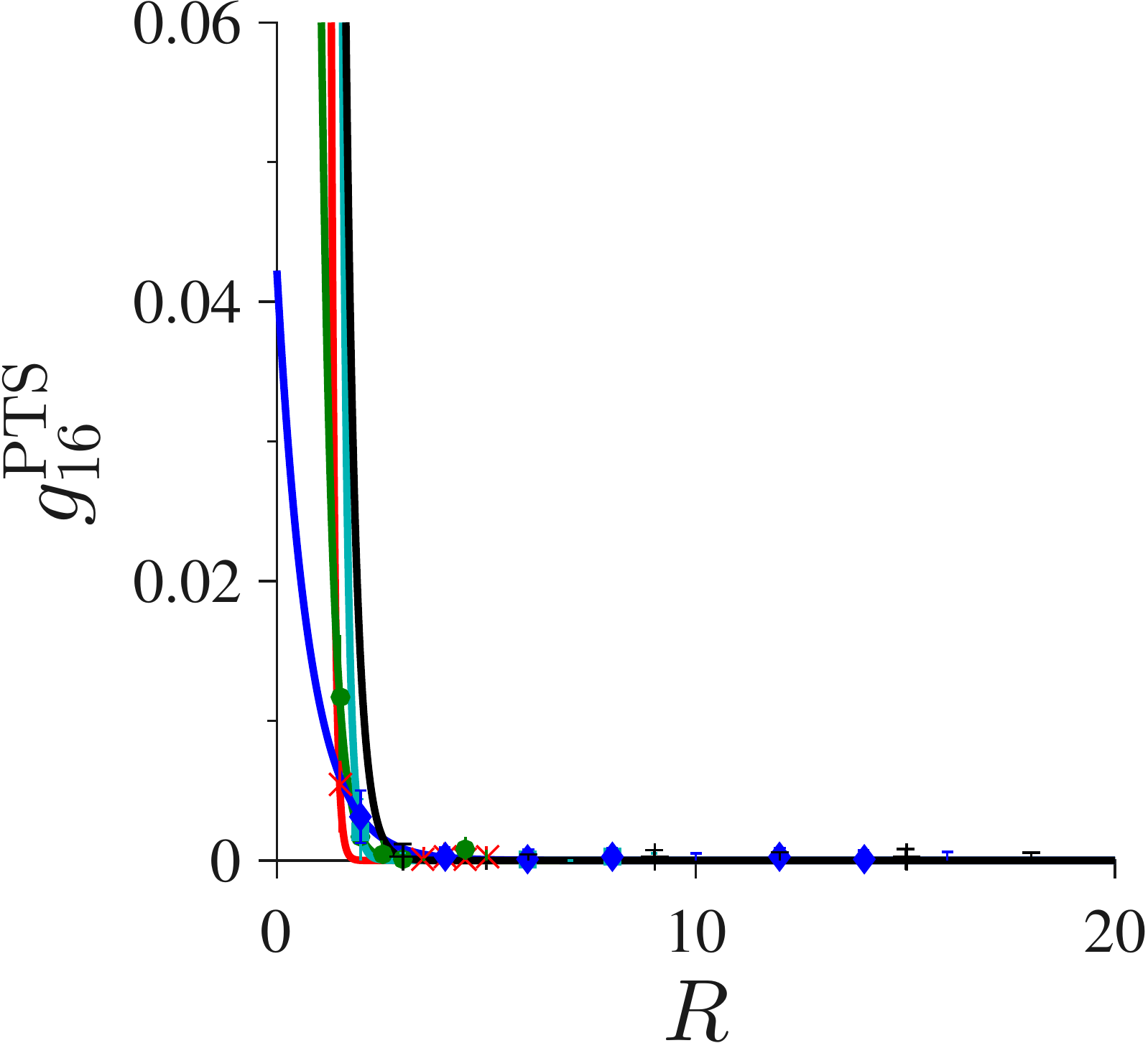}
\caption{Radial decay of the bond-orientational PTS correlations for monodisperse hard disks, $g_{\ell}^{\mathrm{PTS}}(R)$, for $\ell=1,\ldots,16$ at $\PF = 0.600$ (red-cross), $0.650$ (green-circle), $0.680$ (cyan-square), $0.690$ (blue-diamond), and $0.695$ (black-plus).
Note that the vertical axes have the same range for all $\ell$ except for $\ell = 6$.
Solid lines are exponential fits.}
\label{fig_BOoverlap_2D_additional}
\end{figure*}
\subsection{Positional overlap}

Denote a pair of configurations ${\bf X}=\le\{{\bf x}_i\ri\}$ and ${\bf Y}=\le\{{\bf y}_i\ri\}$. Here, ${\bf X}$ corresponds to the frozen initial configuration and ${\bf Y}$ to re-equilibrated configurations. For hard disks, we thus have $2\times200$ such pairs for each cavity. For each particle ${\bf x}_i$, find the nearest particle ${\bf y}_{i_{\rm nn}}$ (of the same species for the KABLJ liquid; no such proviso for mono/polydisperse systems).
Assign an overlap value $Q^{\bf X;Y}_{\rm pos} \le({\bf x}_i\ri)\equiv w\le(\big|{\bf x}_i-{\bf y}_{i_{\rm nn}}\big|\ri)$, where
\begin{equation}
w(z)\equiv {\rm exp}\le[-\le(\frac{z}{b}\ri)^2\ri]
\end{equation}
with $b=0.2$. This choice defines overlap values $Q^{\bf X;Y}_{\rm pos} \le({\bf x}_i\ri)$ at scattered points $\le\{{\bf x}_i\ri\}$. We then define $Q^{\bf X;Y}_{\rm pos} \le({\bf r}\ri)$ to be a continuous function passing through these points. Specifically, we first subdivide space with a Delaunay tessellation and within each simplex associate linearly interpolated values. Similarly we obtain $Q^{\bf Y;X}_{\rm pos} \le({\bf r}\ri)$, and finally define $Q^{\bf XY}_{\rm pos}\le({\bf r}\ri)\equiv \frac{1}{2}\le\{Q^{\bf X;Y}_{\rm pos}\le({\bf r}\ri)+Q^{\bf Y;X}_{\rm pos}\le({\bf r}\ri)\ri\}$.

The overlap around the core of the cavity, 
\begin{equation}
Q^{\bf XY}_{\rm pos}\equiv\frac{\int_{|{\bf r}|<r_{\rm c}}d{\bf r}\ Q^{\bf XY}_{\rm pos}
\le({\bf r}\ri)}{2\pi^{d/2} r_{\rm c}^d/\Gamma\le(d/2\ri)},
\end{equation}
is evaluated through Monte Carlo integration, with $10,000$ points within a ball of size $r_{\rm c}=0.5$ for the $d=3$ KABLJ liquid and $1000$ points with $r_{\rm c}=0.5$ for $d=2$ hard disks.
Here, ${\bf r}={\bf 0}$ denotes the center of the cavity.

The positional PTS correlation function is
\begin{equation}
g^{\rm PTS}_{\rm pos}\le(R\ri)\equiv\le[\langle Q_{\rm pos}\rangle_{J(R)}\ri]-\langle Q_{\rm pos}\rangle_{\rm{bulk}}\, ,
\end{equation}
where $\langle...\rangle_{J(R)}$ denotes the thermal average over re-equilibrated configurations inside the cavity with quenched disorder $J(R)$ set by a pinned external configuration, and $[...]$ the average over disorders ({\it i.e.}, average over cavity centers).
We have subtracted the bulk value corresponding to $R=\infty$ such that the PTS correlation function vanishes at infinity. The bulk value is evaluated by taking $10^5$ pairs of independent configurations in bulk samples for the mono/polydisperse systems and $4000$ pairs for the KABLJ model.

In this notation, the positional PTS susceptibility is 
\begin{equation}
\chi_{\rm pos}^{\rm PTS}\le(R\ri)\equiv \le[\langle Q_{\rm pos}^2\rangle_{J(R)}-\langle Q_{\rm pos}\rangle_{J(R)}^2\ri]\, .
\end{equation}

\subsection{2D orientational overlap for mono/poly-disperse}
In $d=2$, we define an $\ell$-orientational overlap field $Q^{\bf XY}_{\ell}\le({\bf r}\ri)$ by first associating an orientational value $\psi_{\ell,m}^{\mathbf X}\le(j\ri)$ to a particle located at ${\mathbf x}_j$, where $m=\pm1$. From a radical Voronoi tessellation, we find $N^{\mathbf X}_j$ Delaunay neighbors $k$; for each neighbor $k$, we define an angle between them $\theta^{\mathbf X}_{j,k}$ by expressing ${\bf x}_{k}-{\bf x}_j\equiv r_{j,k}({\rm cos}\theta^{\mathbf X}_{j,k}, {\rm sin}\theta^{\mathbf X}_{j,k})$; and to the particle at ${\bf x}_j$ we associate the average value
\begin{equation}
\psi_{\ell,m}^{\mathbf X}\le(j\ri)\equiv\frac{1}{N_j^{\mathbf X}}\sum_{k/j}e^{i m\ell\theta^{\mathbf X}_{j,k}}\, .
\end{equation}
Then, for each point in space ${\bf r}$, we find the nearest particle in ${\bf X}$, $j_{\star}$, and associate its value to the point, $\psi_{\ell,m}^{\mathbf X}\le({\bf r}\ri)=\psi_{\ell,m}^{\mathbf X}\le(j_{\star}\ri)$.
Similarly, we obtain $\psi_{\ell,m}^{\mathbf Y}\le({\bf r}\ri)$. 
We finally define the $\ell$-orientational overlap field as 
\begin{equation}
Q_\ell^{\mathbf X \mathbf Y}(\mathbf r)=\prefac\sum_{m=\pm1}\le\{\psi_{\ell,m}^{\mathbf X}(\mathbf r)\ri\}^*\psi_{\ell,m}^{\mathbf Y}(\mathbf r)\, 
\end{equation}
with $\prefac\equiv1/2$.
Note that this quantity is real and independent of the choice of axis in defining angles.

The $\ell$-orientational PTS correlation functions have a similar definition as the positional ones.
Figure~\ref{fig_BOoverlap_2D_additional} shows the bond-orientational overlap functions $g_{\ell}^{\mathrm{PTS}}(R)\equiv\le[\langle Q_\ell\rangle\ri]( R)$ (for bond-orientational correlation, the bulk value is zero by symmetry) for $\ell=1$ to $16$ for  monodisperse hard disks. (We have also studied values of $\ell$ from $17$ to $24$ but the resulting curves are noisy. Furthermore, for high values of $\ell$, small displacements originating from mere vibrations tend to decorrelate the overlap over short distances no bigger than the interatomic distance.)
Note the growing length scale for $\ell=12$ as the symmetry is compatible with sixfold order.
The bond-orientational correlations for $\ell$ incompatible with the sixfold order, in contrast, do not track its growth and instead their spatial extents stay within the order of positional scale $\xi_{\rm pos}$.

\begin{figure*}
\centerline{
\includegraphics[width=0.18\textwidth]{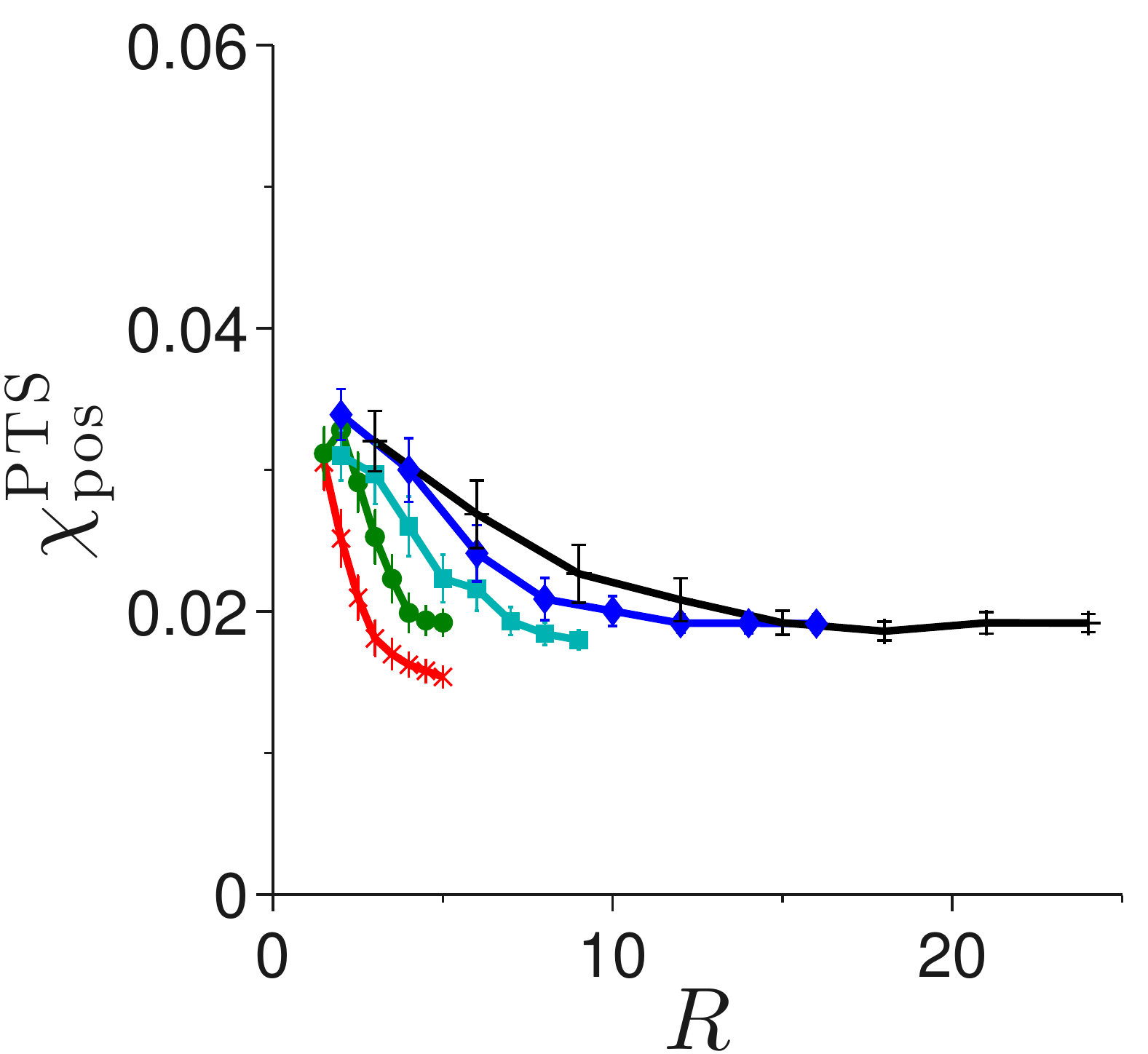}
\includegraphics[width=0.18\textwidth]{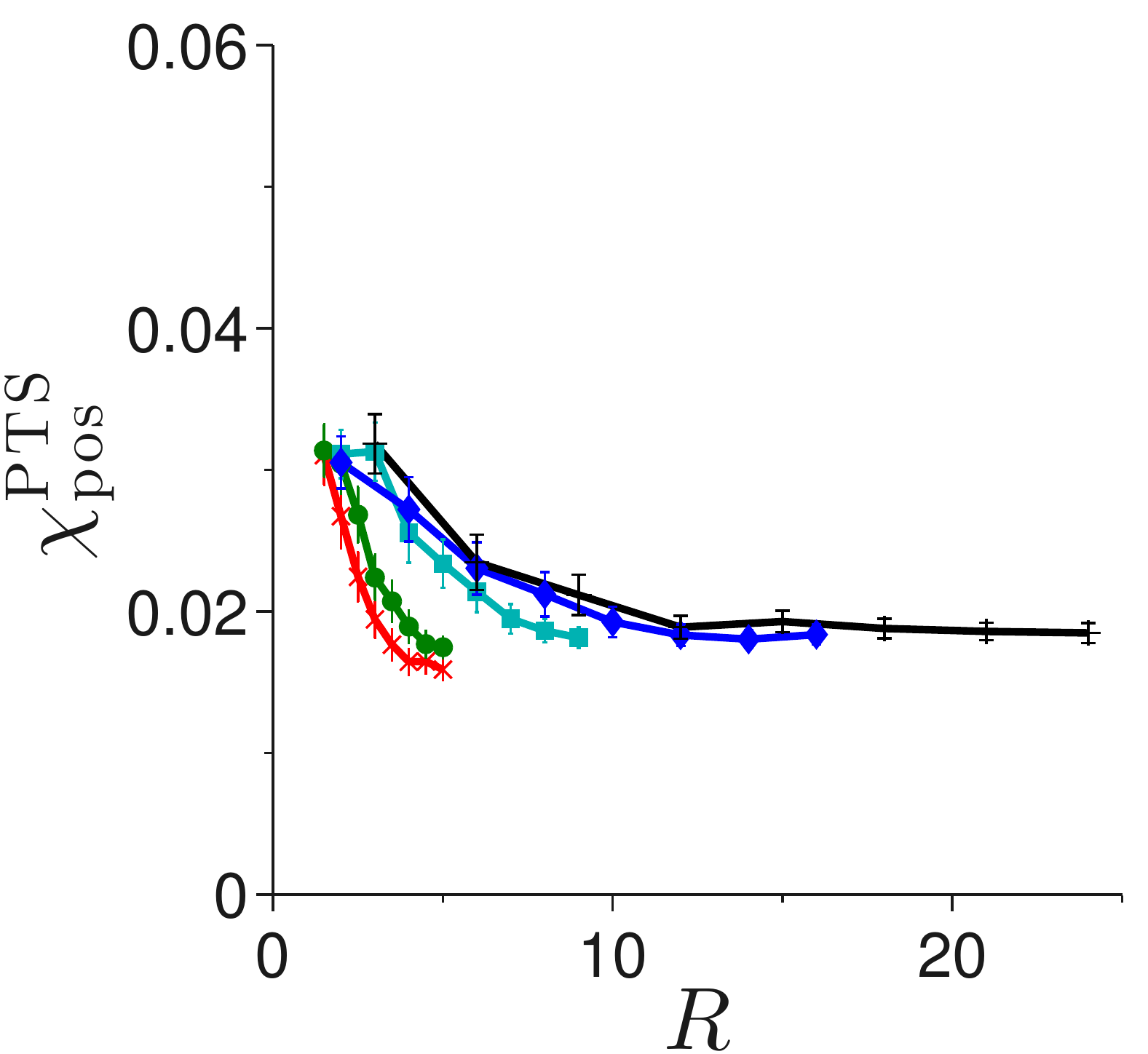}
\includegraphics[width=0.18\textwidth]{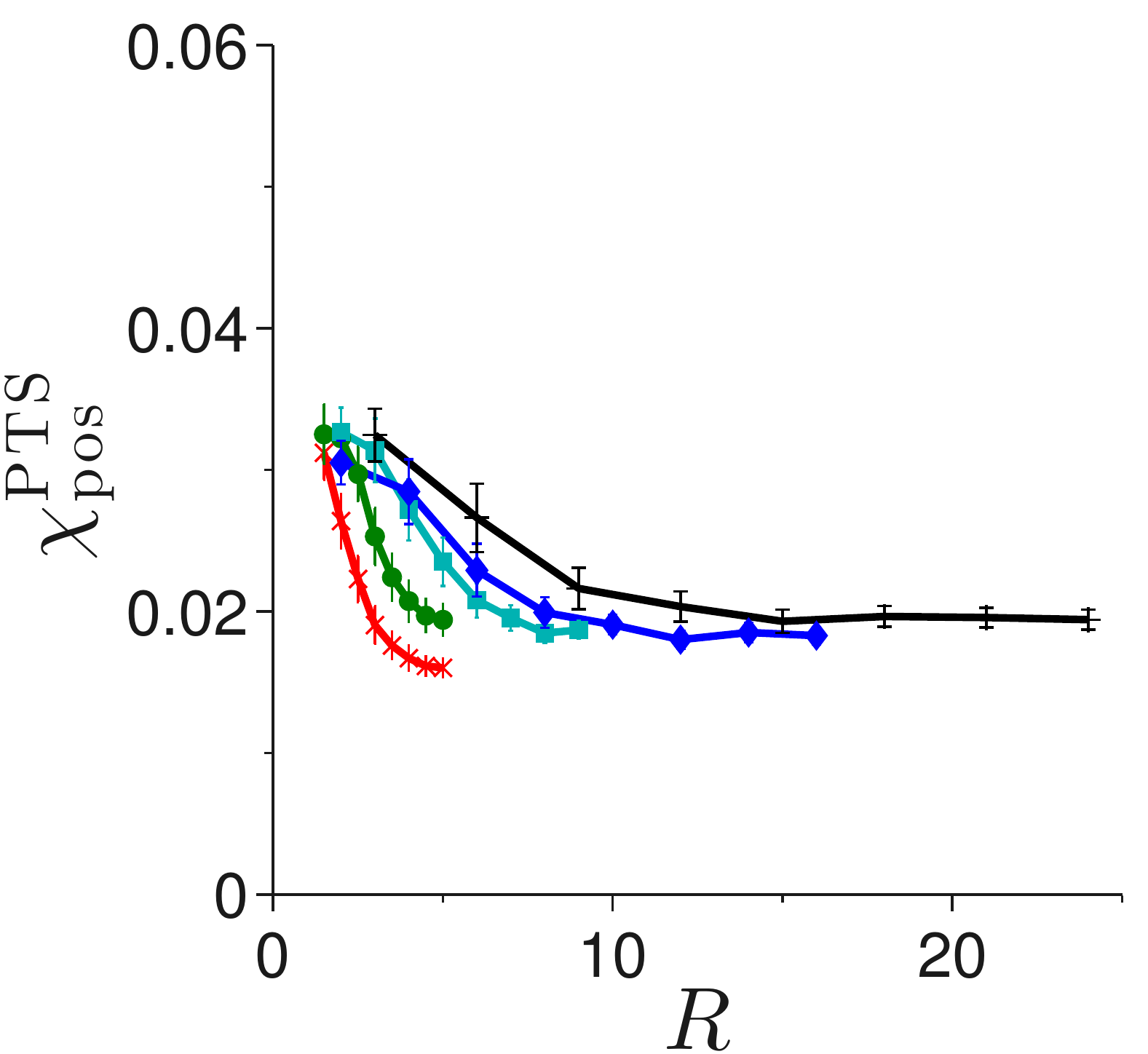}
\includegraphics[width=0.18\textwidth]{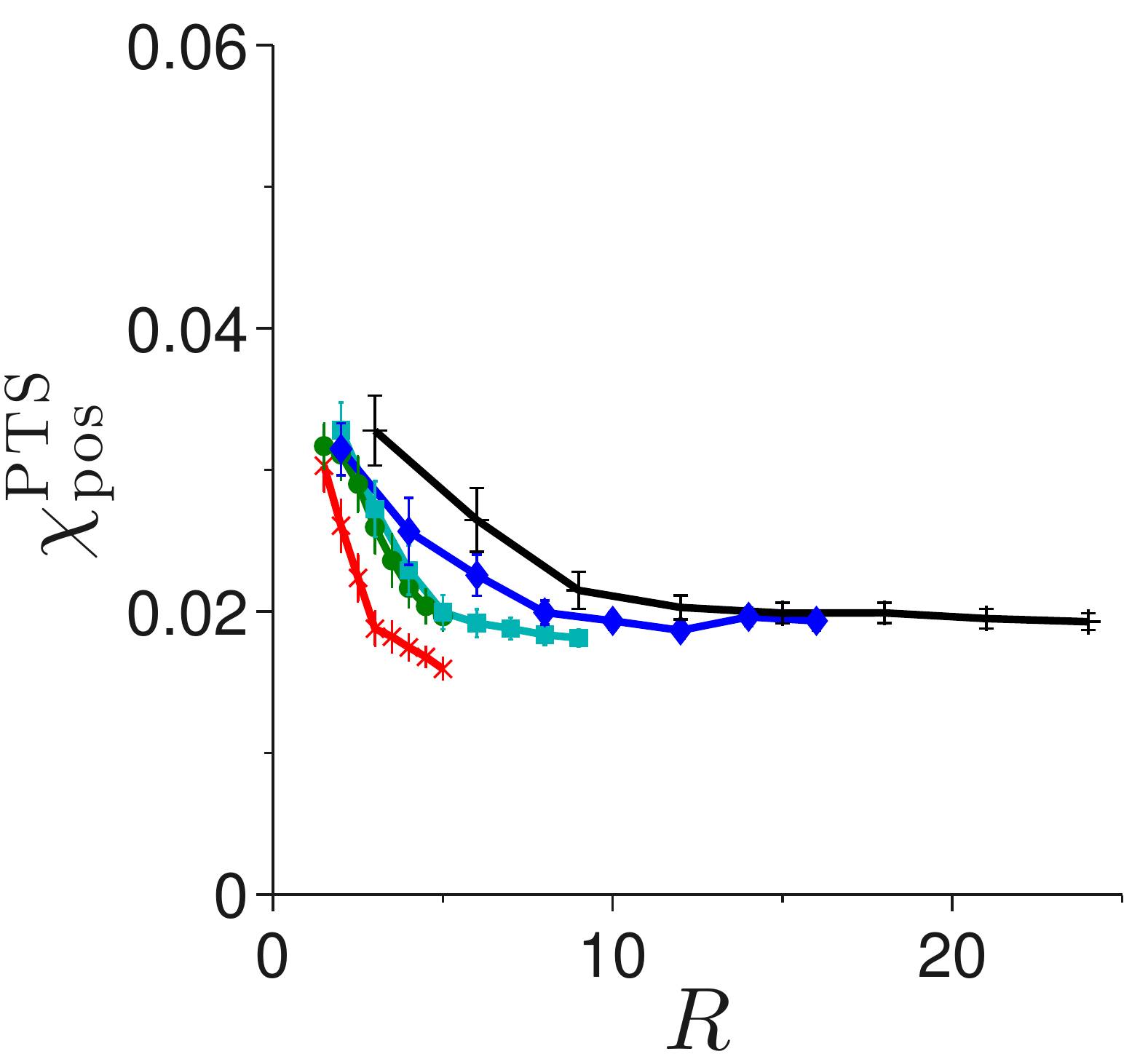}
\includegraphics[width=0.18\textwidth]{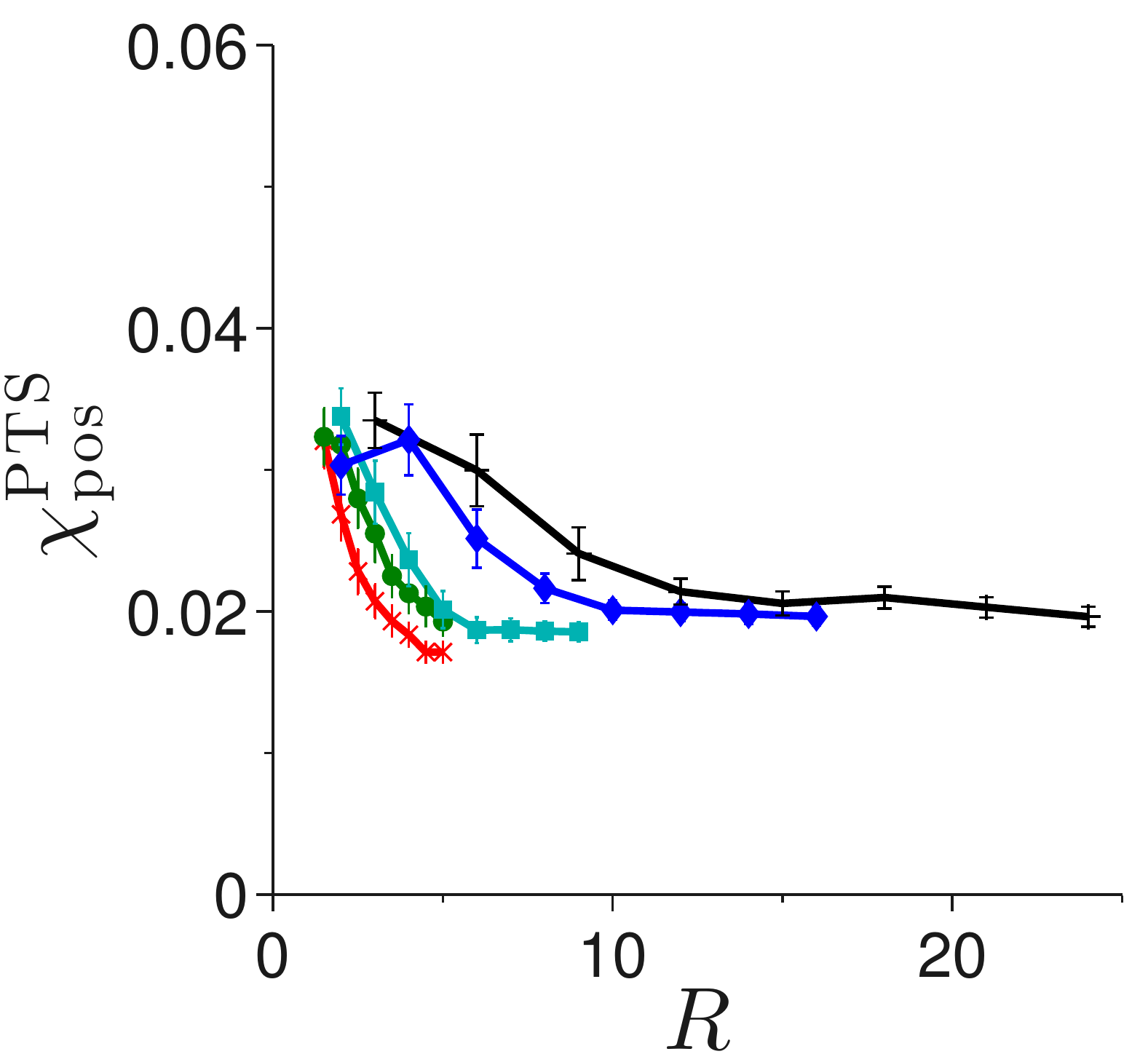}}
\centerline{
\includegraphics[width=0.18\textwidth]{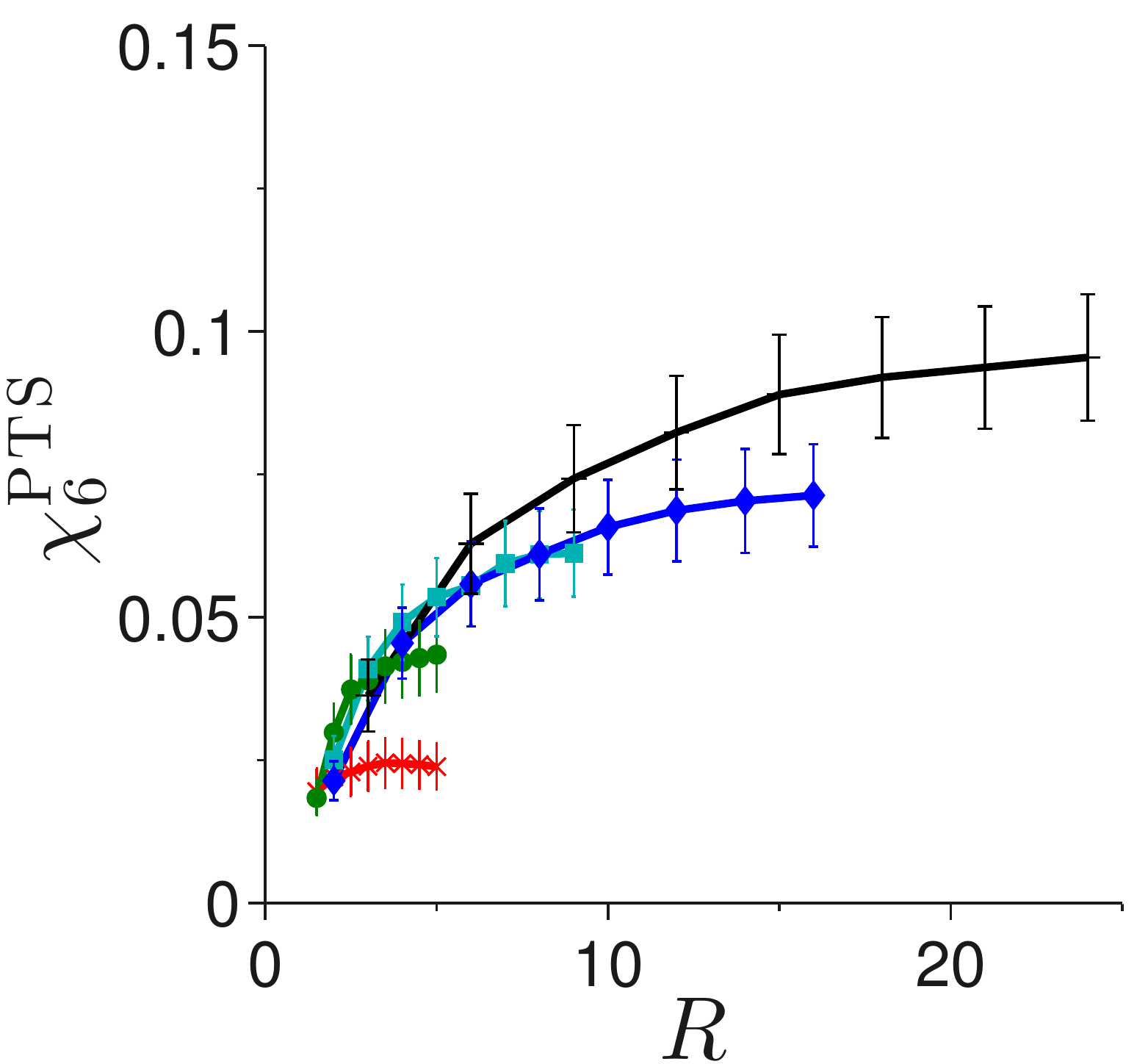}
\includegraphics[width=0.18\textwidth]{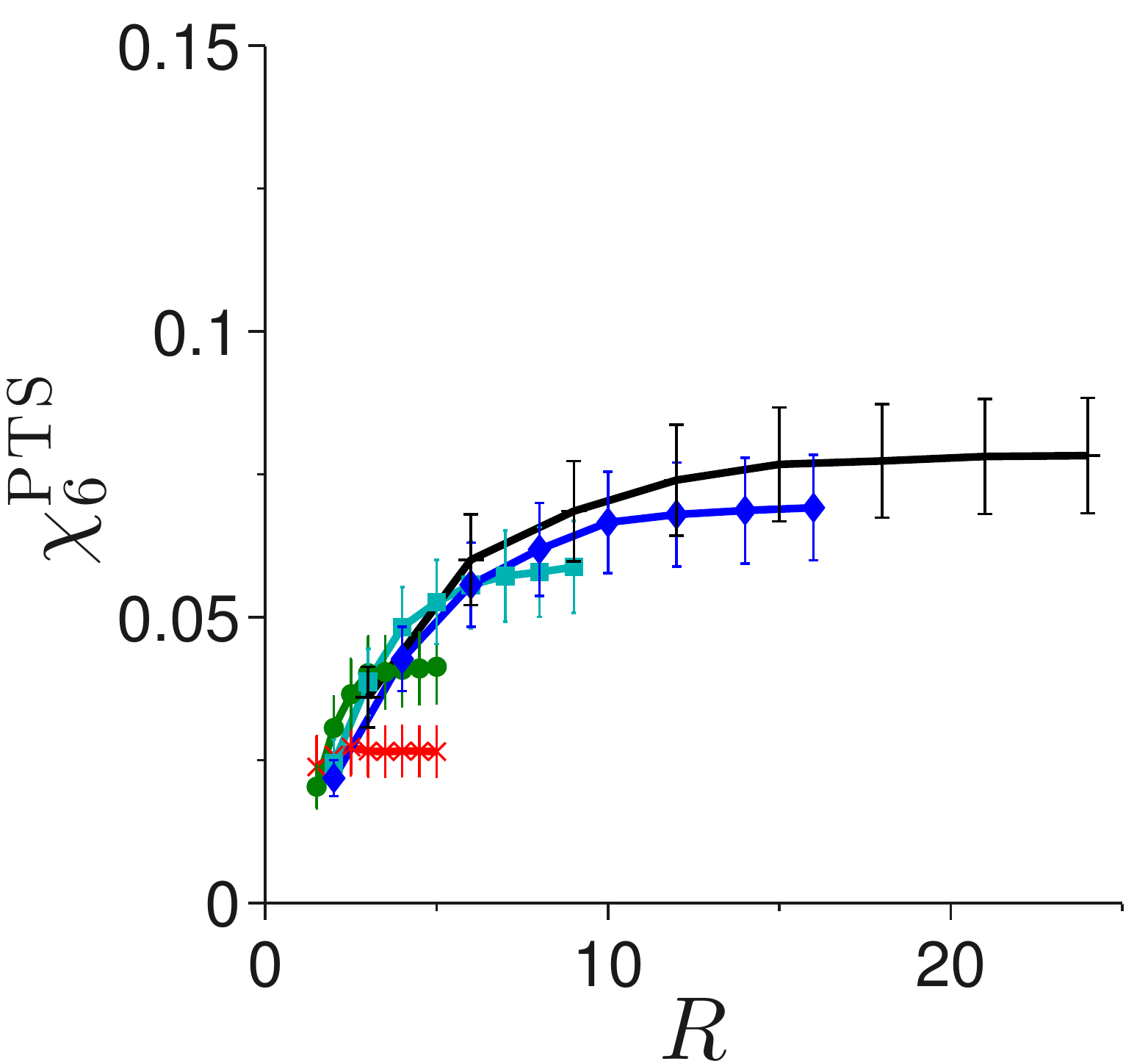}
\includegraphics[width=0.18\textwidth]{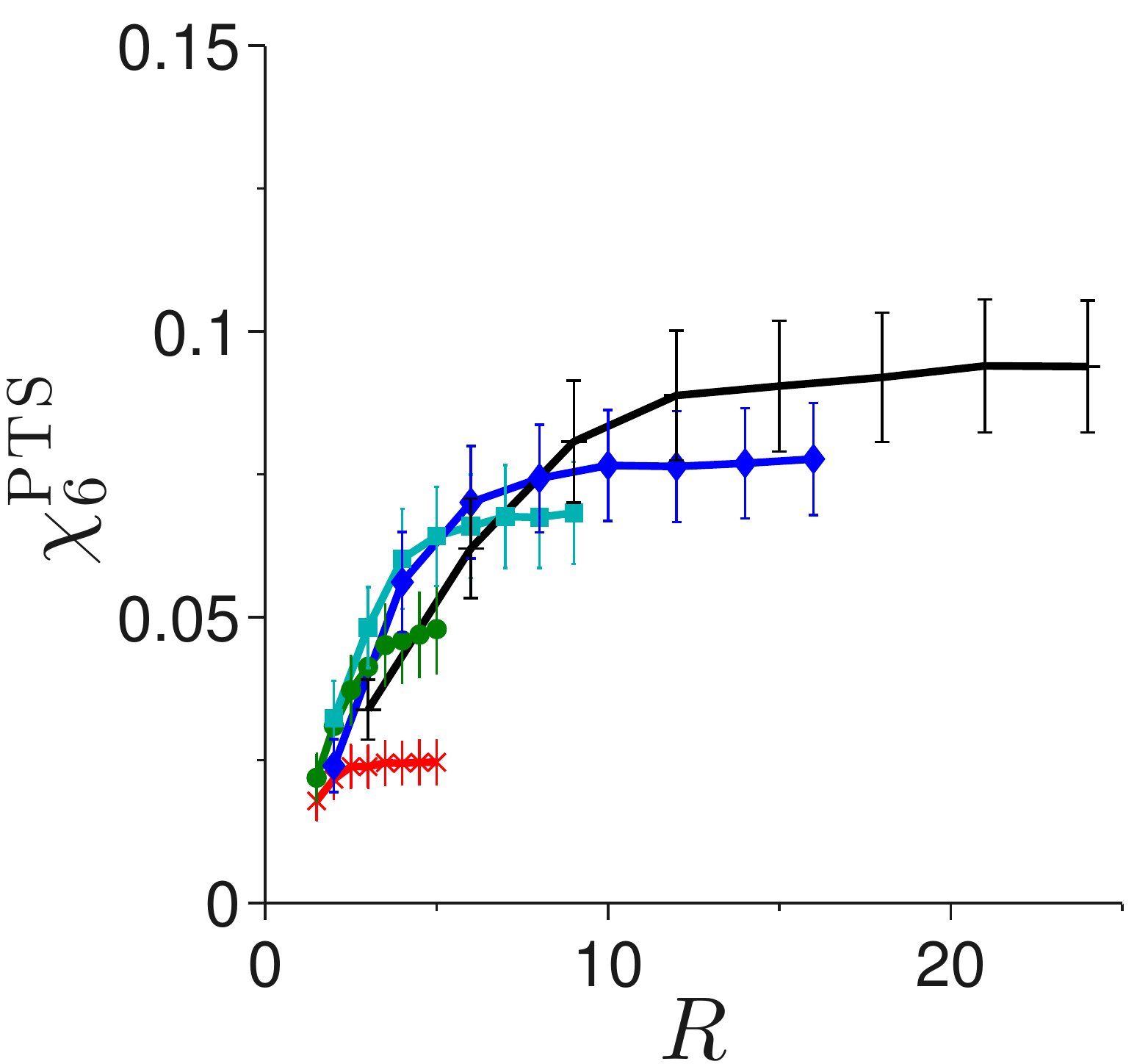}
\includegraphics[width=0.18\textwidth]{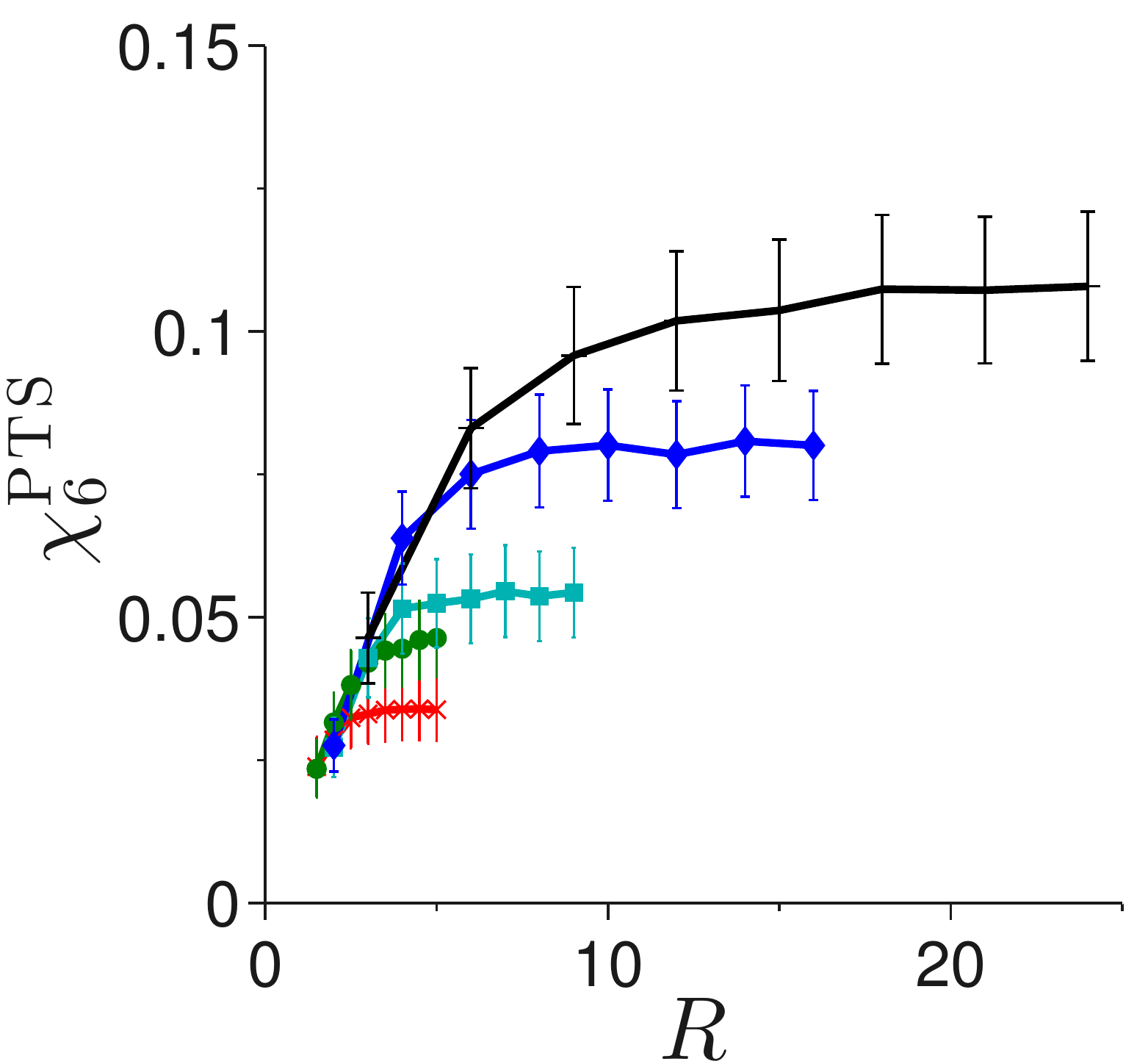}
\includegraphics[width=0.18\textwidth]{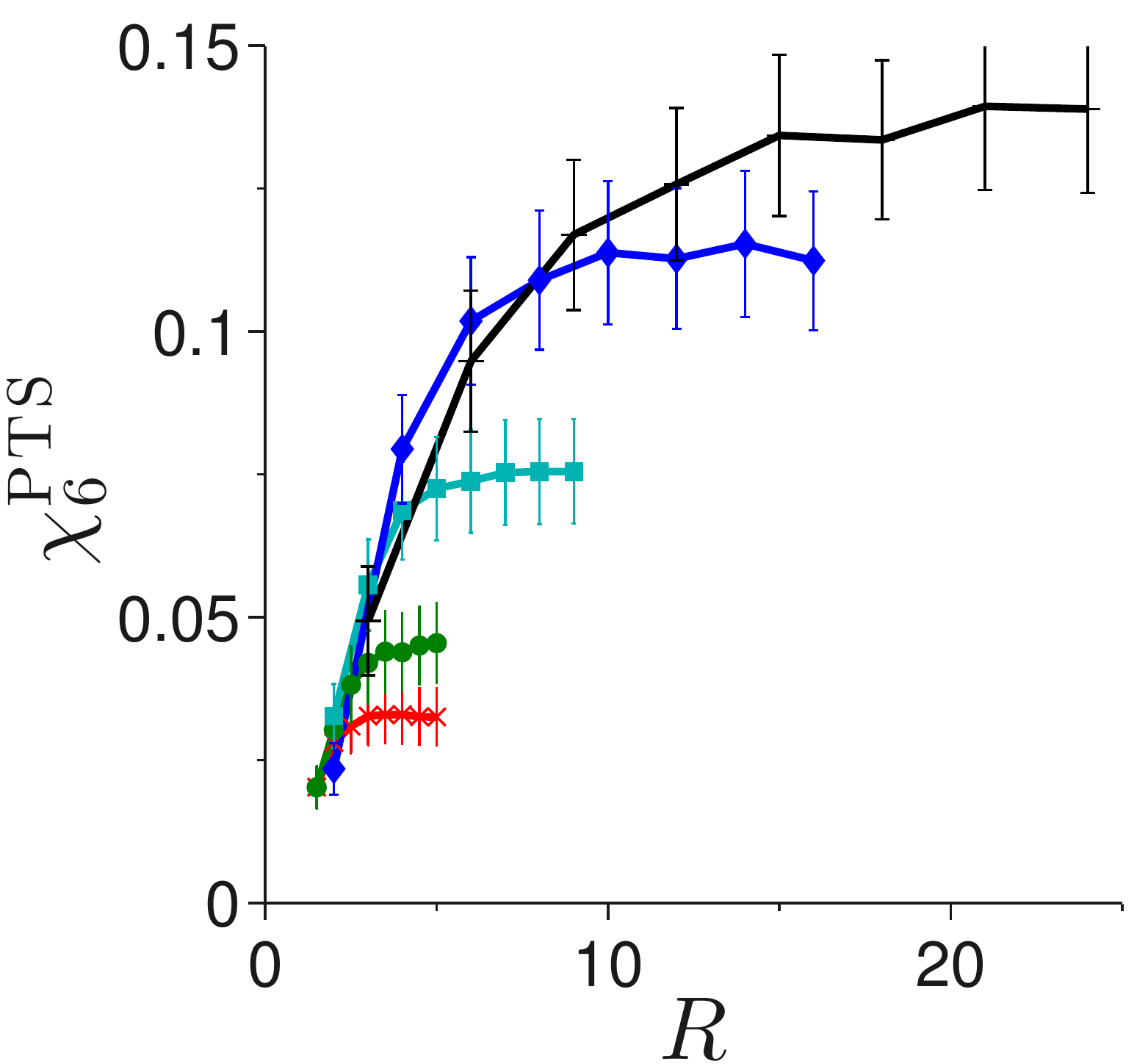}}
\centerline{\hspace{+0.2in}
\includegraphics[width=0.18\textwidth]{PTS_pair0.pdf}
\includegraphics[width=0.18\textwidth]{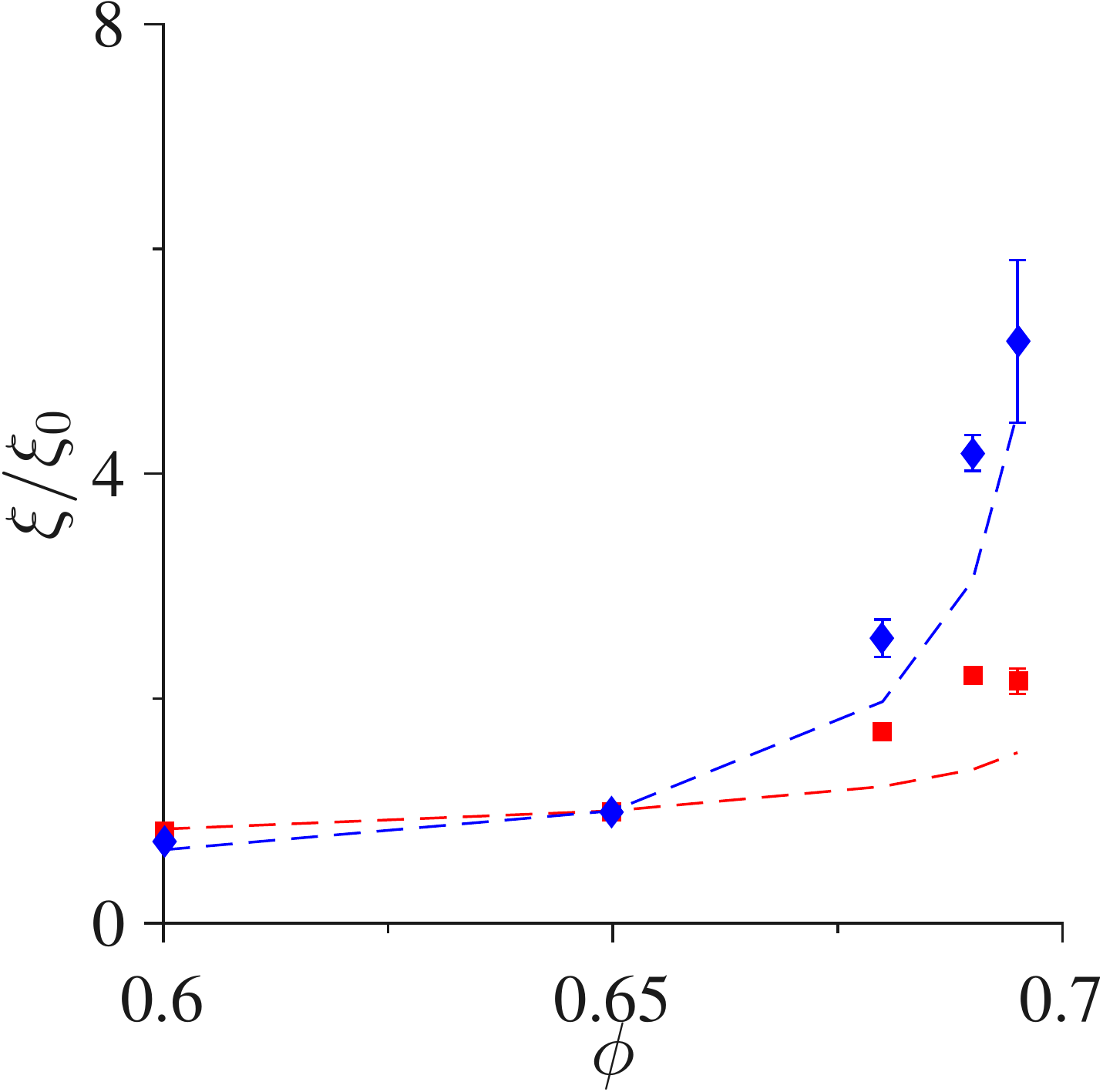}
\includegraphics[width=0.18\textwidth]{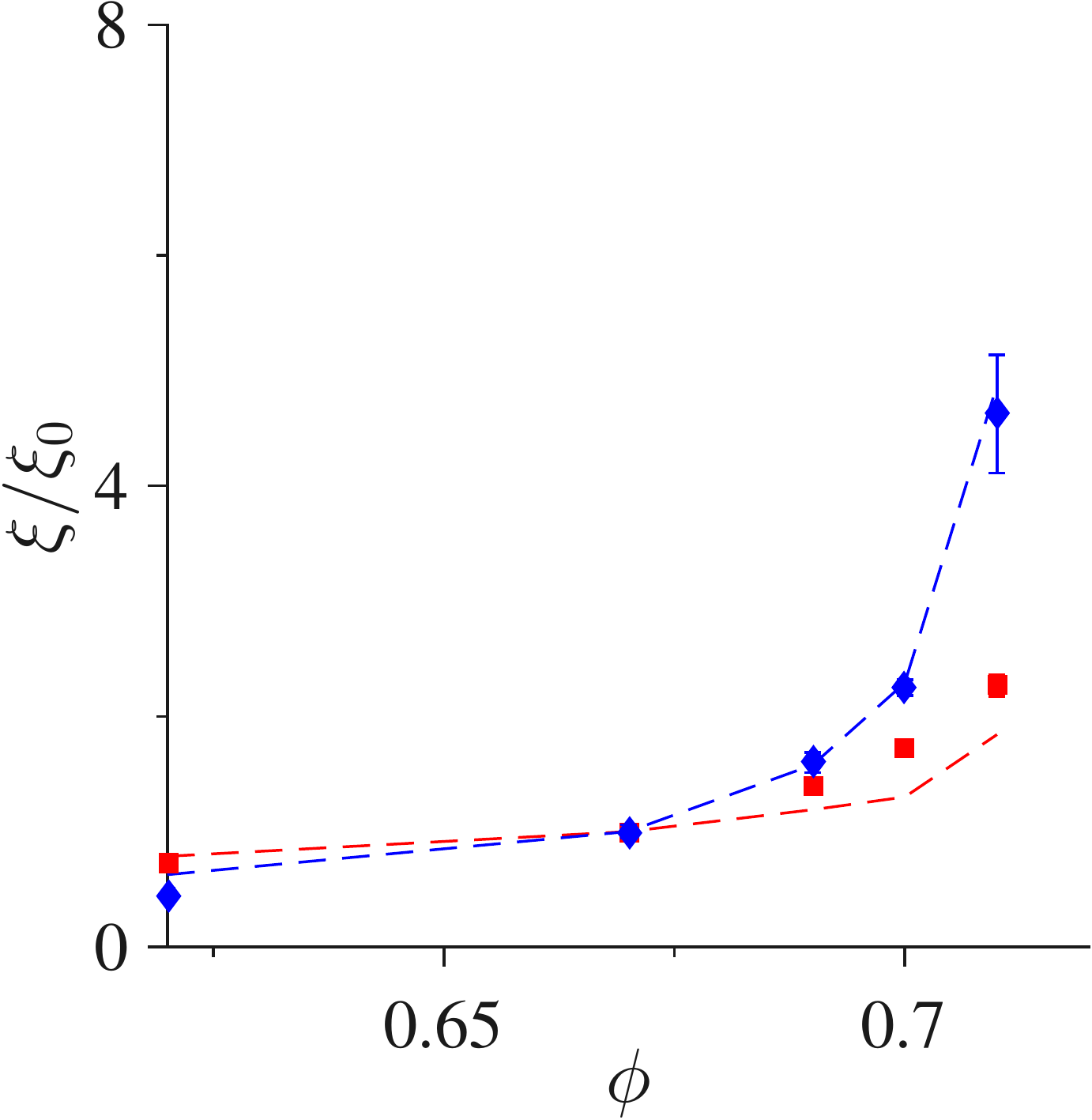}
\includegraphics[width=0.18\textwidth]{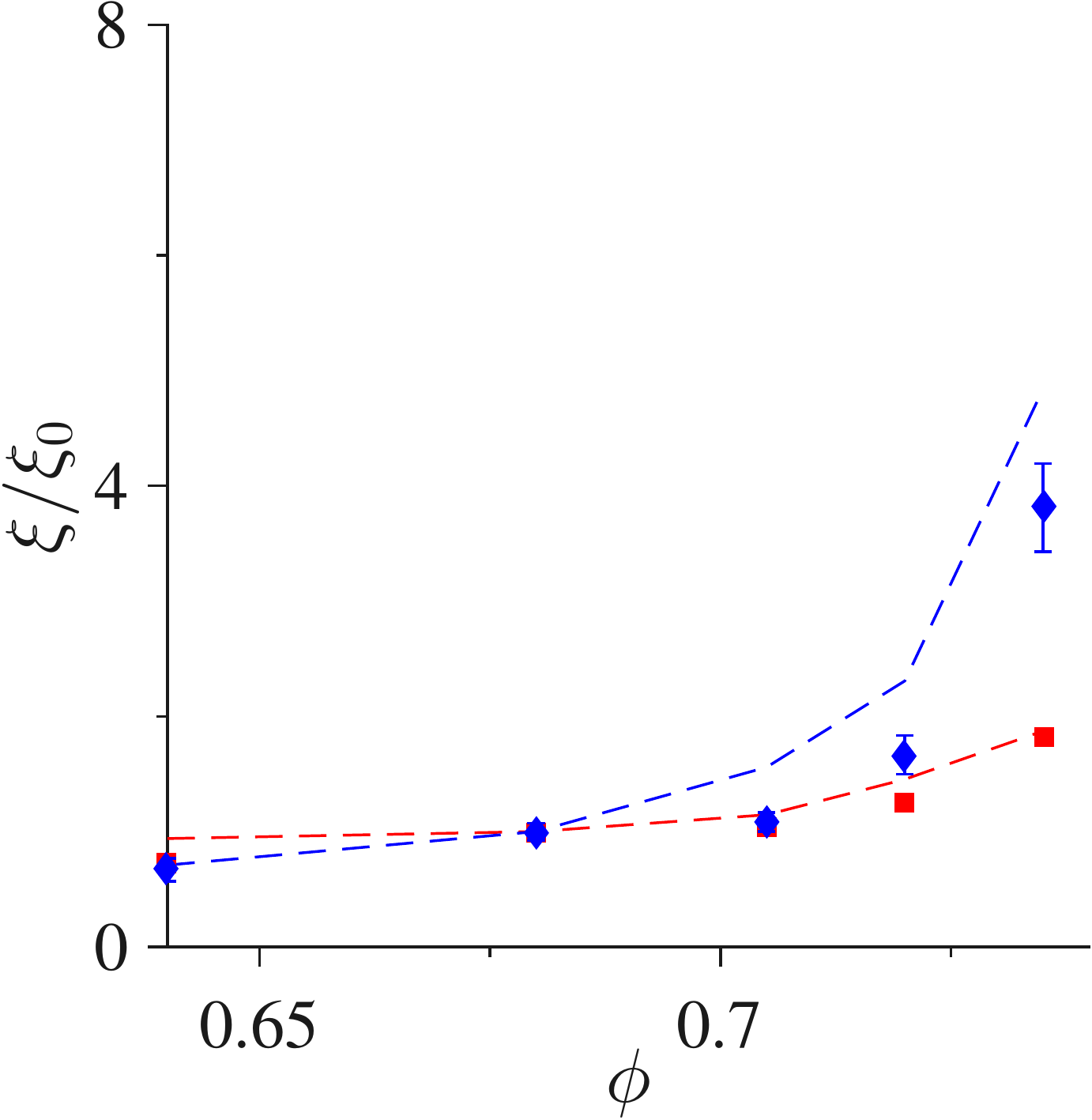}
\includegraphics[width=0.18\textwidth]{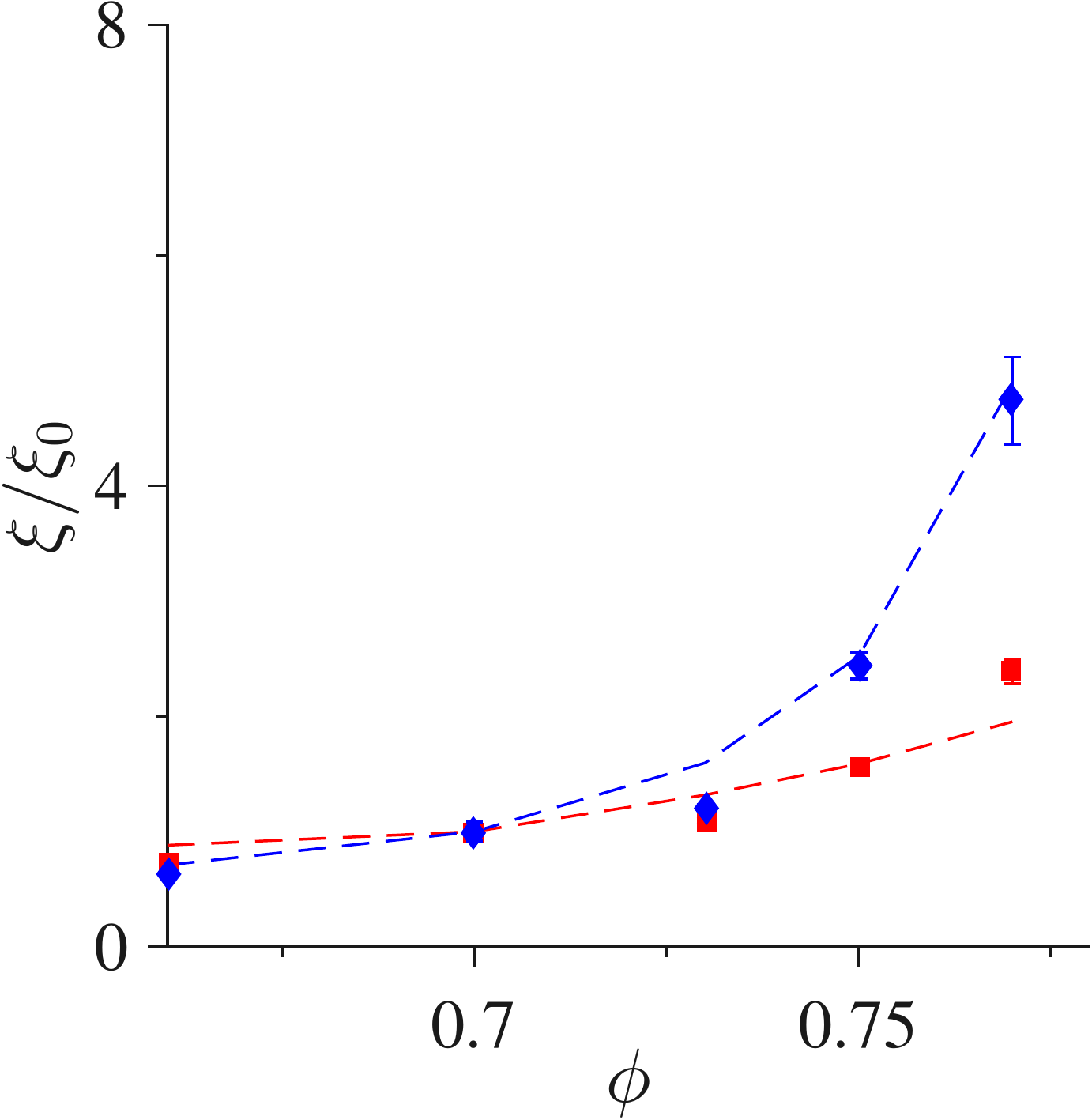}}
\caption{Positional $\chi^{\rm PTS}_{\rm pos}$ (top row) and $\ell=6$ bond-orientational $\chi^{\rm PTS}_{6}$ (middle row) PTS susceptibilities as a function of cavity radius $R$ for hard disks with polydispersity $\Delta=0\%, 3\%, 6\%, 9\%, 11\%$ (from left to right), at $\PF = \PF_1$ (red-cross), $\PF_2$ (green-circle), $\PF_3$ (cyan-square), $\PF_4$ (blue-diamond), and $\PF_5$ (black-plus), where the values of $\PF_i$ at each polydispersity are the same as in Fig.~\ref{phase_diagram}. The bottom row, for the same range of polydispersity, records the growth of positional (red-square) and hexatic (blue-diamond) PTS lengths extracted through the exponential fits to PTS correlations. Dashed lines are positional and sixfold pair correlation lengths. Each length is relative to $\xi_0\equiv\xi(\PF_0=\PF_2)$.}
\label{trends}
\end{figure*}

The $\ell$-orientational PTS susceptibilities can also be defined similarly to the positional ones.
Figure~\ref{trends} compares positional and $\ell=6$ bond-orientational PTS susceptibilities, $\chi_{\rm pos}^{\rm PTS}$ and $\chi_{\ell=6}^{\rm PTS}$, for polydispersity $\Delta=0\%, 3\%, 6\%, 9\%, 11\%$ at various packing fractions, where $0\%$ corresponds to monodisperse hard disks, along with the growth of associated PTS lengths.
We detect no qualitative changes as a function of polydispersity.
Note that there is no distinctive peak structure in susceptibilities and the bulk value of the $\ell=6$ bond-orientational susceptibility has a strong density dependence that tracks the growth of the local sixfold order.
We expect the same behavior to be seen in the simple ferromagnetic Ising model upon approaching its critical temperature from above.

\subsection{3D orientational overlap for Kob-Andersen}
\begin{figure*}
\centerline{
\includegraphics[width=0.2\textwidth]{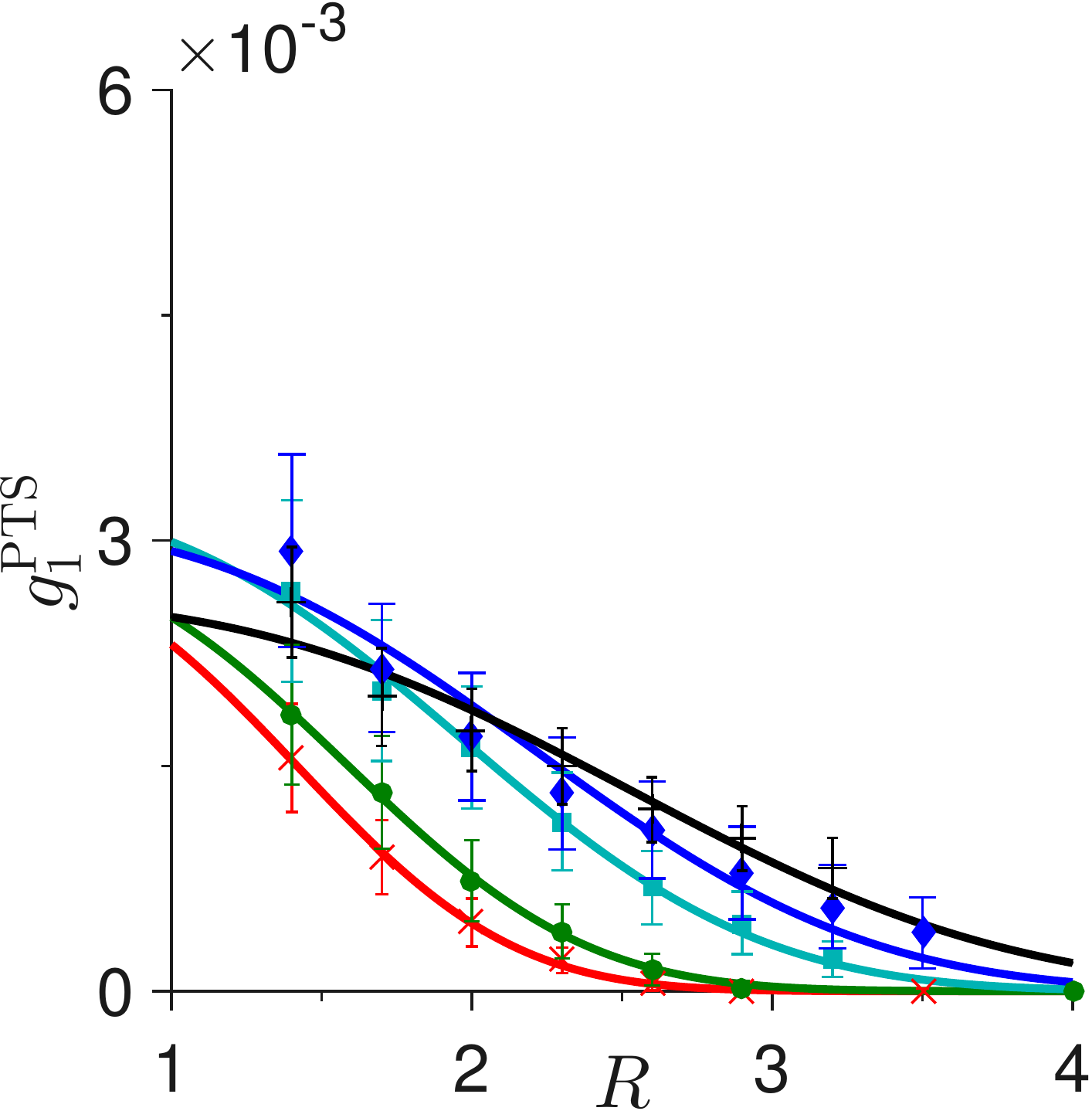}
\includegraphics[width=0.2\textwidth]{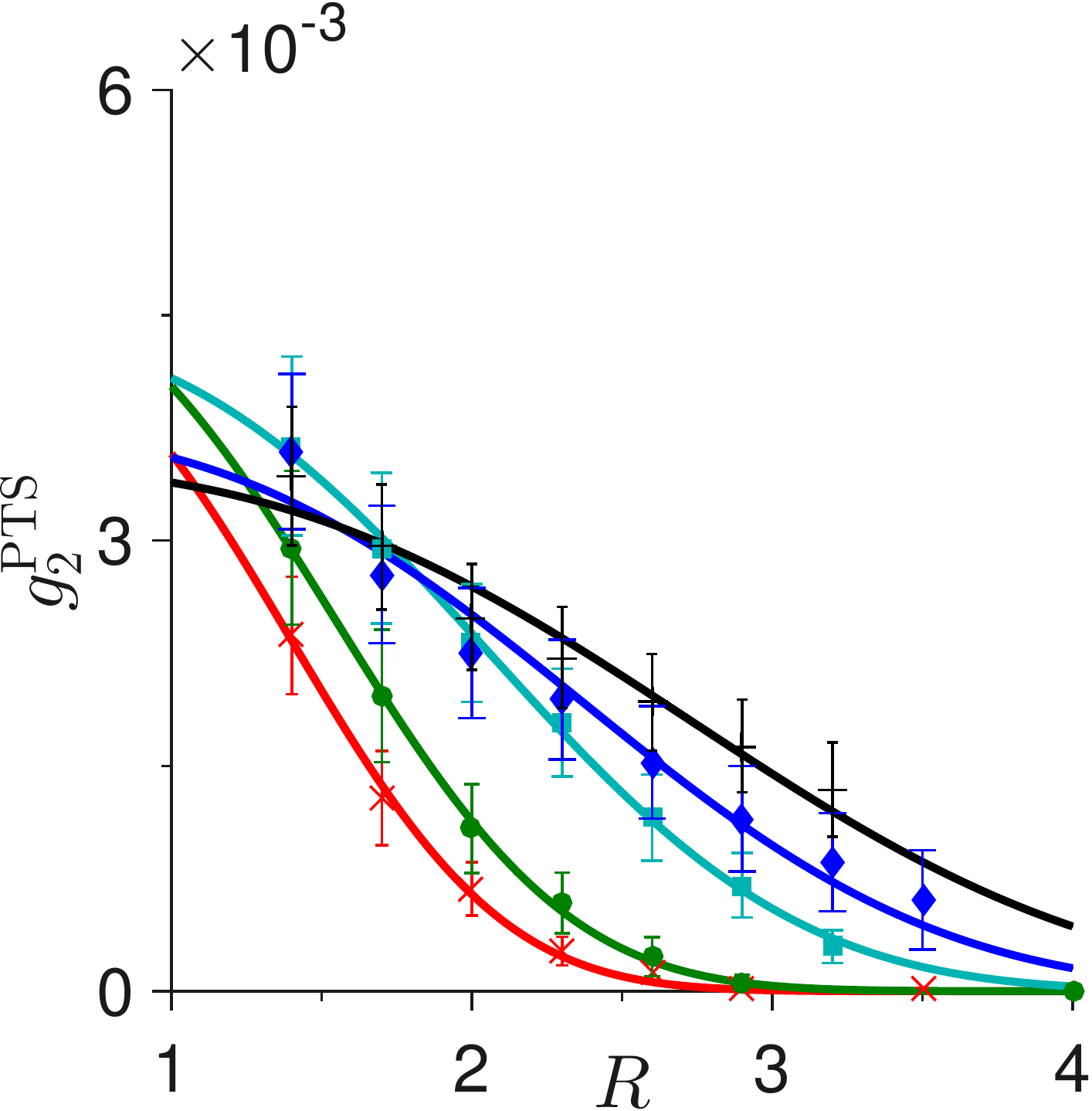}
\includegraphics[width=0.2\textwidth]{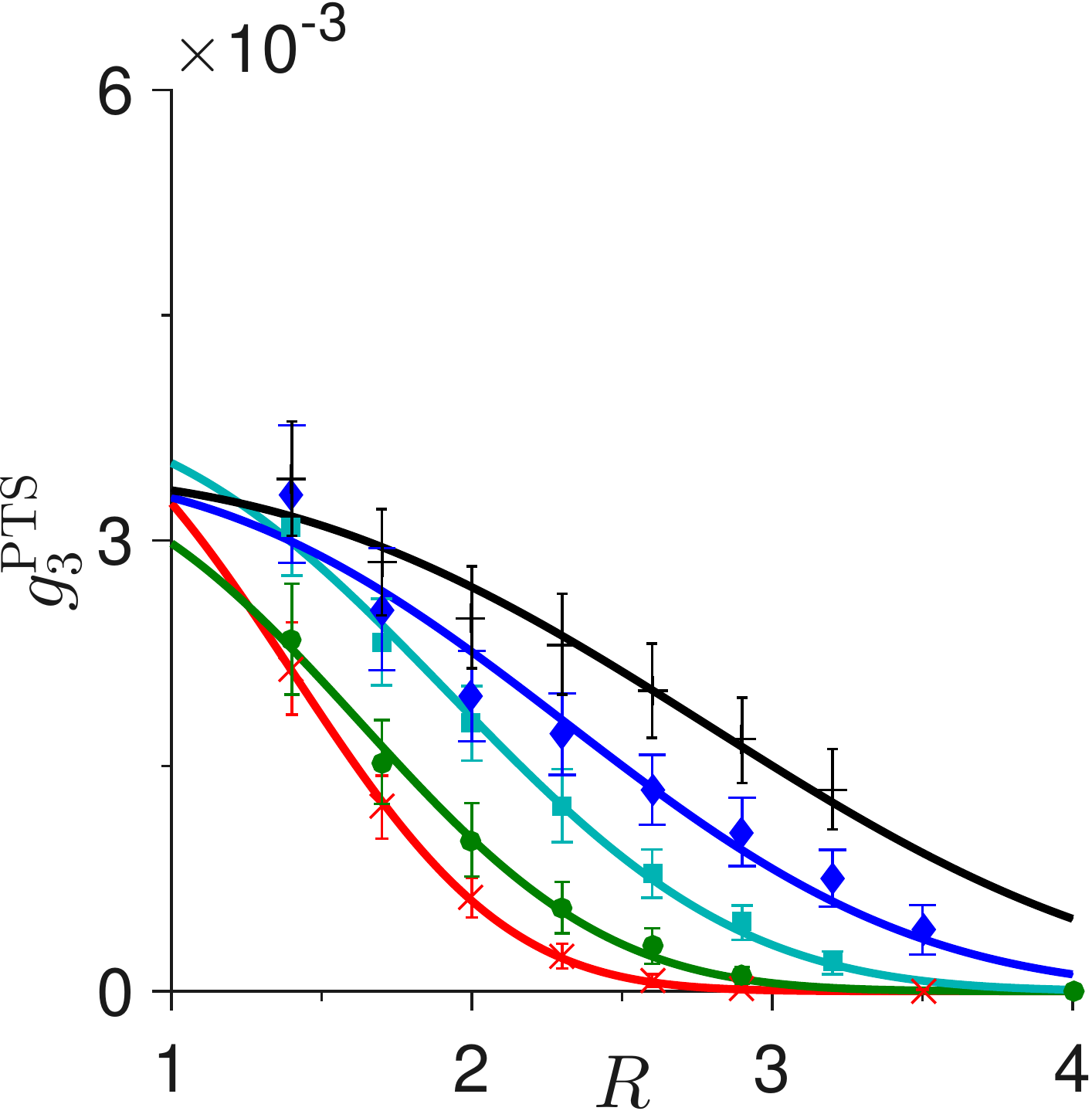}
\includegraphics[width=0.2\textwidth]{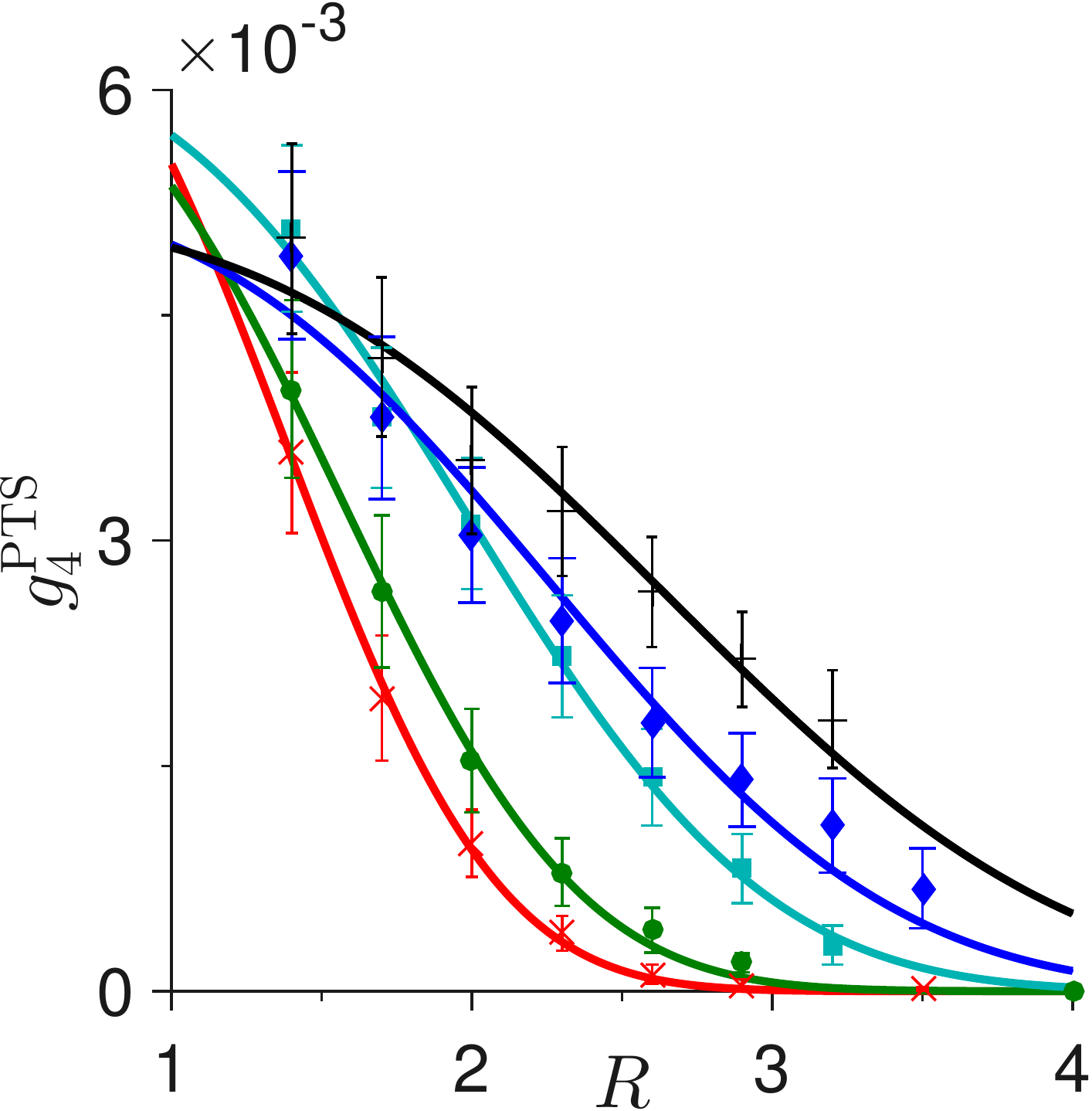}
}
\centerline{
\includegraphics[width=0.2\textwidth]{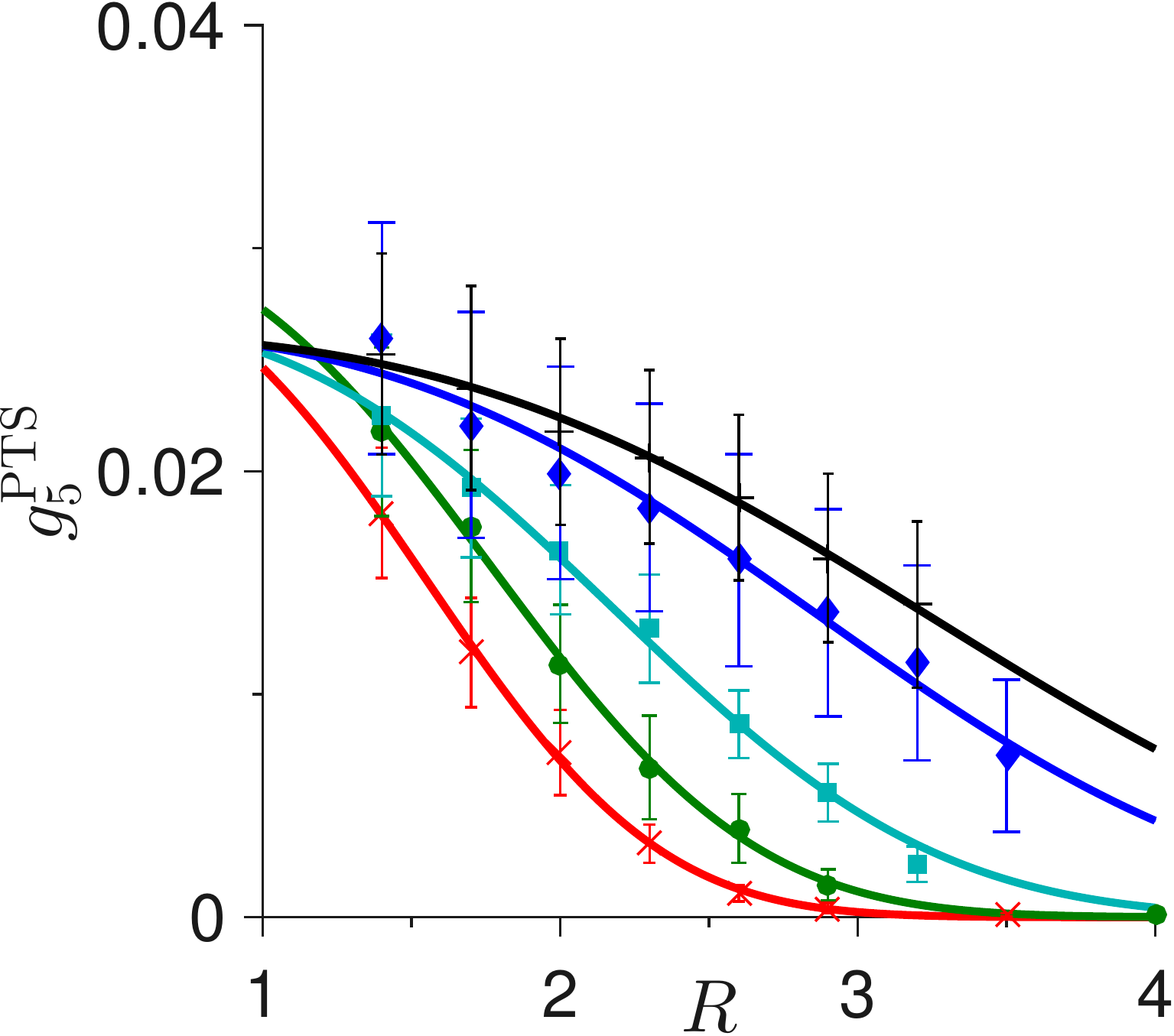}
\includegraphics[width=0.2\textwidth]{PTS_6.pdf}
\includegraphics[width=0.2\textwidth]{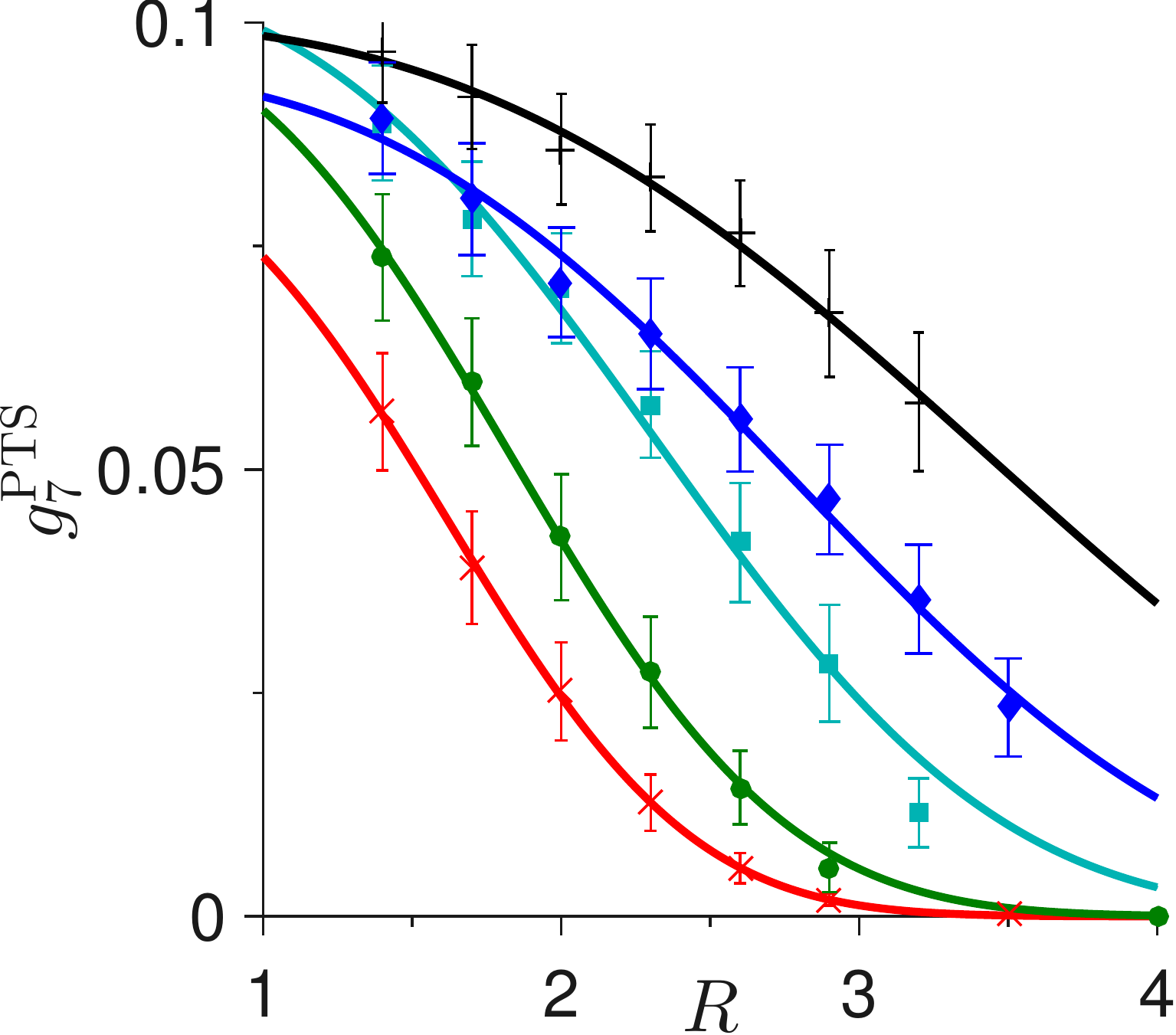}
\includegraphics[width=0.2\textwidth]{PTS_8.pdf}
}
\centerline{
\includegraphics[width=0.2\textwidth]{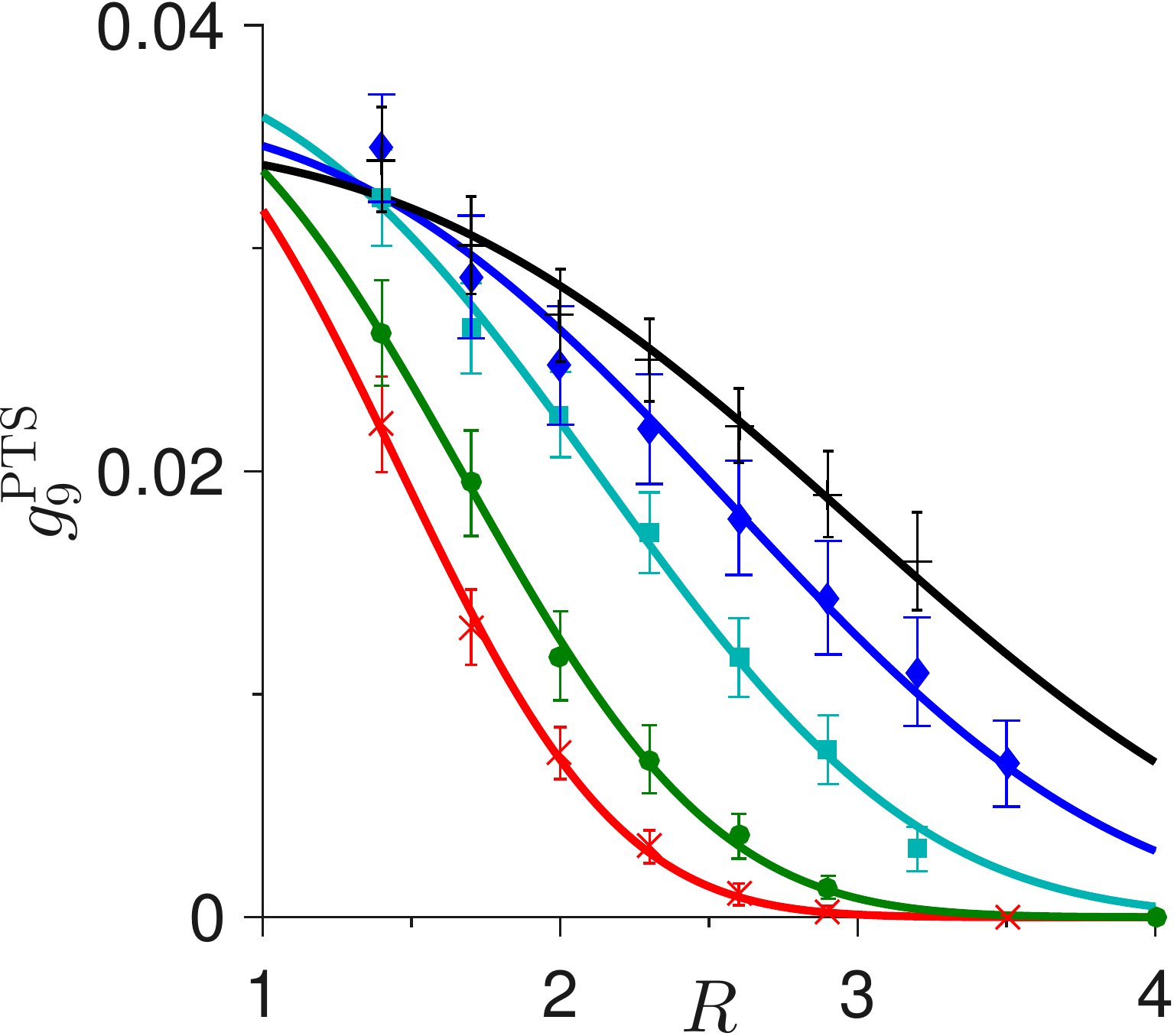}
\includegraphics[width=0.2\textwidth]{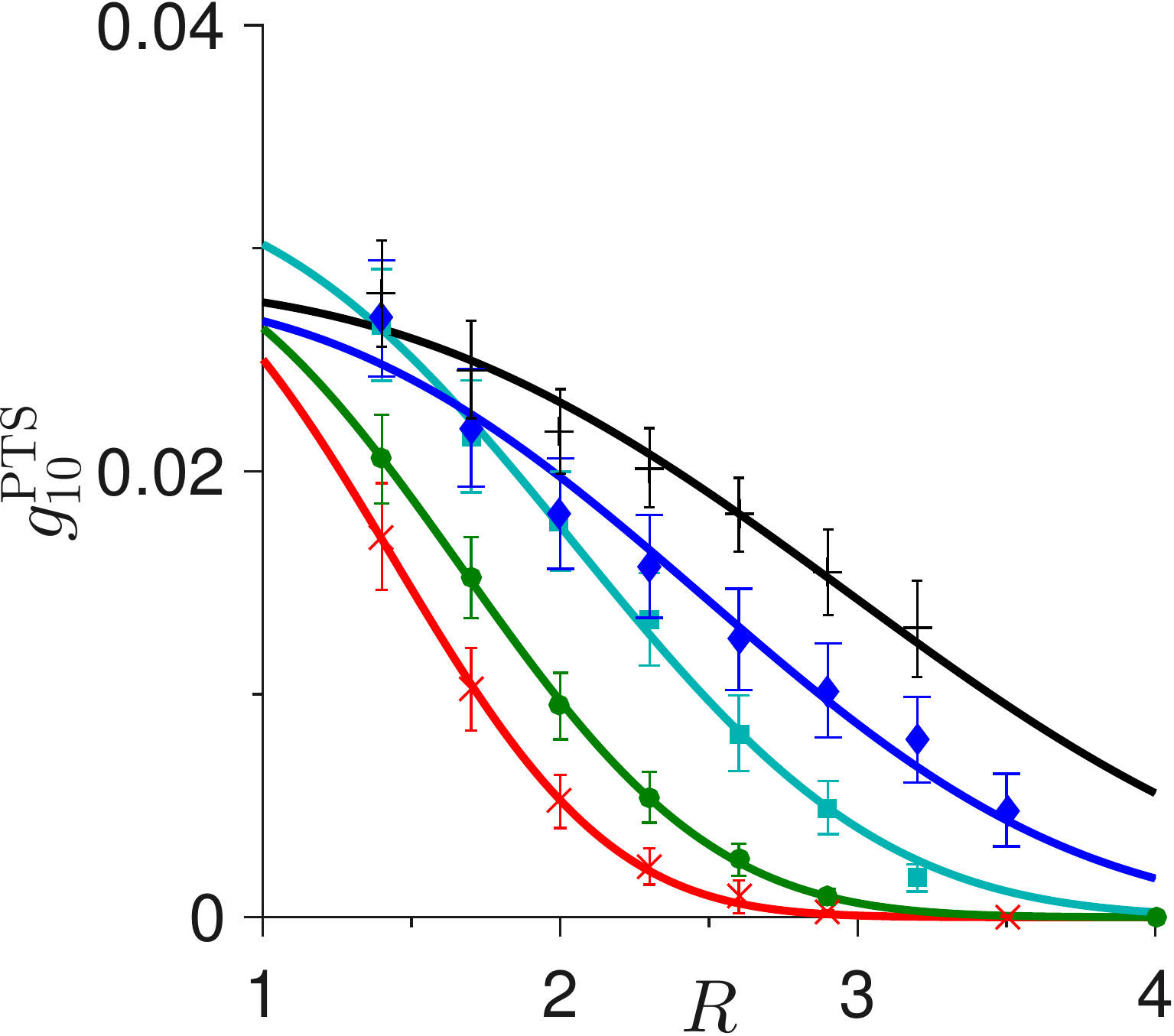}
\includegraphics[width=0.2\textwidth]{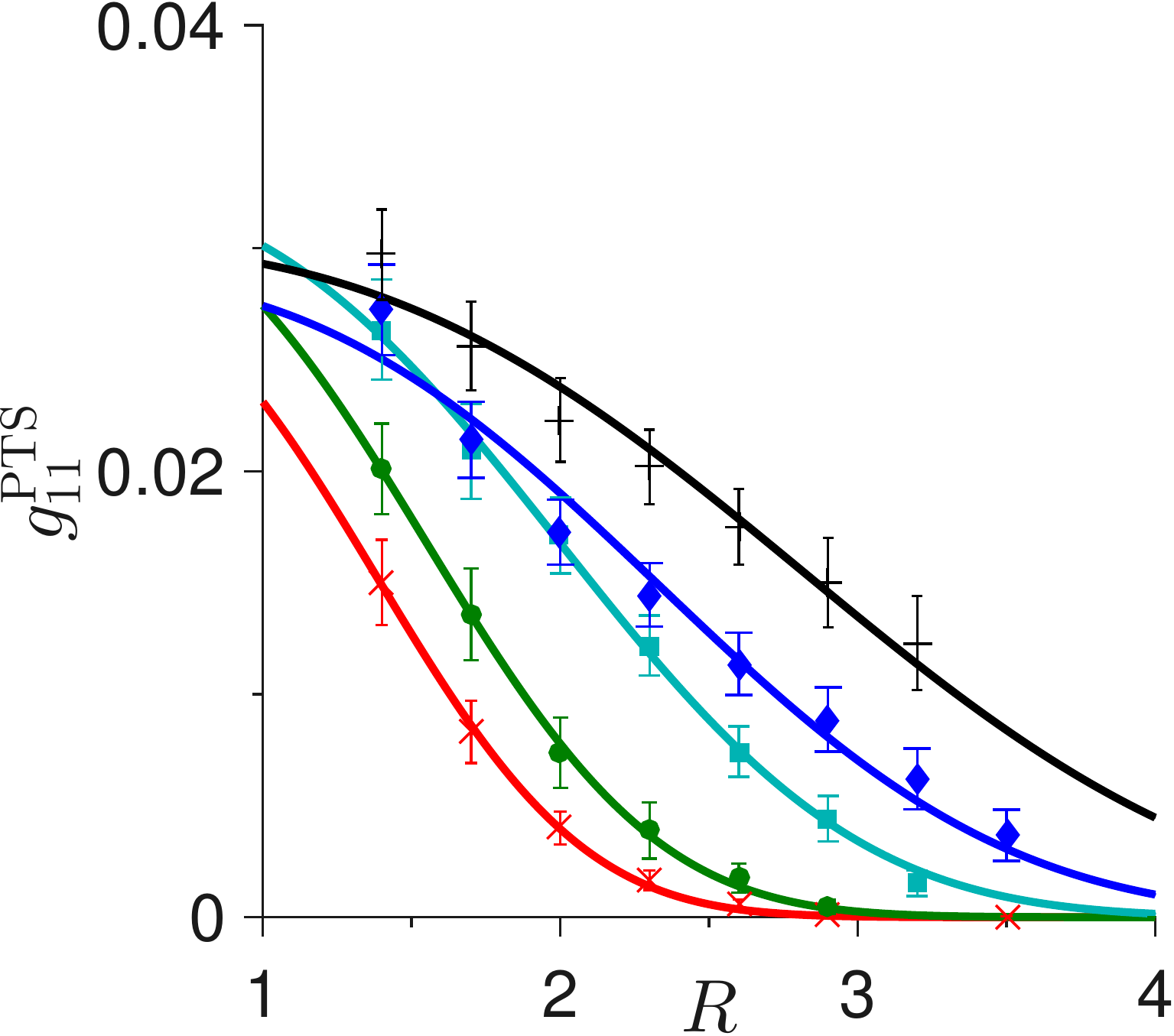}
\includegraphics[width=0.2\textwidth]{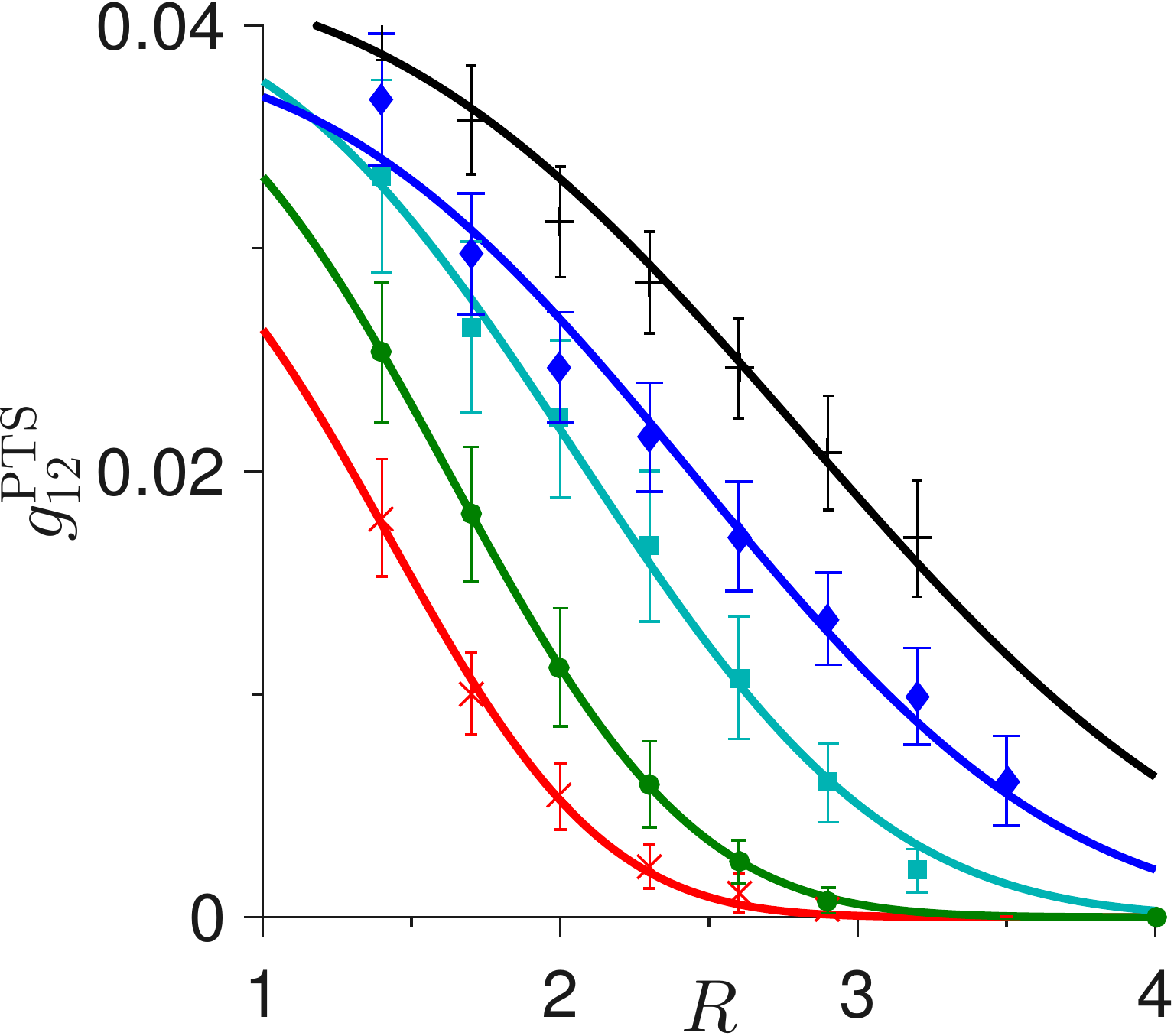}
}
\centerline{
\includegraphics[width=0.2\textwidth]{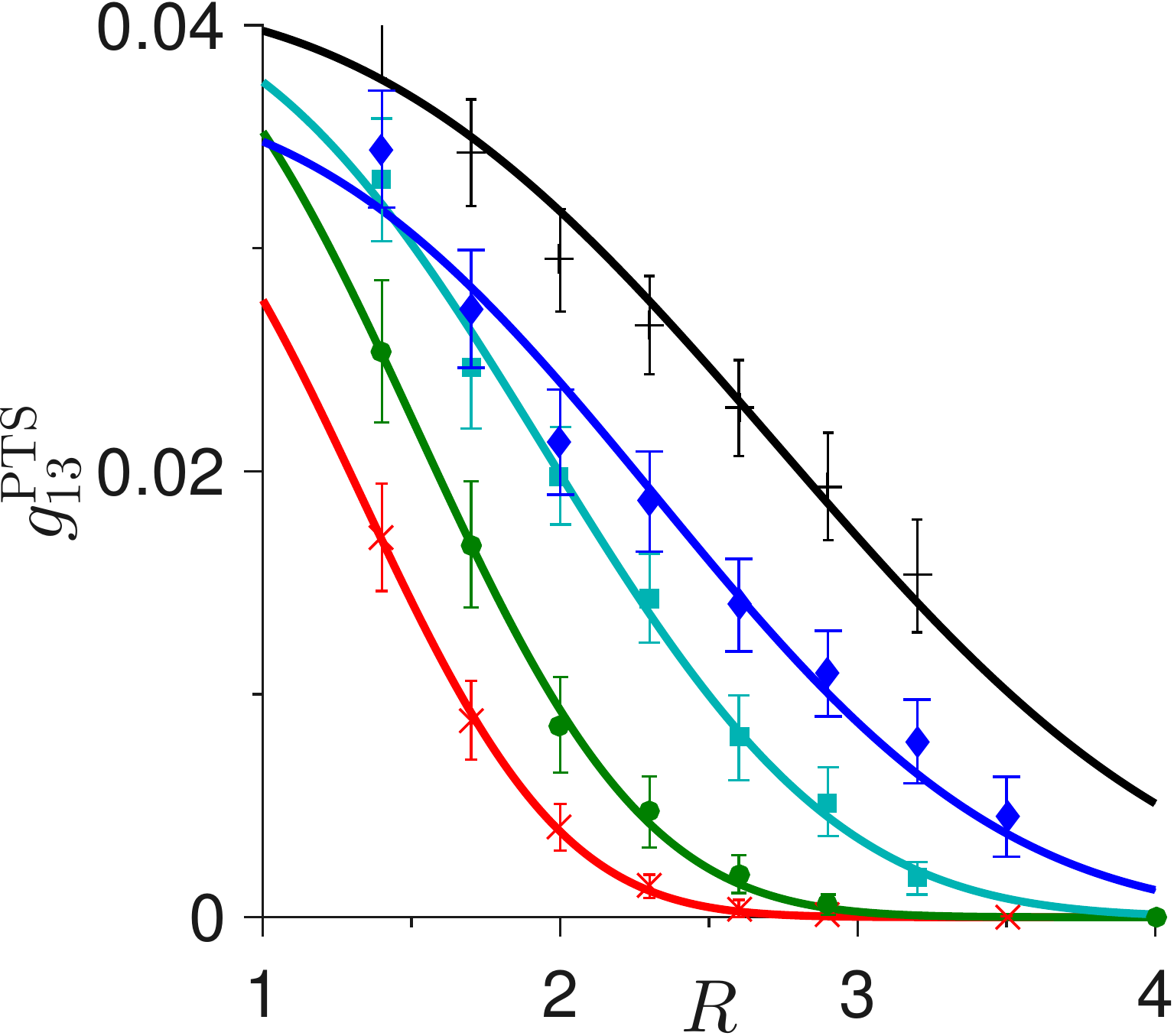}
\includegraphics[width=0.2\textwidth]{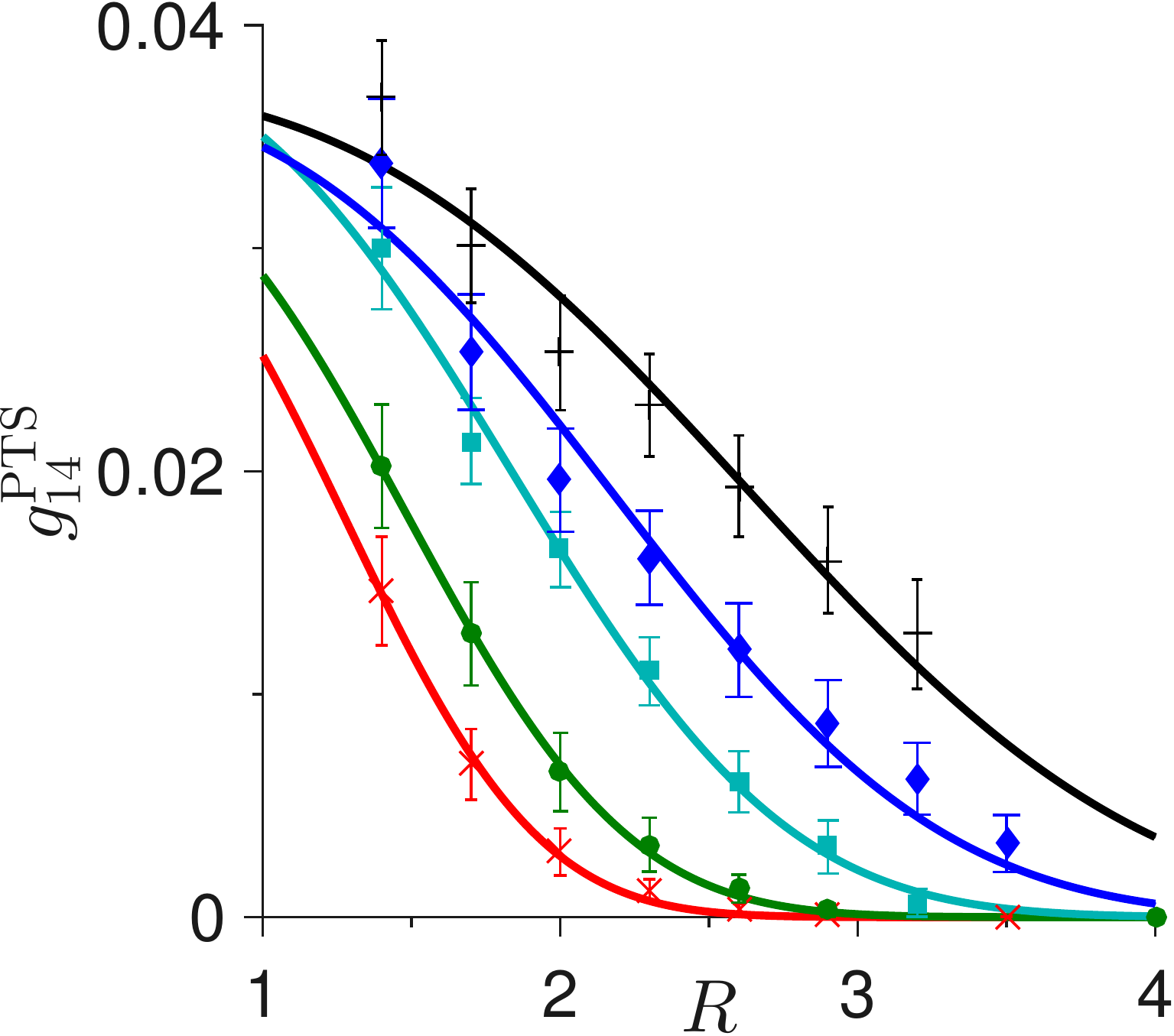}
\includegraphics[width=0.2\textwidth]{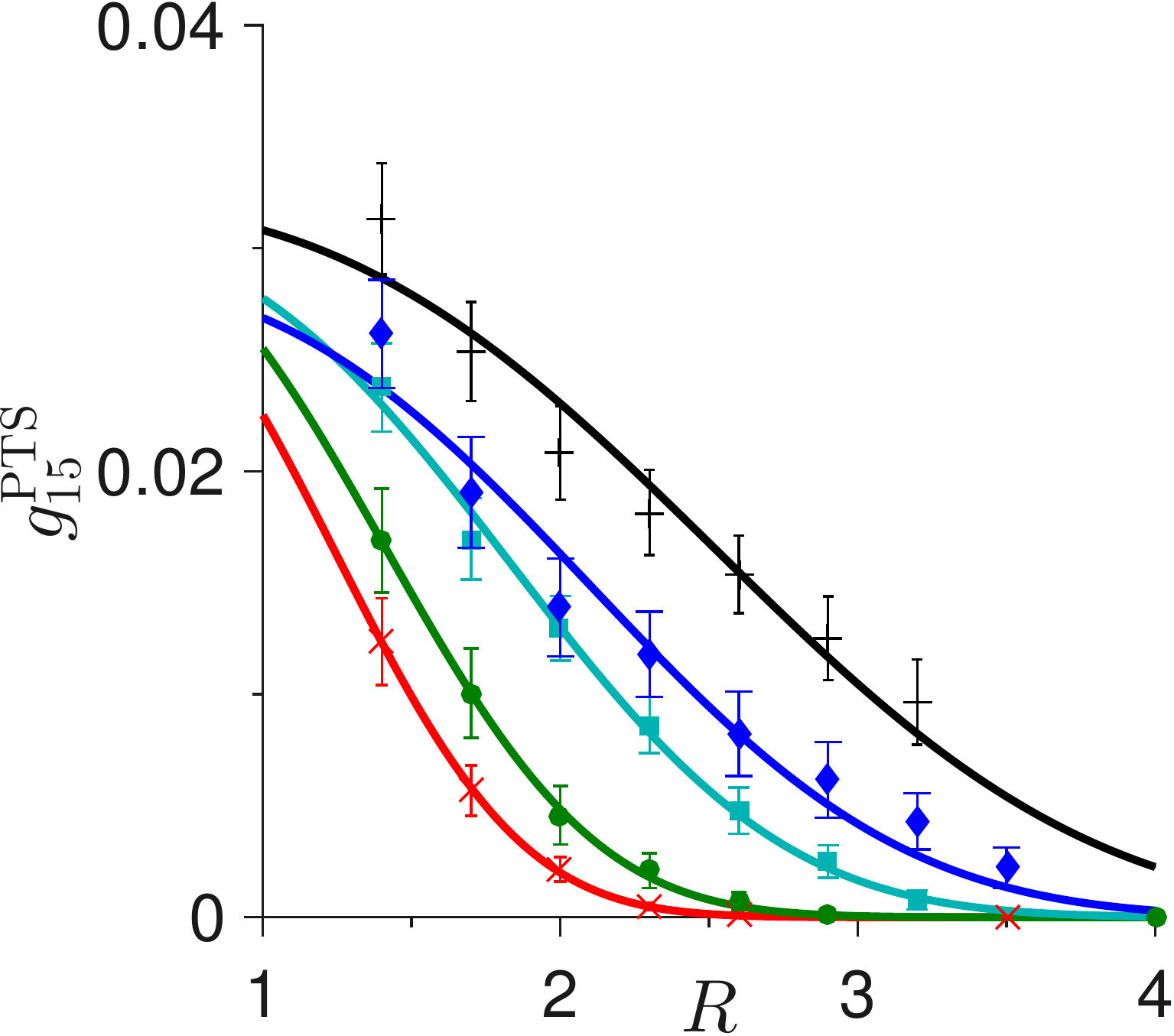}
\includegraphics[width=0.2\textwidth]{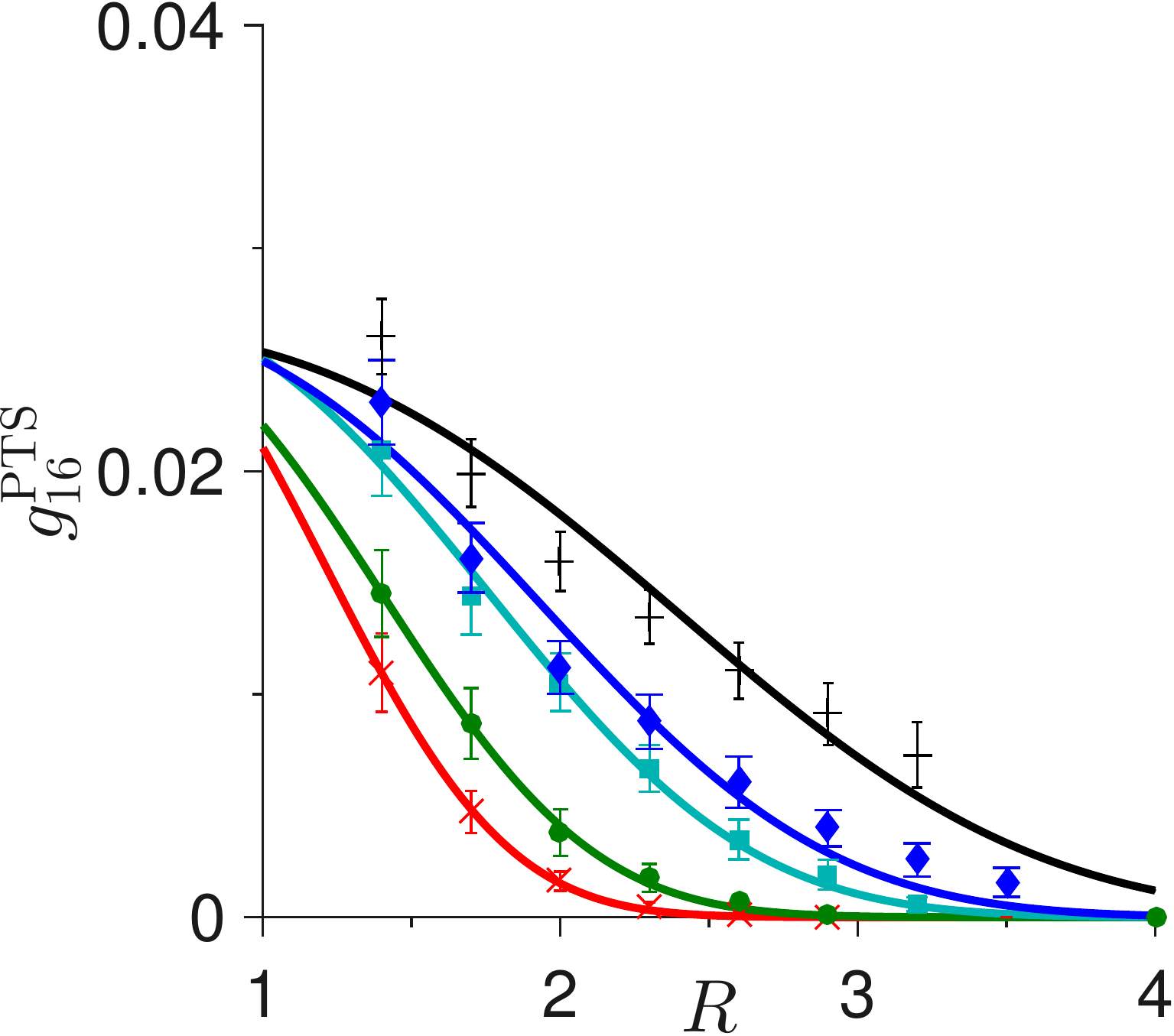}
}
\caption{Radial decay of the bond-orientational PTS correlations for the KABLJ model, $g_{\ell}^{\mathrm{PTS}}(R)$, for $\ell=1,\ldots,16$ at $T = 1.00$ (red-cross), $0.80$ (green-circle), $0.60$ (cyan-square), $0.51$ (blue-diamond), and $0.45$ (black-plus).
Solid lines are compressed exponential fits $g^{\mathrm{PTS}}=A\exp[-(R/\xi^{\rm PTS}_{\rm fit})^{\eta}]$ with $\eta=3$.}
\label{KAPTScatalog}
\end{figure*}

The $d=3$ observables are essentially the same as in the $d=2$ case, except that we employ standard (rather than radical) Voronoi tessellation; angle $\theta$ is replaced by $(\theta,\varphi)$; $e^{im\ell\theta}$ is replaced by spherical harmonics $Y_{l,m}(\theta,\varphi)$; the summation over $m$ is replace by $m=-\ell,\ldots,\ell$; and the prefactor $\prefac\equiv\frac{4\pi}{2\ell+1}$.
This quantity is again real and independent of the choice of axis in defining spherical harmonics. Figure~\ref{KAPTScatalog} lists bond-orientational PTS correlations for $\ell=1$ to $16$ for the KABLJ model.
We see no qualitative differences among different angular components.

\section{Coarse-grained two-point functions}
\label{sec:2ptfunct}
Coarse-grained bond-orientational two-point functions are defined by first taking a point ${\bf r}_1$ in the bulk configuration and another point ${\bf r}_2$, separated by distance $R$ in a randomly-chosen direction and then defining orientational fields $\psi^{{\mathbf I}}_{\ell,m}\le({\bf r}+{\bf r}_1\ri)$ around the first point as before and $\psi^{{\mathbf{II}}}_{\ell,m}\le({\bf r}+{\bf r}_2\ri)$ around the second.
The two-point function is given by
\begin{equation}
\frac{\prefac\sum_{m}}{\le\{2\pi^{d/2} r_{\rm c}^d/\Gamma\le(d/2\ri)\ri\}}\int_{|{\bf r}|<r_{\rm c}}d{\bf r}\ \le\{\psi_{\ell,m}^{\mathbf I}({\bf r}+{\bf r}_1)\ri\}^*\psi_{\ell,m}^{\mathbf{II}}({\bf r}+{\bf r}_2)\, 
\end{equation}
averaged over $200$ different pairs of points for each of $100$ bulk configurations for hard-disk models, and $100$ pairs for $50$ bulk configurations for the KABLJ liquid.

Pair bond-orientational correlation lengths, $\xi_{\ell}$, are extracted through the exponential fit to these coarse-grained orientational two-point functions.
Pair positional correlation length, $\xi_{\rm pos}$, is extracted through the exponential fit to the peak values of the two-point radial correlation function $g(r)$ for $r\geq4$.

Finally, to assess the bulk relaxation time in the hard-disk system, the coarse-grained bond-orientational autocorrelation function, $f_{\ell}(t)$, is defined in the way similar to the corresponding two-point functions defined above. One difference is that, rather than comparing two points separated by distance $R$, we consider two points separated in time $t$. Averages are over $100$ initial bulk configurations and $200$ randomly-chosen points ${\bf r}_1$ within each of them. The relaxation time, $\tau_{\ell}$, is then extracted through the exponential fit to the autocorrelation function $f_{\ell}(t)$. We find $\tau_{\alpha;\ell=6}$ to be the most rapidly growing time scale.

\bibliography{HS,glass}

\begin{thebibliography}{60}
\bibitem [{foo({\natexlab{a}})}]{footnote_HS}%
  \BibitemOpen
  \href@noop\ \bibinfo {note} {A similar,
  activated-like expression holds for fluids of hard objects for which
  temperature is only a trivial scaling factor. The length $\xi$ is then a
  function of density $\rho$ and $1/T$ is replaced by pressure,
  $P(\rho)/T$.}\BibitemShut {Stop}%
\bibitem [{foo({\natexlab{c}})}]{footnote_pinning}%
  \BibitemOpen
  \href@noop \ \bibinfo {note} {Pinning particles at
  random, as often used~\cite{BK12,CCT12,RT15}, may be insufficient to detect
  an incipient bond-orientational order. One should then pin at least pairs of
  nearest neighbors so that bonds are sufficiently constrained by the boundary
  condition.}\BibitemShut {Stop}%
\end{thebibliography}
\end{document}